\documentclass[aps,prx,floats,twocolumn,superscriptaddress,reprint,longbibliography]{revtex4-1}
\usepackage{graphicx,epsfig}% Include figure files
\usepackage{graphics,dcolumn,bm,epic, eepic,fleqn,float}
\usepackage{amssymb,amsmath,amsfonts,multirow,rotate,color}
\usepackage{soul,xcolor}

\usepackage[hidelinks]{hyperref}
\hypersetup{
	colorlinks,
	linkcolor={black!50!black},
	citecolor={blue!50!black},
	urlcolor={blue!80!black}
}	

\usepackage{placeins}
\usepackage[]{algorithm2e}
\usepackage{appendix}

\newcommand{\avg}[1]{\langle #1 \rangle}

\newcommand{\lay}[1]{^{[#1]}}

\newcommand{\complM}[0]{\mathcal{C}(\mathcal{M})}

\def\Erdos{Erd\"os}

\begin{document} 

\title{Algorithmic complexity of multiplex networks}
	
\author{Andrea Santoro}
\affiliation{School of Mathematical Sciences, Queen 
	Mary University of
	London, London E1 4NS, United Kingdom}
\affiliation{The Alan Turing Institute, The British 
	Library, NW1 2DB,
	London, United Kingdom}
\author{Vincenzo Nicosia}
\email[Corresponding author: ]{v.nicosia@qmul.ac.uk}
\affiliation{School of Mathematical Sciences, Queen 
	Mary University of
	London, London E1 4NS, United Kingdom}

\begin{abstract}
  Multilayer networks preserve full information about the different
  interactions among the constituents of a complex system, and have
  recently proven quite useful in modelling transportation networks,
  social circles, and the human brain. A fundamental and still open
  problem is to assess if and when the multilayer representation of
  a system provides a qualitatively better model than the classical
  single-layer aggregated network. Here we tackle this problem from
  an algorithmic information theory perspective. We propose an
  intuitive way to encode a multilayer network into a bit string,
  and we define the complexity of a multilayer network as the ratio
  of the Kolmogorov complexity of the bit strings associated to the
  multilayer and to the corresponding aggregated graph. We find that
  there exists a maximum amount of additional information that a
  multilayer model can encode with respect to the equivalent
  single-layer graph.  We show how our complexity measure
  can be used to obtain low-dimensional representations of
  multidimensional systems, to cluster multilayer networks into a
  small set of meaningful super-families, and to detect tipping
  points in the evolution of different time-varying
  multilayer graphs. Interestingly, the low-dimensional
  multiplex networks obtained with the proposed method also retain
  most of the dynamical properties of the original systems, as
  demonstrated for instance by the preservation of the epidemic
  threshold in the multiplex SIS model. These results suggest
  that information-theoretic approaches can be effectively employed
  for a more systematic analysis of static and time-varying
  multidimensional complex systems.
\end{abstract}

\maketitle 

\section{Introduction}

The success of network science in modelling real-world complex
systems~\cite{Newman_book2010,Latora_Nicosia_Russo_book2017} relies
on the hypothesis that the interconnections among the elementary
units of a system --i.e., the network of their interactions-- are
responsible for the emergence of complex dynamical
behaviours~\cite{Pastor-Satorras2015,Arenas2008}.  Traditionally,
relevant contributions towards a better understanding of complex
networks have come from statistical
physics~\cite{Jaynes1957information,Bianconi_2008,Anand2009entropy},
where the main aim is to characterise the ensembles of random graphs
comparable with an observed real-world network.  However, really
interesting results have also come from information theory. A quite
prolific line of research in this area aims at adapting classical
concepts and methods from information theory to networks
analysis~\cite{Dehmer2008information,Passerini2009,Mowshowitz2012entropy}.
Some other studies have focused instead on the definition of entropy
measures on empirical
networks~\cite{Dehmer_history_entropy_2011,Cimini2019statistical},
and on the quantification of the significance of structural
indicators based on algorithmic information
theory~\cite{Morzy2017,Zenil2018review}.

Multi-layer and multiplex networks, which take into account
different kinds of relations among the same set of nodes at the same
time~\cite{Arenas_2013,Boccaletti_2014,Bianconi2018multilayer}, are
a currently hot research topic in network science.  The main idea
behind the investigation of high-dimensional network representations
is that retaining full information about the structure of a system
under study is often fundamental to fully understand its
behaviour. Indeed, multi-layer networks have helped unravelling
interesting structural properties in transportation
systems~\cite{Cardillo_2013,Gallotti2016} and
neuroscience~\cite{DeDomenico2016mapping,Battiston2017multilayer},
and have revealed qualitatively new emerging phenomena, including
abrupt cascading failures~\cite{Havlin_2010},
super-diffusion~\cite{Arenas_Diffusion_2013}, explosive
synchronisation~\cite{Nicosia_Skardal_2017}, hyperfast
spreading~\cite{Soriano_Gardenes_2018,Granell2013dynamical,Gleeson_Moreno_2016,DeDomenico2016physics}.

These encouraging results have transformed our understanding of many
physical systems, but an overarching question remains about whether
it is necessary to incorporate all the available data about a system
in order to fully characterise its
behaviour~\cite{Lacasa_multiplex_2018}. Some recent studies have
indeed shown that the multi-layer version of some dynamical
processes cannot be reduced to the corresponding single-layer
process on any simple combination of the existing
layers~\cite{Diakonova_2016}. Nevertheless, determining
whether a lower-dimensional multi-layer network can exhibit the
same structural and dynamical richness of the full multi-layer
graph is still an open question. Some concrete attempts
to solve this problem have come from a formalisation of
multi-layer dimensionality reduction in terms of a quantum
information problem~\cite{DeDomenico_Nicosia_2015}, and from other
approaches relying on mesoscopic similarity between
layers~\cite{Iacovacci_Wu_2015,Kao2018layer,Stanley2016clustering,DeBacco2017community}.
However, we still lack a convincing method to quantify the amount of
information contained in a multi-layer network model, and to compare
the information content of different multi-layer networks.

\begin{figure}[!ht]
	\centering
	\includegraphics[width = 3.1in]{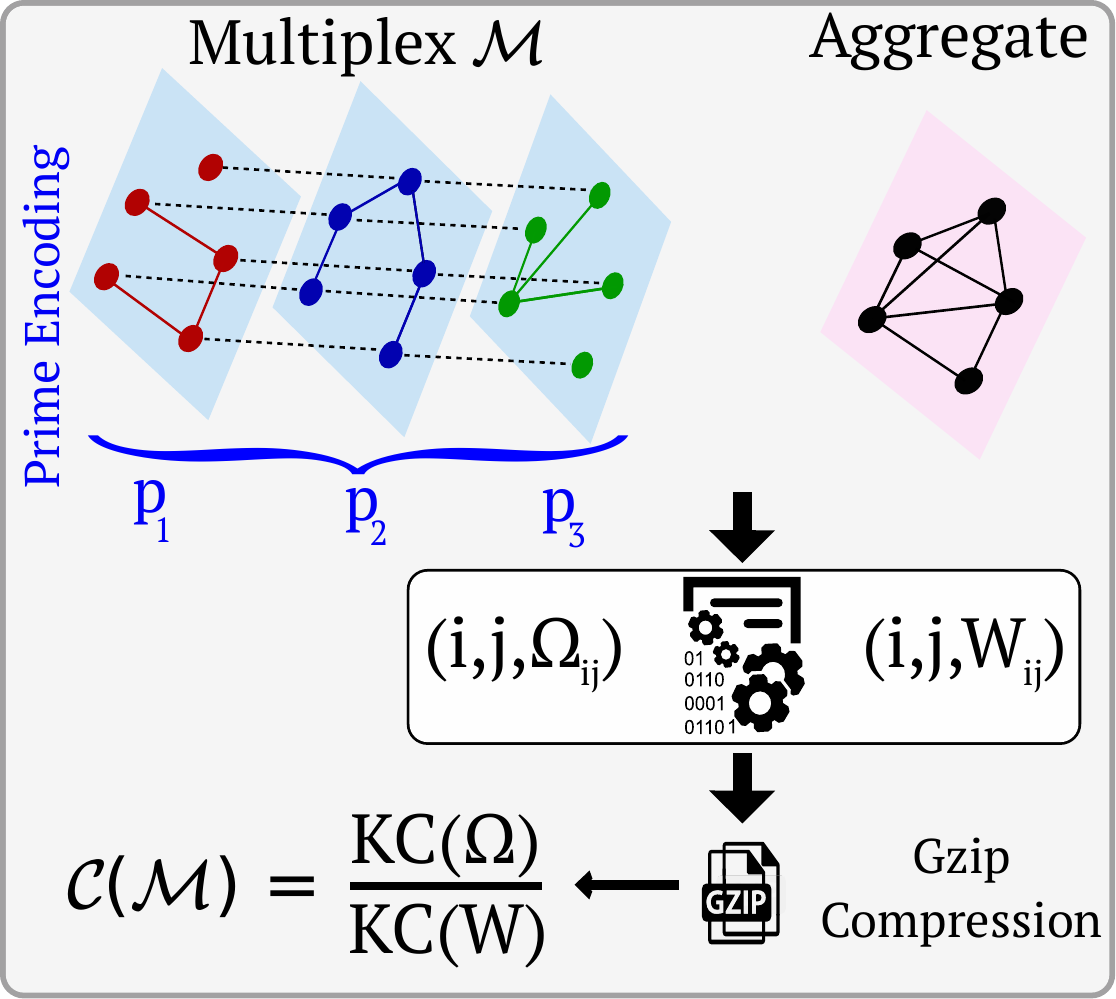}
	\noindent
	\caption{(Colour online) The complexity $\complM$ of a multiplex
    network $\mathcal{M}$ is defined as the ratio between its
    Kolmogorov complexity and the Kolmogorov complexity of the
    associated single-layer aggregated graph. The multiplex is
    transformed into a string of bits by means of the prime-weight
    matrix. Since Kolmogorov complexity is not computable, we rely
    on an upper-bound based on the size of the compressed string of
    bits associated to each object. A common way to obtain an upper
    bound is by computing the length of the string compressed
    through the gzip algorithm.}
\end{figure}

In this paper, we take an algorithmic complexity perspective on this
problem, and we define the complexity measure
$\mathcal{C}(\mathcal{M})$ to quantify the amount of information
contained in a multiplex network $\mathcal{M}$. The measure
leverages the classical concept of Kolmogorov
complexity~\cite{Kolmogorov_1998}, according to which the complexity
of a bit string is equal to the length of the shortest possible
program that can produce that string as its output. In particular,
we show that $\complM$ is quite useful in determining the optimal
number of layers needed to represent a multi-layer network, and in
detecting structural and dynamical similarities among
multi-layer networks from different domains.

\section{Results}
We propose here a formalism to quantity the complexity of a
multiplex network over $N$ nodes and $M$ layers, based on the
comparison of the Kolmogorov complexity of the multiplex and of
the corresponding aggregated graph. We start by encoding the
unweighted multiplex network $\mathcal{M}$ into the $N\times N$
\textit{prime-weight matrix} $\Omega$ defined as follows:
\begin{equation}
  \Omega_{ij} =
	\begin{cases}
	  \displaystyle \prod_{\alpha: a_{ij}\lay{\alpha}=1}
    p\lay{\alpha} \qquad\\ \quad\;0 \quad \qquad \,\textrm{if }
    a_{ij}\lay{\alpha}=0\quad \forall \alpha=1,\ldots, M
	\end{cases}
\end{equation}
The prime-weight matrix is obtained by assigning a distinct prime
number $p\lay{\alpha}$ to each of the $M$ layers of the multiplex,
and then setting each element $\Omega_{ij}$ equal to the product of
the primes associated to the layers where an edge between node $i$
and node $j$ actually exists. This procedure can be easily
generalised to the case of weighted multiplex networks with integer weights, as
explained in Appendix~\ref{appendix:prime_weight_encoding}.  Note
that, given a certain assignment of prime numbers to the $M$ layers,
the matrix $\Omega$ is uniquely determined. Moreover, thanks to the
unique factorisation theorem, the prime-weight matrix preserves full
information about the multiplex network $\mathcal{M}$, i.e., about
the placement of all its edges.

\subsection{Complexity of multiplex networks} 
We define the complexity of a multiplex network $\mathcal{M}$ with
$N$ nodes and $M$ layers as the ratio:
\begin{equation}
	\mathcal{C}\left(\mathcal{M}\right) =
	\frac{KC\left(\Omega\right)}{KC\left(W\right)},
	\label{eq:information_complexity}
\end{equation}
where the numerator is the Kolmogorov
complexity~\cite{Kolmogorov_1998} of $\mathcal{M}$ and the
denominator is the Kolmogorov complexity of the weighted aggregated
graph associated to $\mathcal{M}$. In particular, the matrix
$\Omega$ is the \textit{prime-weight matrix} representation of
$\mathcal{M}$, while $W$ is the single-layer network obtained by
aggregating all the $M$ layers. We compute an approximation of the
Kolmogorov complexity of a matrix by looking at the size of the
compressed weighted edge list (see
Appendix~\ref{appendix:multiplex_complexity} and Supplementary
Information S-1.5 for details). The measure of
complexity in Eq.~(\ref{eq:information_complexity}) effectively
quantifies the relative amount of additional algorithmic information
needed to encode the multiplex network with respect to the amount
needed to encode the corresponding single-layer aggregated graph. As
a particular case, $\complM = 1$ if the multiplex network consists
of $M$ identical layers, but in general $\complM\ge 1$, since the
different possible arrangements of edges across the layers require
more than one symbol to be encoded. The main hypothesis is that the
higher the value of $\complM$, the larger the amount of information
lost when representing the multiplex as a single-layer graph. In
practice, if $\complM \approx 1$ it would not make much difference
to represent the multiplex as a single-layer graph, since the
multiplex representation is not adding much more information.
Conversely, when $\complM > 1$ the aggregation of the multiplex into
a single-layer graph would discard relevant information, and the
larger the value of $\complM$ the more important it is to retain the
full multiplex model. The code
for computing the complexity of a multiplex network is available at \cite{Paper_code}.

\subsection{Synthetic multiplex networks}
The fundamental ingredients contributing to the complexity of a
multiplex network $\mathcal{M}$ as quantified by $\complM$ are the
number of distinct pairs of nodes connected by an edge, and the
actual number of distinct symbols present in the prime-weight matrix
$\Omega$. In fact, both a larger number of connected pairs of nodes
and a larger number of distinct symbols will in general result in a
larger encoding, and a larger value of $\complM$. Indeed, the number
of symbols present in $\Omega$ is equal to the number of different
multiplex motifs with two nodes~\cite{Battiston2017multilayer}.
\begin{figure}[!ht]
	\begin{center}
    \includegraphics[width=3in]{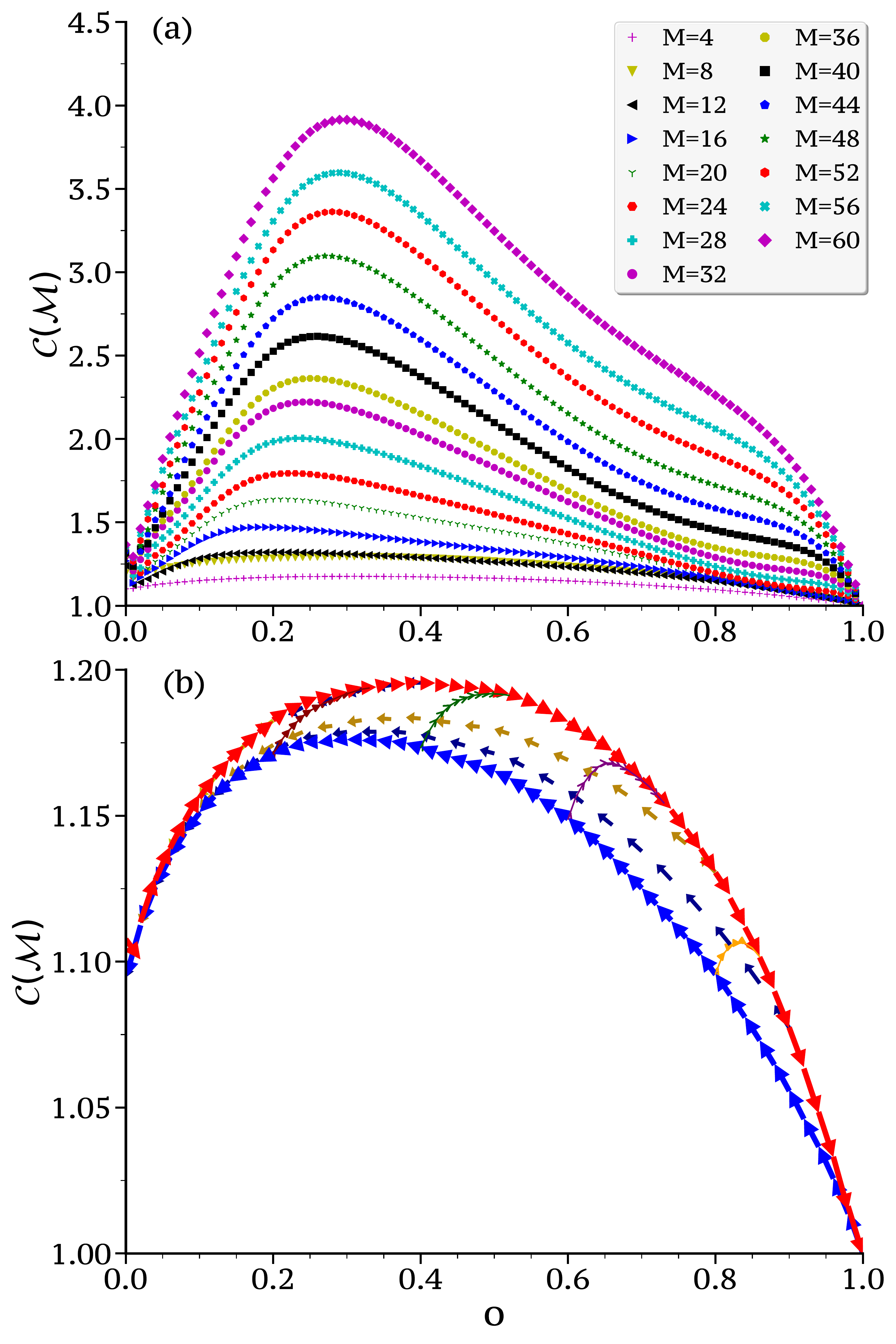}
  \end{center}
	\caption{ (Colour online) (a) Complexity $\complM$ of ensembles of
    synthetic multiplex networks with a variable number of layers
    $M$ and tunable structural overlap $o$. Each layer is an
    \Erdos-Renyi graph with $N = 10000$ nodes and average degree
    equal to $\avg{k} = 6$. Irrespective of the number of layers
    $M$, the complexity $\complM$ always has a maximum for values of
    structural overlap in the interval $[0.15, 0.4]$, which are
    compatible with the typical values of overlap measured in
    real-world networks~\cite{Diakonova_2016} (see SI
    Section S-1.3 for additional results on different synthetic
    networks). (b) Hysteresis loop of $\complM$ for a multiplex
    with $N = 10000$ nodes, with $M = 4$ layers having each average
    degree $\langle k \rangle = 6$. Red arrows represent the
    trajectory observed when the structural edge overlap is reduced,
    while the blue arrows indicate the trajectories observed when
    the structural edge overlap is increased.}
	\label{fig:synthetic_networks}
\end{figure}

The structural edge overlap $o$ is a an easy-to-compute proxy for
the variety of different multi-edge configurations in a multiplex
network (see Appendix~\ref{appendix:structural_edge_overlap}). In
order to understand the effect of edge overlap on $\complM$, we
considered ensembles of synthetic multiplex networks with different
number of layers, where the total number of nodes and the average
node degree on each layer are kept fixed ($N=10000$, $\langle k
\rangle = 6$), while the structural edge overlap $o$ is tuned as
explained in Appendix~\ref{appendix:synthetic_networks}. The results
are shown in Fig.~\ref{fig:synthetic_networks}(a). As expected,
$\complM=1$ in multiplex networks with $M$ identical layers
($o=1$). Indeed, when we start rewiring the edges of a multiplex
with $M$ identical layers, thus reducing the value of structural
overlap, we expect $\complM$ to increase, since the prime-weight
matrix contains a larger number of symbols. Conversely, when $o
\approx 0$ each edge exists on exactly one of the $M$ layers,
meaning that the number of distinct edges in the multiplex is
roughly equal to the number of distinct edges in the aggregated
graph. As a consequence, we expect the values of their Kolmogorov
complexity to not differ too much, and the corresponding value of
$\complM$ to be somehow close to $1$. This is exactly what we
observe in Fig.~\ref{fig:synthetic_networks}(a), respectively for
$o\approx 1$ and for $o \approx 0$.  However, the most interesting
result is that $\complM$ is a non-monotonic function of the
structural edge overlap, for any value of $M$. In particular, it is
evident from Fig.~\ref{fig:synthetic_networks}(a) that $\complM$
always has a maximum for $o \,\in \, [0.15, 0.4]$, indicating that
there exists indeed a maximum amount of additional information that
a multiplex can encode with respect to the corresponding aggregate
graph. We find it quite remarkable that the range at which $\complM$
peaks is compatible with the typical values of structural edge
overlap observed in many real systems~\cite{Diakonova_2016}.

\textit{Structural hysteresis. -- } In order to fully explore the
behaviour of the complexity $\mathcal{C}$, we considered an ensemble
of synthetic multiplex networks where we iteratively decrease and
increase the structural overlap $o$ (see
Appendix~\ref{appendix:synthetic_networks} for details).
Interestingly, we found two robust and distinct trajectories when
the structural overlap is decreased (resp. increased), characterised
by a hysteresis loop (Fig.~\ref{fig:synthetic_networks}(b)). In the
simulations, we start from a multiplex network with $N=10000$ nodes
and $\langle k \rangle = 6$, where the $M=4$ layers are identical
\Erdos-Renyi random graphs, and we iteratively rewire the links in
order to decrease the total edge overlap until we obtain a multiplex
network with $o=0$. After that, we successively increase the
structural overlap until the system results in a multiplex network
with $o=1$.  Remarkably, $\complM$ remains a non-monotonic function
of the structural edge overlap, but the plot of $\complM$ as a
function of $o$ reveals the presence of a structural hysteresis,
i.e., the trajectory leading from $o=1$ to $o=0$ is different from
the one obtained when the structural overlap is increased from $o=0$
to $o=1$. This indicates that the procedures used to decrease and
increase edge overlap are not ergodic, due to the intrinsic
difference between the way overlap is created and destroyed. Indeed,
if we start from a multiplex $\mathcal{M}$ with identical layers,
the total number of ways in which a random rewiring can reduce the
overlap is significantly larger than the total number of ways in
which the edge overlap can be increased through a random
move. Similar results are found when the graph on each
layer is a regular or scale-free degree distribution, as shown in
Supplementary Figure S-3. This result
indicates that some caution is required when rewiring the edges of
a multi-layer graph, which is a problem we will explore in a
future work.
\begin{figure*}[!ht]
	\begin{center}
    \includegraphics[width=6in]{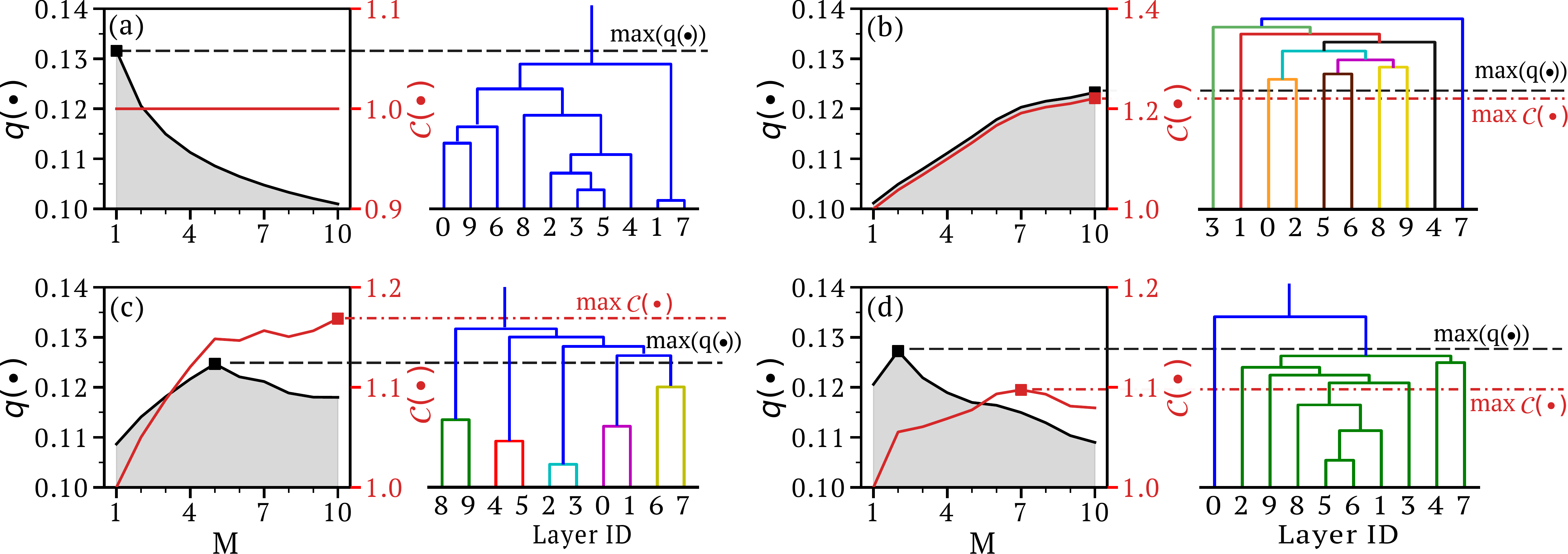}
  \end{center}
  \caption{(Colour online) Reducibility of four synthetic
    multiplex networks with $N=1000$ nodes, $\avg{k}=4$ and $M=10$
    layers. The four panels correspond to: (a) ten identical
    layers; (b) ten distinct layers, (c) five pairs of identical
    layers; (d) 9 identical layers and 1 distinct layer. Notice
    that all the distinct layers are independent realisations of
    \Erdos-R\'{e}nyi random graphs.  For each synthetic multiplex
    we show both the global quality function $q(\bullet)$ and the
    complexity $\mathcal{C}(\bullet)$ as a function of the number
    of layers $M$ (left panel) and the dendrogram resulting from
    the greedy aggregation steps of the reducibility procedure
    (right panel). Notice that the maximum of the quality function
    $q$ corresponds to the true partition of layers, while the
    complexity $\mathcal{C}$ generally fails to identify the
    correct partition.}
	\label{fig:reducibility_synthetic_multiplex}
\end{figure*}

\subsection{Multiplex complexity and reducibility}

One of the main issues of multi-dimensional data sets is that they
normally contain redundant information. Consequently, the direct
transformation of each type of relation available in a
multi-dimensional data set into a distinct layer of a multiplex
network will possibly result in a structurally redundant
representation of the original system. However, dealing with
parsimonious models is always desirable, and is especially important
in the case of multi-dimensional systems, where additional model
complexity usually yields additional computational costs and
raises some questions about the interpretability of the
results. The ``multiplex reducibility problem'',
originally formulated in Ref.~\cite{DeDomenico_Nicosia_2015}, is the
problem of finding low-dimensional representations of a multiplex
network which preserve as much structural information as possible
about the original system.

The Multiplex complexity $\complM$ that we have defined provides a
natural and meaningful way of obtaining reduced (low-dimensional)
versions of a multiplex networks over $M$ layers. If we start from
the original multiplex network $\mathcal{M}$ and we aggregate some
of its layers, we obtain a reduced multiplex network $\mathcal{X}$
with $X\le M$ layers, which will have a multiplex complexity
$\mathcal{C}(\mathcal{X})$. We propose to quantify the normalised
information content of the reduced multiplex network $\mathcal{X}$
as:
\begin{equation}
	q(\mathcal{X}) = \frac{\mathcal{C}(\mathcal{X})}{\log K_\mathcal{X}}
	\label{eq:quality_function}
\end{equation}
where $K_\mathcal{X}$ is the number of distinct links in the
multiplex $\mathcal{X}$. The normalisation by $\log K_{\mathcal{X}}$
is necessary, since in general the Kolmogorov complexity of a bit
string of length $n$ is not smaller than $c + \log n $, for some
$c\ge 0$~\cite{Delahaye2012numerical}. The length of the bit string
associated to a multiplex network is proportional to the number of
distinct links in the multiplex hence, on average, a multiplex with
a larger number of edges is expected to have a higher multiplex
complexity. This is an inconvenience for the multiplex
reducibility problem, since a network with a larger number of
layers would in principle have a larger number of edges as
well. The normalisation used in $q(\mathcal{X})$ allows us to
safely compare alternative low-dimensional multiplex networks
obtained from the same original system.
\begin{figure*}[!ht]
	\centering
  \includegraphics[width=6in]{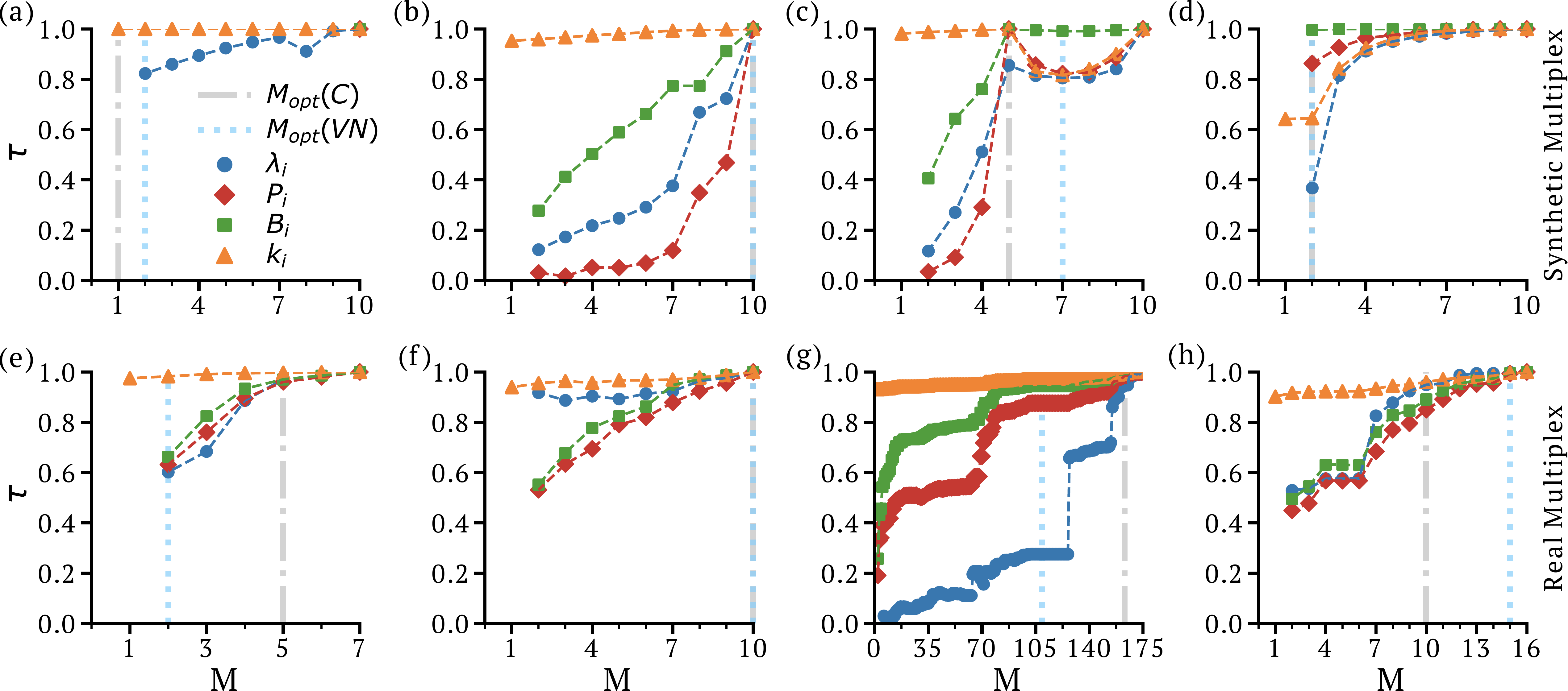}
	\caption{ (Colour online) Kendall's $\tau$ correlation between the
    rankings induced by four structural descriptors in
    synthetic and real multiplex networks and the corresponding
    reduced networks with M layers. Panels (a)-(d) are for the
    same synthetic benchmarks shown in
    Fig.~\ref{fig:reducibility_synthetic_multiplex}. The panels on
    the bottom row correspond to real-world systems, namely (e)
    Saccharomyces Pombe, (f) APS scientific
    collaborations among countries, (g) European Airports, and
    (h) Pierre Auger scientific collaboration. The multiplex
    structural descriptors considered here are total node degree
    ($k_i$), node activity ($B_i$), participation coefficient
    ($P_i$), and node interdependence ($\lambda_i$).  For each
    network we also indicate the number of layers in the optimal
    reduced system identified by multiplex complexity
    ($M_{opt}(C)$, grey dash-and-dot line) and by von Neumann
    entropy ($M_{opt}(VN)$, blue dotted
    line)~\cite{DeDomenico_Nicosia_2015}. In all the synthetic
    cases, the quality function $q(\bullet)$ in
    Eq.~\ref{eq:quality_function} correctly identifies the optimal
    partition as the one where identical layers are aggregated,
    thus preserving the structural properties of the original
    multiplex. Conversely, in some cases the partitions found
    using the von Neumann entropy are different from the expected
    ones. Notice that for the benchmark made of only identical
    layers (a), the correlation coefficient is not defined for two
    structural descriptors (constant behaviour). For all the
    real-world networks the multiplex complexity provides a more conservative
    representation of the system and retains most of the properties
    of the original graph ($\tau > 0.8$ for all the
    structural descriptors), while the reduction based on von
    Neumann entropy might probably discard important information
    and/or create structural artefacts.}
	\label{fig:structmeasure_reducibility}
\end{figure*}

Notice that $q(\mathcal{X})$ behaves as a quality function, meaning
that larger values of $q(\mathcal{X})$ indicate that the (possibly
reduced) multilayer network $\mathcal{X}$ encodes a relatively
larger amount of information with respect to the corresponding
weighted aggregated graph $W_\mathcal{X}$. Hence, our goal is to
find argmax $[q(\mathcal{X})]$, which represents the optimally
reduced multiplex $\mathcal{X}_\textrm{max}$ yielding the maximum
value of information with respect to the aggregated graph. In
particular, if all the layers of the multiplex network $\mathcal{X}$
are identical, then the maximum of $q$ will always be at $1=X \leq
M$ layers, since the multiplex network and the aggregate graph are
equivalent.

Maximising $q(\mathcal{X})$ by enumerating all the possible
partitions of the $M$ layers is not feasible, since that number
increases super-exponentially with $M$. Hence, here we employed a
classical agglomerative greedy algorithm to approximate the optimal
solution (see Appendix \ref{appendix:reducibility}). The code for obtaining lower-dimensional representations of a multiplex network using $q(\bullet)$ is available at \cite{Paper_code}.

We started by testing the algorithm on ad-hoc synthetic multiplex
networks where some of the $M$ layers are identical, thus reducing
the number of truly distinct layers by construction.  In
Fig.~\ref{fig:reducibility_synthetic_multiplex} we report the
results of the greedy reduction on four different synthetic
benchmarks. In particular, we plot the global quality function
$q(\bullet)$ and the complexity $\mathcal{C}(\bullet)$ as a
function of the number of layers $M$ (left panels), and the
dendrogram corresponding to the greedy aggregation steps (right
panels). In all the cases considered, the maximum of the quality
function $q(\bullet)$ correctly identifies the partition made of
truly distinct layers, while in general the complexity
$\mathcal{C}(\bullet)$ fails to identify the correct partition
(see Supplementary Information Section S-2 for results
on a wide set of synthetic benchmarks). This confirms our
intuition that, by taking into account differences in the total
number of edges of the multiplex, the quality function
$q(\bullet)$ does a better job at discriminating between essential
and redundant information. Indeed, while the multiplex with $M=10$
identical layers is always aggregated into a single-layer graph
(Fig.~\ref{fig:reducibility_synthetic_multiplex}(a)), in the
multiplex with all distinct layers the maximum of $q(\bullet)$ is
attained by the initial configuration with ten layers
(Fig.~\ref{fig:reducibility_synthetic_multiplex}(b)).

After having checked that $q(\bullet)$ identifies
meaningful layer partitions in synthetic multiplex networks, we
extended our analysis to real-world multiplex data sets. The
results are reported in Table~\ref{table:reducibility}. Notice that
most of the technological and biological multiplex networks in the
Table admit reduced representations which have only a slightly
smaller number of layers than the original systems. This is in
agreement with the observation that in technological systems
structural redundancy is purposedly avoided. Similarly, the poor
redundancy observed in biological multiplex networks is in line with
the functionally different role played by each layer (protein
interaction, functional dependence, mechanical interaction, and so
on). However, technological systems exhibit consistently larger
values of multiplex complexity than biological systems. A comparison
with the reducibility algorithm proposed in
Ref.~\cite{DeDomenico_Nicosia_2015} shows that our definition of
multiplex complexity is in general more conservative, and often
yields an optimal partition that has a slightly larger number of
layers.

\begin{figure*}[!ht]
	\centering
  \includegraphics[width=6in]{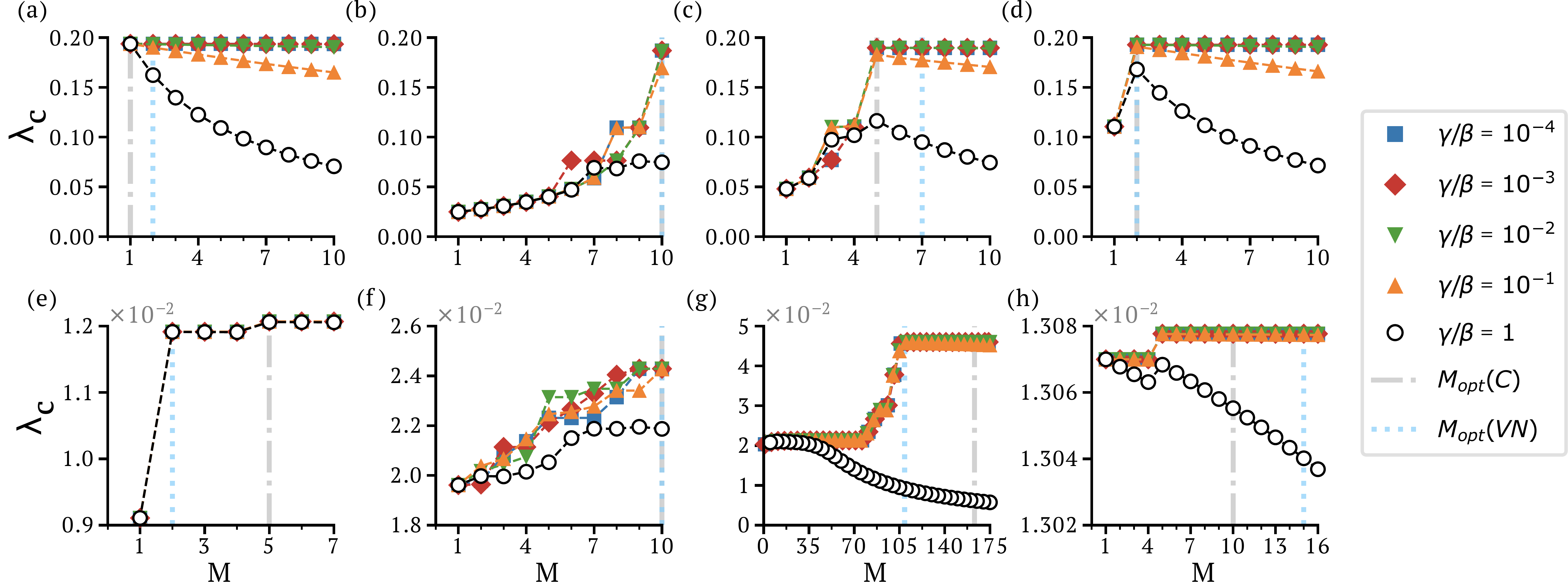}
	\caption{(Colour online) Epidemic threshold
    $\lambda_c$ for the multiplex SIS
    dynamics~\cite{Cozzo2013_SIS} as a function of the number of
    layers $M$ and for different values of the contagion parameter
    $\gamma/\beta$, on the same synthetic and real multiplex
    networks shown in Fig.~\ref{fig:structmeasure_reducibility}.
    For each network we indicate the number of layers in the
    optimal reduced system identified by multiplex complexity
    ($M_{opt}(C)$, grey dash-and-dot line) and by von Neumann
    entropy ($M_{opt}(VN)$, blue dotted
    line)~\cite{DeDomenico_Nicosia_2015}. Interestingly, drops in
    the epidemic threshold correspond to dramatic changes in the
    structure of the multiplex, e.g., the formation of
    new structural patterns. This is clearly observed when
    considering the synthetic benchmarks (a)-(d), where the drops
    occur when truly distinct layers are aggregated. The
    reducibility procedures based on von Neumann entropy and
    multiplex complexity provide similar results in real-world
    networks [panels (e)-(h)], even though they yield different
    optimal partitions.}
	\label{fig:SIS_reducibility}
\end{figure*}
\begin{figure*}[!ht]
	\centering
	\includegraphics[width=6in]{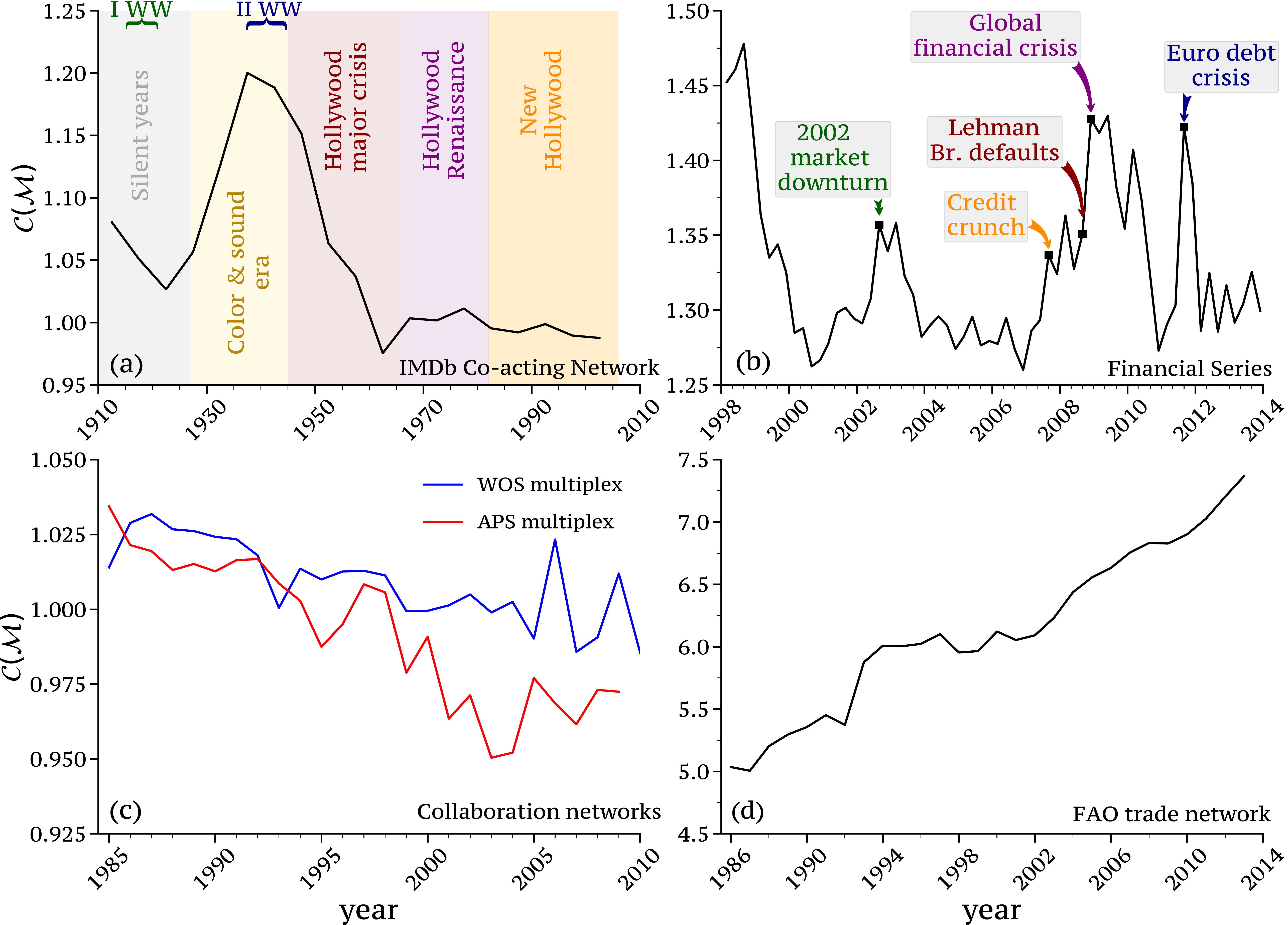}
	\noindent
	\caption{(Colour online) Multiplex complexity as a function of
    time for four different time-varying multiplex networks, namely,
    (a) the IMDb co-starring network, (b) the
    financial multiplex constructed from price time series of 35
    major assets in NYSE and NASDAQ, (c) the co-authorship
    multiplex of collaboration in American Physical Society (APS)
    journals and Web of Science (WOS), and (d) the FAO
    food import/export multiplex network. Notably, the most
    pronounced peaks of the complexity function in (a) and
    (b) correspond to periods of instability and
    crisis. Conversely, the values of complexity in the physics
    collaboration multiplex, for both the APS and WOS data sets
    (c), have remained pretty stable over time, and reveal
    that those systems indeed benefit only marginally from a
    multi-layer representation. Finally, in the FAO food
    import/export multiplex network the complexity has kept
    increasing considerably over time (d), reflecting the
    relevant role played by globalisation in the last twenty years
    in re-shaping the international food market. More
    details and additional results on these four data sets are
    reported in Supplementary Information Section S-3.}
	\label{fig:temporal}
\end{figure*}

\subsection{Structural and dynamical properties of reduced
  multiplex networks}

It is important to note that a layer reduction procedure is
expected to remove redundancies while maintaining as much
information as possible about the original system. However, there
is in general no a-priori guarantee that the reduced multiplex
obtained by aggregating some of the layers actually preserves any
of the structural or dynamical properties of the original
multiplex network to a given level of accuracy. To explore this
aspect of layer reduction, we compared the distributions of four
structural indicators in the original multiplex networks and in
the networks obtained by using the aggregation procedure described
above. In Figure~\ref{fig:structmeasure_reducibility}, we report
the Kendall's $\tau$ correlation coefficient of the rankings
induced by total node degree, node activity, participation
coefficient, and node interdependence in both synthetic and real
multiplex networks (see
Appendix~\ref{appendix:structural_measures_multiplex} for a formal
description of those structural measures).  Interestingly, in
almost all the multiplex networks considered, the optimal
partition identified by the multiplex complexity preserves most of
the structural properties of the original system, as confirmed by
the relatively high values of correlation ($\tau >
0.8$). Conversely, the optimal aggregations based on von Neumann
entropy~\cite{DeDomenico_Nicosia_2015} often correspond to
relatively lower values of correlation. We argue that this is a
very desirable feature of the definition of complexity we have
proposed.  Indeed, a decrease of structural correlation is a clear
indication that aggregation is creating structural artefacts. At
the same time, the fact that the configuration found by using
$q(\bullet)$ always yields high values of correlation with the
original multiplex network confirms that the procedure is removing
only truly redundant information, preserving most of the salient
properties of the system.  This is clearly visible when
considering synthetic benchmarks (Fig.
\ref{fig:structmeasure_reducibility}(a-d)), where the optimal
partition of the multiplex made of only distinct layers is known
by construction. Despite the value of the Kendall's correlation
decreases as the number of layers diminishes, the method based on
multiplex complexity correctly identifies the optimal partition in
all the cases considered.

Although the presence of high correlation between the
structural properties of a multiplex network and its reduced
counterpart indicates that the two systems are structurally
similar, this will not guarantee in general that a dynamical
process happening on the reduced multiplex network will exhibit a
phenomenology similar to that observed on the original
multiplex. As an example, we considered a multiplex SIS
epidemic~\cite{Cozzo2013_SIS} and we computed the epidemic
threshold $\lambda_c$ of the system at each step of the
greedy aggregation procedure. In this process, the epidemic
threshold depends on the contagion parameter $\gamma/\beta$, that
represents the ratio of intra-layer vs inter-layer contagion. In
Fig.~\ref{fig:SIS_reducibility} we report the results of our
analysis for both synthetic and real-world multiplex networks for
different values of the contagion parameter $\gamma/\beta$.
Notice that any drop in the value of the critical threshold
corresponds to an important change in the structure of the reduced
multiplex, e.g., to the formation of new (possibly artificial)
structural patterns. This is easily observable in synthetic
benchmarks (Fig.~\ref{fig:SIS_reducibility}(a)-(d)), where the
epidemic threshold of the reduced multiplex remains the same as
that of the original multiplex up to the point where $q(\bullet)$
is optimal, and then decreases abruptly. This means that, with
respect to epidemic spreading, the reduced multiplex obtained by
optimising the quality function $q(\bullet)$ has basically the
same dynamical behaviour as the original multiplex, while further
aggregations yield a system with different dynamics. As a
consequence, the optimal reduced multiplex network (which has a
smaller number of layers) can be used to make meaningful
predictions about the dynamics of spreading of the original
system. These results provide further evidence that the multiplex
reduction based on Kolmogorov complexity somehow outperforms the
reduction based on von Neumann entropy. See Figure S-6
and S-7 in the Supplementary Information for additional evidence
on several other synthetic benchmarks.

We obtain a similar but more intriguing picture for
real-world multiplex networks, as shown in
Fig.~\ref{fig:SIS_reducibility}(e)-(h). We notice that both
methods preserve most of the information of the original
multiplex, yielding approximately the same
performance. Nevertheless, by looking at both the structural and
dynamical features over the aggregation steps, it appears that the
method based on $\mathcal{C}(\mathcal{M})$ is a bit more
conservative, and finds a reduced multiplex that simultaneously
preserves as much as possible of both the structural and dynamical
features of the original system. Indeed, the best layer partition
identified by the method proposed here has high values of
Kendall's correlation of structural properties ($\approx 0.8$) and
small variation of the epidemic threshold.

\begin{figure*}[!ht]
	\includegraphics[width=6in]{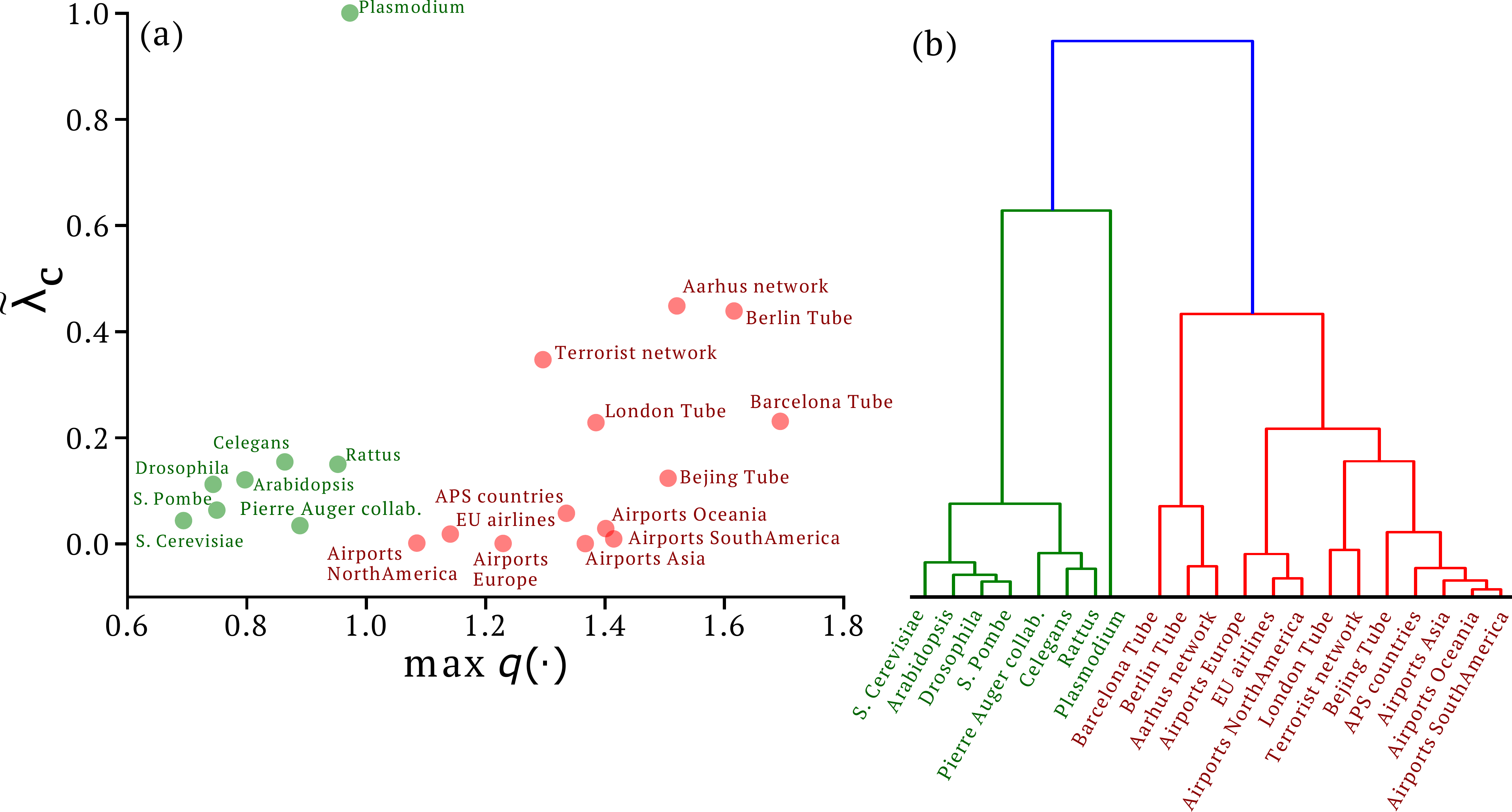}
	\noindent
	\caption{(Colour online) Multiplex cartography of
    real-world systems in the plane $max \; q(\bullet) -
    \tilde{\lambda}_c$ (a) and the corresponding
    dendrogram obtained through Ward's hierarchical
    agglomerative clustering (b). It looks like these two
    structural descriptors alone are able to identify two large
    classes of real-world multiplex networks, namely, biological
    networks on one side (green cluster) and techno-social systems
    on the other side (red cluster).}
	\label{fig:multiplex_cartography}
\end{figure*}

\subsection{Complexity of time-varying multiplex networks}
The multiplex complexity $\complM$ can be used to track the temporal
evolution of the structure of time-varying multiplex networks. In
Fig.~\ref{fig:temporal} we show how the complexity of five
large-scale multiplex networks has changed over time (see
Appendix \ref{appendix:data_sets} for details about the data
sets). Interestingly, $\complM$ provides a very good picture of
the alternating behaviour of the IMDb movie co-starring network over
about a century (Fig.~\ref{fig:temporal}(a)), and of the
network of 35 major assets in the NYSE and NASDAQ financial markets
in the period 1998-2013~(Fig.~\ref{fig:temporal}(b)). In
both cases, local maxima of complexity are consistent with the most
notable periods of crisis in each data set, while local minima of
complexity seem to be precursors of renaissance in IMDb and of
stability in the financial market. The value of complexity of
scientific collaboration networks (APS and Web of Science,
Fig.~\ref{fig:temporal}(c)) has remained stable around
$\complM=1$ over the last 35 years. This is mainly due to the fact
that in these multiplex networks each layer represents a different
field or sub-field of science, and authors normally tend to publish
in one or at most a couple of fields. In fact, the structural
overlap of those multiplexes is always very small, and the large
majority of pairs of nodes are connected in at most two layers. As a
result, there is not indeed much benefit in considering the
multiplex representation, since the information encoded in the
different layers is comparable to that contained in the
corresponding aggregated graph.

It is worth noting that the complexity of the FAO multiplex network
of food exchange has kept increasing steadily in the last 30 years
(Fig.~\ref{fig:temporal}(d)). This is most probably linked
to the globalisation of commercial exchanges in general, which is
reflected also in a more intricate pattern of relations among
countries across a wide range of products. See section S-3 of
Supplementary Information and figures therein for further analyses
of the same data-sets.

\subsection{Mapping multiplex networks}

Finally, in Fig.~\ref{fig:multiplex_cartography} we show how
multiplex complexity can be used to obtain a planar embedding of
multiplex networks of different kind, and to reveal the presence of
interesting clusters. For each multiplex network, we used the
maximum value of the quality function $\max q(\bullet)$ as one of
the coordinates, and the normalised epidemic threshold
$\tilde{\lambda}_c=\lambda_c/M_{opt}(\mathcal{C})$ of the SIS
dynamics~\cite{Cozzo2013_SIS} with contagion parameter
$\gamma/\beta=1$ as the other one. Notice that $\tilde{\lambda}_c$
removes the dependence on the number of layers of the multiplex,
making it possible to compare reduced multiplex networks with
different numbers of layers.  Moreover, since the epidemic
threshold is intimately connected to the spreading dynamics on a
graph, the information it provides is somehow orthogonal to that
captured by multiplex complexity, which is instead a purely
structural quantity. In
Fig.~\ref{fig:multiplex_cartography}(a) we indicated with
different colours the two largest groups obtained through
hierarchical clustering in the ($\tilde{\lambda}_c$, $\max
q(\bullet)$) plane, while in
Fig.~\ref{fig:multiplex_cartography}(b) we show the corresponding
dendrogram, highlighting all the aggregation steps, where at each
step of the procedure we merge two clusters based on the minimum
increase in total within-cluster variance over all possible
pairs~\cite{Ward1963hierarchical}.  Interestingly, these two
descriptors are already sufficient to cluster multiplex networks
with different functions, so that all the biological multiplex
networks appear in the same cluster and social and technological
systems are put in another cluster. A more intriguing
picture, where biological, social, and technological networks are
put in three distinct clusters, is obtained when the normalised
epidemic threshold is replaced by another dynamical descriptor,
i.e., the maximal entropy rate per
node~\cite{Battiston_exploration_2016} (results reported in
Supplementary Information Section S-4).

\section{Discussion}

Quantifying the structural information encoded in a network is of
fundamental importance to identify the key components of the system
it represents. Indeed, information theory has already proven quite
successful at extracting meaningful structural information from
graphs~\cite{Rosvall2008,Passerini2009,DeDomenico_Nicosia_2015} and
at providing sound null-models for different network-related
tasks~\cite{Peixoto2015,Peixoto2018,Godoy-Lorite2016}.  In the case
of multi-dimensional data sets, and in particular of the multi-layer
networks constructed from them, assessing the actual amount of
information contributed by each additional layer is of fundamental
importance. Despite the current trends in Data Science seem to
suggest otherwise, more data is not always a bliss. Not all
additional data available is indeed informative, and quite often
more data implies more redundancy and more noise.

The algorithmic information approach proposed in this paper
condenses the structural properties of a multiplex in a number --
the multiplex complexity $\complM$.  But since $\complM$ is defined
as the ratio between the Kolmogorov complexity of the multiplex and
that of the corresponding aggregated graph, its values allow us to
assess to which extent a given multiplex representation of a system
is more informative than a single-layer graph (e.g., if $\complM$ is
larger than 1). Consequently, it is meaningful to rank different
multiplex networks according to their value of $\complM$, since
those values are indeed telling us how much a multiplex
representation deviates from the corresponding null-model
hypothesis, i.e., that the system can be represented instead as a
single-layer graph.

One of the most appealing aspects of $\complM$ is that it can be
successfully employed to detect redundancy in a multiplex network,
and to obtain meaningful lower-dimensional representations of a
system.  In particular, it is worth noting that the quality function
$q(\bullet)$ provides a consistent way to quantify the differences
in complexity between alternative low-dimensional representations of
the same system. The main difference between $\complM$ and
$q(\bullet)$ is that the former quantifies the relative information
encoded in a multiplex with respect to the aggregated graph, while
the latter appropriately takes into account the fact that multiplex
networks with different numbers of edges, even if obtained by
reducing the same original multiplex network, will in general be
associated to different aggregated graphs.

The results about multiplex reducibility shown in the paper have a
double-pronged significance. On the one hand, the fact that almost
all the multiplex networks analysed admit a more compact
lower-dimensional version is a warning against the quest to obtain
more and more detailed data. There is a clear indication that not
just more data points (edges), rather more informative data points
are needed to complement the information already provided by
existing layers. On the other hand, the possibility of reducing the
number of layers of a multiplex has a lot of practical
implications. Even simple multilayer structural descriptors, such as
clustering coefficient, average shortest path or any centrality
measure based on paths scale super-linearly or exponentially as a
function of the number of layers. Hence, a sound procedure to reduce
the dimensionality of a network, without sacrificing information,
would considerably speed-up most of the computations on multiplex
networks, without compromising on accuracy. Indeed, the optimal
aggregations found by the quality function $q(\mathcal{M})$ often
offer a substantial reduction in the number of layers needed to
represent the system while still retaining most of the structural
complexity of the original system (i.e., without introducing
structural artefacts) as well as the salient characteristics
determining the behaviour of dynamical processes happening on it
(see the example of the preservation of the epidemic threshold in
reduced multiplex networks).  This is confirmed by synthetic
benchmarks, where the optimal partitions is known, but also true for
real systems, where $q(\mathcal{M})$ efficiently balances both the
structural and dynamical features in the aggregation procedure.  In
a way, these results represent an important step towards finding
minimal higher-order models that best compress information while
preserving the original structural and dynamical
patterns~\cite{Lambiotte2019}.

We believe it is quite remarkable that a reduction of layers
based on the multiplex complexity proposed here usually produces
reduced graphs that are sensibly different from those obtained
using the classical reduction method based on von Neumann
entropy~\cite{DeDomenico_Nicosia_2015}. We recall here that the
definition of von Neumann entropy of a graph relies on a purely
formal parallel between the quantum mixing operator and the
rescaled Laplacian of the graph. It is true that the rescaled
Laplacian is somehow related to the diffusive properties of the
system, thus providing a concise description of the global
properties of the graph. However, it is relatively difficult to
pin-point a specific graph property as responsible for a change of
value of the von Neumann entropy. This fact was also noted by the
authors of Ref.~\cite{DeDomenico_Nicosia_2015} in the SI of the
same paper, where they showed by simulation that a difference in
the placement of a single edge of the graph can frequently result
in relatively large fluctuations of the value of von Neumann
entropy. Conversely, the multiplex complexity $\complM$ proposed
here links quite closely to the traditional meaning of complexity
of a system as the amount of information needed to fully describe
it. This link is made possible by the prime-weight matrix, which
encodes the full structure of the system in a string of bits. It is
true that, in principle, the prime-weight matrix encoding depends
on the chosen assignment of labels to nodes and primes to
layers. However, it is remarkable that the assignment of primes to
layers in increasing order of total number of edges provides a
consistent approximation of Kolmogorov complexity, although a quite
conservative one.  Due to the way $\complM$ is defined, its value
varies in a somehow predictable way if the edges of the graph are
re-organised. In particular, if we add a single edge to an existing
multiplex, then we can expect the value of multiplex complexity to
change only slightly. Moreover, if the newly added edge increases
the structural overlap of the multiplex, then the value of
$\complM$ will increase if the original multiplex had a small
structural overlap, or decrease if the multiplex had a large
structural overlap. In this sense, $\complM$ is more closely
associated to the structure of the system, and its changes provide
information that are more readily interpretable.

\section*{Code availability}
Implementations of the algorithms to compute the complexity $\mathcal{C}(\mathcal{M})$ and to obtain reduced representations of a multiplex network based on the function $q(\bullet)$ are available at~\cite{Paper_code}: \url{https://github.com/andresantoro/ALCOREM}.  All the data sets analysed in the paper can be downloaded from the same URL.

\section*{Acknowledgments} 
The authors thank Lucas Lacasa for helpful conversations. A.S.
acknowledges support from The Alan Turing Institute under the
EPSRC Grant No. EP/N510129/1. This work made use of the MidPLUS
cluster, EPSRC Grant EP/K000128/1.\\

\begin{appendices}
\section{Prime-weight matrix encoding} 
\label{appendix:prime_weight_encoding}
An unweighted multiplex network $\mathcal{M}$ over $N$
nodes is a set of $M$  unweighted graphs (layers), each
representing one type of interaction among the $N$ nodes. In this
framework, each node has a replica on each of the $M$ layers, and
the structure of each of the layers is in general distinct.  The
classical way to represent an unweighted multiplex network is by
means of a vector of adjacency matrices $\mathcal{A} =
\{a_{ij}\lay{\alpha}\},\quad\alpha = 1, \ldots,
M$~\cite{Battiston_2014}. The generic element $a\lay{\alpha}_{ij}$
of the adjacency matrix $A\lay{\alpha}$ at layer $\alpha$ is equal
to $1$ if and only if node $i$ and node $j$ are connected by a link
at that layer, and zero otherwise. If we assign a distinct prime
number $p\lay{\alpha}$ to each of the $M$ layers, we can define the
\textit{prime-weight} matrix $\Omega$ whose elements are:
\begin{equation}
	\centering
	\Omega_{ij} =
	\begin{cases}
	  \displaystyle \prod_{\alpha: a_{ij}\lay{\alpha}=1} p\lay{\alpha}
    \qquad\\ \quad\;0 \qquad \quad \qquad \,\textrm{if }
    a_{ij}\lay{\alpha}=0\quad \forall \alpha=1,\ldots, M
	\end{cases}
	\vspace*{0.2 cm}
\end{equation}
The matrix $\Omega\in \mathbb{R}^{N\times N}$ is a compact encoding
of the vector of adjacency matrices $\mathcal{A}$. In fact, thanks
to the unique factorisation theorem, the adjacency matrix of a
generic layer $\alpha$ can be obtained from $\Omega$ by considering
all the elements $\Omega_{ij}$ which are divisible by the
corresponding prime $p\lay{\alpha}$. Notice that this encoding works
also for graphs with integer weights on the links, e.g.,
by associating to each pair of nodes $(i,j)$ the number $\Omega_{ij}
= \prod_{\alpha=1}^{M}
\left(p\lay{\alpha}\right)^{w\lay{\alpha}_{ij}}$, where
$w\lay{\alpha}_{ij}$ is the weight of the edge $(i,j)$ on layer
$\alpha$. Nevertheless, in this paper we always consider
the case of unweighted multiplex networks, therefore all the
weights $w_{i,j}\lay{\alpha}$ are equal to 1.

Although the actual set of primes associated to the layers does not
impact the construction of $\Omega$, for practical reasons it makes
sense to always use the sequence of the first $M$ primes $\{2, 3,
5,...\}$, since the actual number of bits required to store the
matrix $\Omega$ is $O\left(N^2M \log_2 \left[
  \max_\alpha\{p\lay{\alpha}\}\right]\right)$. Notice that,
given a multi-layer graph with $M$ layers and a set of $M$
distinct primes, we can construct $M!$ distinct prime-weight
matrices, one for each of the possible permutations of the primes
associated to the $M$ layers. In this paper we choose a canonical
prime association, that is the one that associates prime numbers to
layers in increasing order of their total number of edges. In
practice, we assign the prime $2$ to the layer with the smallest
total number of edges $K\lay{\alpha}=1/2\sum_{ij}
a\lay{\alpha}_{ij}$, the prime $3$ to the layer with the
second-smallest total number of edges, and so on (see
section S-1 of the Supplementary Material for a detailed
discussion regarding the canonical prime association).

\section{Multiplex Complexity}
\label{appendix:multiplex_complexity}
The Kolmogorov complexity $KC(S)$ of a bit string $S$ is defined as
the length of the shortest computer program that generates $S$ as
output~\cite{Kolmogorov_1998}. However, it is easy to prove that
$KC(S)$ is a non-computable function~\cite{Chaitin_1969}, thus it is
only possible to approximate it. A common approach is to compress
$S$ using a given compression algorithm, and to consider the length
of the compressed string $S'$ as an estimate of $KC(S)$. In fact, it
is possible to obtain $S$ from the compressed string $S'$ by using
the decompression routine corresponding to the compression algorithm
used to obtain $S'$. Thus, the concatenation of $S'$ and the
decompression routine is a program able to generate $S$, and its
length is an upper bound for $KC(S)$. We associate a bit
string $S(\mathcal{M})$ to a given multiplex network $\mathcal{M}$
by considering the bit string of the edge list associated to the
prime-weight matrix $\Omega$, where edges are listed in
lexicographic order and each edge reports the corresponding entry
of $\Omega$~(see SI section S-1.2 for a discussion about
fluctuations due to node labelling). We define the Kolmogorov
complexity $KC(\mathcal{M})$ of the multiplex $\mathcal{M}$ as the
length of the bit string $S'(\mathcal{M})$ obtained by compressing
$S(\mathcal{M})$ with gzip (https://www.gzip.org). Notice
that this is not the only feasible choice, as any other
compression algorithm can be used instead of gzip for computing an
upper bound of $KC$~(see SI section S-1.4 and figures
therein for additional comparisons between different compression
algorithms).

The complexity $\complM$ of a multiplex network $\mathcal{M}$ is
equal to the Kolmogorov complexity of its prime-weight matrix
$\Omega$ divided by the Kolmogorov complexity of the single-layer
weighted matrix $W$, obtained by considering the aggregate
binary matrix multiplied by the largest entry of $\Omega$. In
other terms, we express $W$ as:
\begin{equation}
	\centering W_{ij} =
	\begin{cases}
	  \displaystyle \max_{i,j}\{\Omega_{ij}\} \quad \quad\;\;\,\textrm{if }
    \Omega_{ij} \neq 0 \qquad\\ \quad\;0 \qquad \quad \qquad
    \,\textrm{otherwise}
	\end{cases}
\end{equation} Thus, the complexity 
$\mathcal{C}({\mathcal{M}})$ of the multiplex $\mathcal{M}$ is
defined as:
\begin{equation}
	\mathcal{C}\left(\mathcal{M}\right) =
  \frac{KC\left(\Omega\right)}{KC\left(W\right)}.
	%%\label{eq:information_complexity_m}
\end{equation}
Notice that alternative representations of $W$ are
possible. One possibility is to set the weight of each existing
edge $W_{ij} = o_{ij}$, where
$o_{ij}=\sum_{\alpha}a_{ij}\lay{\alpha}$. Another option is to set
$W_{ij}=2^{o_{ij}}$. Nevertheless, our definition of $W$ is the
only one that guarantees that $\mathcal{C}\left(\mathcal{M}\right)
= 1$ when all the layers of the multiplex $\mathcal{M}$ are
identical~(see SI Section S-1.5 and Figure S-5 for comparisons
between different representations of $W$).

In general, the complexity of a multiplex might depend on the
association of prime numbers to layers and on the actual node
labelling. Numerical evidence confirms that the value of multiplex
complexity obtained using the canonical prime association is always
in the right-most tail of the corresponding distribution
(see SI section S-1.1 and table therein for further
details).  As a consequence, the canonical prime association
represents a conservative upper-bound for the actual value of
Kolmogorov complexity.

To reduce the effect of the other source of variability in the
values of $\complM$ (i.e., the actual labelling of nodes, which
affects the lexicographic ordering of the edge list), we define
$\complM$ as the average of the multiplex complexity obtained by
using the canonical prime association on $10^3$ realisations of node
relabelling on the same multiplex graph (see SI S-1.2 for
details on the distribution of $\complM$ as a function of node
relabellings).

\section{Structural edge overlap}
\label{appendix:structural_edge_overlap}
Given a multiplex $\mathcal{M}$ and a pair of nodes $(i,j)$, the
overlap $o_{ij}$ of the pair is defined as the number of layers in
which an edge exists between node $i$ and node $j$. The matrix
$O=\{o_{ij}\}$ is the overlapping matrix associated to the multiplex
$\mathcal{M}$~\cite{Battiston_2014}.  The edge overlap $o_s$ of a
multiplex network is the expected number of layers in which a pair
of nodes is connected by an
edge~\cite{Battiston_2014,Battiston_Axelrod_2017}:
\begin{equation}
	o_s = \frac{\sum_{i,j}^{N} o_{ij}}{ M\, \sum_{i,j}^{N}
    \Theta(o_{ij})}
\end{equation}
where $\Theta(x)$ is the Heaviside function, i.e.  $\Theta(x)=1$ if
$x>0$, and $1/M\leq o_s\leq 1$. In particular, $o_s = 1/M$ when
there is no edge appearing in more than one layer, while $o_s=1$
when all the $M$ layers are identical. We define the structural
overlap of a multiplex as:
\begin{equation}
	o = \frac{M}{M-1} \left(o_s - \frac{1}{M}\right)
\end{equation}
where the linear transformation $f(o_s) := \frac{M}{M-1} (o_s -
\frac{1}{M})$ maps $o_s$ onto $[0, 1]$.

\section{Synthetic networks}
\label{appendix:synthetic_networks}
The results shown in Fig.~\ref{fig:synthetic_networks}(a) correspond
to multiplex networks with $N=10000$ nodes. The plots are obtained
by starting from a multiplex network with $M$ identical
\Erdos-R\'{e}nyi random graphs as layers (thus having structural
overlap $o=1$), and then iteratively rewiring the edges on each
layer in order to decrease the structural overlap to $o=0$. Edge
rewiring is performed by selecting a pair of edges and swapping
their end-points uniformly at random. This rewiring procedure is
similar to the one used in Ref.~\cite{Diakonova_2016}, and preserves
the degree sequence at each layer.  Consequently, the layers of all
the multiplex networks obtained through relabelling are
\Erdos-R\'{e}nyi random graphs belonging to the same ensemble. The
value of multiplex complexity corresponding to a certain value of
structural overlap is obtained by averaging over $10^2$ distinct
realisations.

The algorithm to increase the structural overlap of the multiplex is
similar to that used to decrease it, with the only difference that a
rewiring is accepted only if results in the increase of the edge
overlap of at least one of the two edges involved in the rewiring.

\section{Reducibility} 
\label{appendix:reducibility}
Computing the global maximum of the quality function $q(\bullet)$
is in general computationally unfeasible, since it requires to
enumerate all the possible partitions of $M$ objects. This is a
NP-hard problem that requires a number of operations that scales
super-exponentially with $M$~\cite{Bell1934}. In order to avoid
this problem, we used instead a greedy algorithm, which reduces
the time complexity to $\mathcal{O}(M^2)$. The algorithm starts
from the original multiplex with $M$ layers and at each step
computes the complexity of the two-layer multiplex networks
corresponding to all the possible pairs of layers.  We
call $\bar{D}$ the pair of layers with the maximum value of
complexity, and we consider the set of pairs of layers whose
overlap is larger than or equal to that of $\bar{D}$. Then, we
aggregate the pair of layers $D$ of that set yielding the
smallest value of complexity $\mathcal{C}(D)$. Aggregation is
performed by considering the union of the edges in the two
layers.  The rationale behind this choice is that if two layers
form a duplex with relatively high overlap and small
complexity, then they are similar enough and can be thus
flattened in a single layer.  The iteration of this procedure will
result in a sequence of multiplex networks with
$\{M,M-1,M-2,\ldots,2,1\}$ layers.  Among those $M$ reduced
multiplex networks, we choose the one yielding the largest value
of $q(\bullet)$.
%  Notice that unlike the process described in
%  Ref.~\cite{DeDomenico_Nicosia_2015}, we use the same quality measure
%  $q(\bullet)$ both to perform layer aggregation and to select the
%  best (sub-)optimal partition.

\section{Structural multiplex measures }
\label{appendix:structural_measures_multiplex}
To analyse the structural properties of both synthetic and real
multiplex networks, we considered four different structural
descriptors~\cite{Battiston_2014, Nicosia_Latora_2015}.

\noindent\textit{Total degree --}
$$
k_i = \sum_{\alpha} k_i\lay{\alpha} =\sum_{\alpha}\sum_{j}
a_{ij}\lay{\alpha},
$$
i.e., the total number of links incident on node $i$ across all
the layers.

\noindent\textit{Node participation coefficient --}
$$
P_i = \frac{M}{M-1}\left[ 1 - \sum_{\alpha} \left(\frac{ k_i\lay{\alpha}}{k_i}
  \right)^2 \right],
$$
which measures the heterogeneity of the number of neighbours of
node $i$ across the layers.

\noindent\textit{Node activity --}
$$
B_i = \sum_{\alpha} \theta\left(k_i\lay{\alpha}\right),
$$
i.e., the number of layers on which node $i$ has at least one
neighbour. Here $\theta$ represents the Heaviside step
function.

\noindent\textit{Node interdependence --}
$$
\lambda_i=\frac{1}{N-1}\sum_{\substack{j \in N \\ j \neq i}}
\frac{\psi_{ij}}{\sigma_{ij}}.
$$
In the expression, $\psi_{ij}$ is the number of shortest paths
between $i$ and $j$ that span across more than one layer, while
$\sigma_{ij}$ is the total number of shortest paths between $i$
and $j$. If $\lambda_i \approx 1$ then $i$ fully exploits the
multiplex structure of the system to reach other nodes, while if
$\lambda_i \approx 0$ node $i$ reaches other nodes through
shortest paths whose edges are on just one layer. In
SI section S-5 we describe an algorithm to compute node
interdependence that exploits the prime-weight matrix
introduced in this paper.

\section{Multiplex data sets} 
\label{appendix:data_sets}
The data sets introduced in this paper for the reducibility
comparisons are: (i) the undirected routes of the 11 lines of the
Barcelona tube network (https://www.tmb.cat/), (ii) the 9 lines of
the Berlin tube (https://www.berlin.de/en/public-transportation/),
(iii) the 17 lines of the Beijing subway
(https://www.bjsubway.com/), and (iv) the scientific collaboration
among countries (APS countries) obtained considering the papers
published in the journals of the American Physical Society. For the
latter, starting with the multiplex data set introduced in
\cite{Nicosia_Latora_2015}, we constructed a weighted multiplex
collaboration network, in which nodes represent countries and a link
connects two countries if scientists based in those countries
co-authored a paper together. Authors having multiple affiliations
were considered as belonging to multiple countries. The weight on
each link represents the number of co-authorship relations between
the corresponding two countries. In our analysis the unweighted
version of such system has been used.

In addition, the time-varying data sets used in this paper are: (i)
the IMDb co-starring network~\cite{Nicosia_Latora_2015}, (ii) the
financial multiplex network constructed from price time series of 35
major assets in NYSE and NASDAQ~\cite{financial_series}, (iii) the
physics collaboration multiplex network of the American Physical
Society (APS) and Web of Science (WOS)~\cite{Nicosia_Latora_2015},
and (iv) the FAO food import/export multiplex network
(http://www.fao.org/statistics/databases). From each original data
set (i,iii,iv), we constructed a time varying multiplex network by
partitioning the original system in temporal windows of one year. In
this process, we associate to each time window the corresponding
static multiplex network containing all the links registered in that
year.

\begin{table*}[!ht]
	\hspace*{-0.5 cm}	
	\small
	\centering
	\begin{tabular}{|lccccccc|}
		\hline
		Multiplex &    $M$ & $o$ & $M_{opt}(C)$ & $[max 
			\,q(\cdot)_{C}]$ & $C_{opt}$ & $M_{opt}(VN)$ & 
		$q(\cdot)_{VN}$ \\
		\hline
		London Tube~\cite{DeDomenico_Nicosia_2015}&   
		13 & 0.006810 &          11 &         0.183 
		&             1.125 &    2  &         0.499 \\
		Barcelona Tube                      &   11 & 
		0.002367 &          11 &         0.224 
		&             1.152 &    11 &         0.513 \\
		Bejing Tube                         &   17 & 
		0.000197 &          15 &         0.199 
		&             1.140 &    17 &         0.528 \\
		Berlin Tube                         &   9  & 
		0.001359 &          8 &         0.214 
		&             1.110 &     9 &         0.461 \\
		Airports North America~\cite{Nicosia_Latora_2015} &  143 &  0.003958 
		&       129 &         0.143 &       1.271 
		&      93 &         0.697 \\
		Airports Europe~\cite{Nicosia_Latora_2015}        &  175 &  
		0.003185 &       163 &         0.162 &       
		1.413 &      109 &        0.675 \\
		Airports Asia~\cite{Nicosia_Latora_2015}          &  213 &  
		0.005477 &       209 &         0.180 &       
		1.636 &      146 &        0.291\\
		Airports South America~\cite{Nicosia_Latora_2015} &   58 &  0.014244 
		&        53 &         0.187 &       1.325 
		&       41 &         0.682 \\
		Airports Oceania~\cite{Nicosia_Latora_2015}       &   37 &  
		0.014532 &        27 &         0.185 &       
		1.192 &       31 &         0.665 \\
		Airports Africa~\cite{Nicosia_Latora_2015}        &   84 &  
		0.006876 &        74 &         0.191 &       
		1.274 &       65 &         0.719 \\
		EU airlines~\cite{Cardillo_2013}                  &   37 &  
		0.005964 &        37 &         0.151 &       
		1.233 &       37 &         0.411 \\
		Train UK~\cite{Santoro_2018}                      &   41 
		&  0.002687 &        24 &         0.120 &       
		1.019 &       15 &         0.225 \\
		APS countries                                     & 
		10 &  0.451138 &        10 &         0.176 
		&       1.618 &        2 &         0.047 \\
		Aarhus network~\cite{Dedomenico_website}    &    5 &  0.189093 &         
		5 &         0.201 &       1.291 &        2 
		&         0.158 \\
		Terrorist network~\cite{Nicosia_Latora_2015}                              &
		4 &  0.153558 &         4 &         0.171 
		&       1.166 &        2 &         0.239 \\
		Pierre Auger collab.~\cite{DeDomenico_Arenas_Rosvall_2015} &   16 & 
		0.006901 &          10 &         0.117 
		&             1.018 &   15 &         0.423\\
		Arabidopsis~\cite{DeDomenico_Nicosia_2015}             &    
		7 &  0.007690 &         6 &         0.105 
		&       1.023 &        7 &         0.421 \\
		Candida~\cite{DeDomenico_Nicosia_2015}                 &
		7 &  0.007892 &         5 &         0.177 
		&       1.030 &        3 &         0.620 \\
		Celegans~\cite{DeDomenico_Nicosia_2015}                & 
		6 &  0.003095 &         6 &         0.114 
		&       1.023 &        5 &         0.430 \\
		Drosophila~\cite{DeDomenico_Nicosia_2015}              &   
		7 &  0.004389 &         5 &         0.098 
		&       1.011 &        6 &         0.379 \\
		Gallus~\cite{DeDomenico_Nicosia_2015}                  &
		6 &  0.012923 &         5 &         0.179 
		&       1.043 &        5 &         0.577 \\
		Human Herpes-4~\cite{DeDomenico_Nicosia_2015}          &    4 
		&  0.042056 &         2 &         0.196 &       
		1.063 &        4 &         0.353 \\
		Human HIV-1~\cite{DeDomenico_Nicosia_2015}             &    
		5 &  0.022294 &         5 &         0.150 
		&       1.073 &        4 &         0.353 \\
		Mus~\cite{DeDomenico_Nicosia_2015}                     &
		7 &  0.010776 &         7 &         0.106 
		&       1.041 &        6 &         0.375 \\
		Oryctolagus~\cite{DeDomenico_Nicosia_2015}             &    
		3 &  0.019231 &         3 &         0.209 
		&       1.026 &        2 &         0.500 \\
		Plasmodium~\cite{DeDomenico_Nicosia_2015}              &   
		3 &  0.000206 &         2 &         0.128 
		&       0.987 &        3 &         0.611 \\
		Rattus~\cite{DeDomenico_Nicosia_2015}                  &
		6 &  0.012401 &         6 &         0.126 
		&       1.040 &        5 &         0.472 \\
		S. Cerevisiae~\cite{DeDomenico_Nicosia_2015}           &    7 
		&  0.017603 &         5 &         0.092 &       
		1.122 &        3 &         0.135 \\
		S. Pombe~\cite{DeDomenico_Nicosia_2015}                & 
		7 &  0.007070 &         5 &         0.099 
		&       1.067 &        2 &         0.206 \\
		Xenopus~\cite{DeDomenico_Nicosia_2015}                 &
		5 &  0.025692 &         5 &         0.169 
		&       1.071 &        4 &         0.410 \\
		\hline
	\end{tabular}
	\caption{\textbf{Reducibility of technological, social, and
      biological multiplex networks.} From left to right, the
    columns report the number of layers in the original system
    ($M$), the structural edge overlap ($o$), the number of optimal
    layers ($M_{opt}(C)$) obtained when maximising the quality
    function $q(\bullet)$, the value $max\; q(\bullet)$, and the
    optimal value of complexity $C_{opt}$ observed. The last two
    columns show the optimal number of layers $M_{opt}(VN)$ and the
    corresponding value of the quality function $q(\cdot)_{VN}$
    obtained when using the multiplex structural reducibility
    procedure introduced in
    Ref.~\cite{DeDomenico_Nicosia_2015}. Although the two methods
    yield different results, they share similar features,
    i.e. technological multiplex networks are less likely to be
    reduced compared to biological and social systems.}
	\label{table:reducibility} 
\end{table*}

\end{appendices}

\clearpage

\renewcommand\theequation{{S-\arabic{equation}}}
\renewcommand\thetable{{S-\Roman{table}}}
\renewcommand\thefigure{{S-\arabic{figure}}}
\renewcommand\thesection{{S-\arabic{section}}}

\setcounter{section}{0}
\setcounter{table}{0}
\setcounter{figure}{0}
\setcounter{equation}{0}

\onecolumngrid

\begin{center}
  \Large
  \textbf{Supplementary Material: Algorithmic complexity of multiplex
	  networks}
\end{center}

\maketitle

\section{Algorithmic complexity}

In the main text, we introduced the complexity measure $\complM$ to
evaluate the relative amount of additional information needed to
encode a multiplex network with respect to the amount needed to encode
the corresponding single-layer aggregated graph. This quantity relies
on the approximation of the Kolmogorov Complexity (KC) of a string and
it is formally defined as:
\begin{equation}
	\mathcal{C}(\mathcal{M}) = \frac{ KC (\Omega)} {KC(W)}
	\label{eq:complexity}
\end{equation}
where $KC(\Omega)$ represents the KC of the prime-weight matrix
$\Omega$, while $KC(W)$ is the KC of the single-layer weighted matrix
$W$, obtained by considering the aggregate binary matrix multiplied by
the largest entry of $\Omega$. Notice that 
among the possible definition of $W$, we consider the 
following representation:
\begin{equation}
	\centering
	W_{ij} =
	\begin{cases}
		\displaystyle max (\Omega_{ij}) \quad 
		\quad\;\;\,\textrm{if 
		}\exists\,\alpha \,:\, a_{ij}\lay{\alpha}=1
		\qquad\\ \quad\;0 \qquad
		\quad \qquad \,\textrm{if } 
		a_{ij}\lay{\alpha}=0\quad 
		\forall
		\alpha=1,\ldots, M
	\end{cases}
	\vspace*{0.2 cm}
	\label{eq:weighted_aggregate}
\end{equation}
so that the  complexity $\mathcal{C}$ of multiplex 
consisting of all identical layers  is exactly equal to 
one.
In general, the numerical computation of the complexity 
measure
$\mathcal{C}$ is based on the approximation of the Kolmogorov
Complexity by means of the compression algorithm gzip. In particular,
this quantity will, in general, depend on two factors,
namely the actual assignment of prime numbers to the layers and the
actual labelling of the $N$ nodes. Here we investigate how $\complM$
depends on these two factors, and we show that the results reported in
the main text are robust with respect to prime-layer association and
node labellings. Lastly,  we present extensive numerical
simulations to estimate the impact of (i) alternative representations
of the weighted network $W$, and (ii) different compression algorithms
when computing the complexity $\complM$ of a multiplex network.

\begin{figure}[!htb]
	\includegraphics[width=1\textwidth]{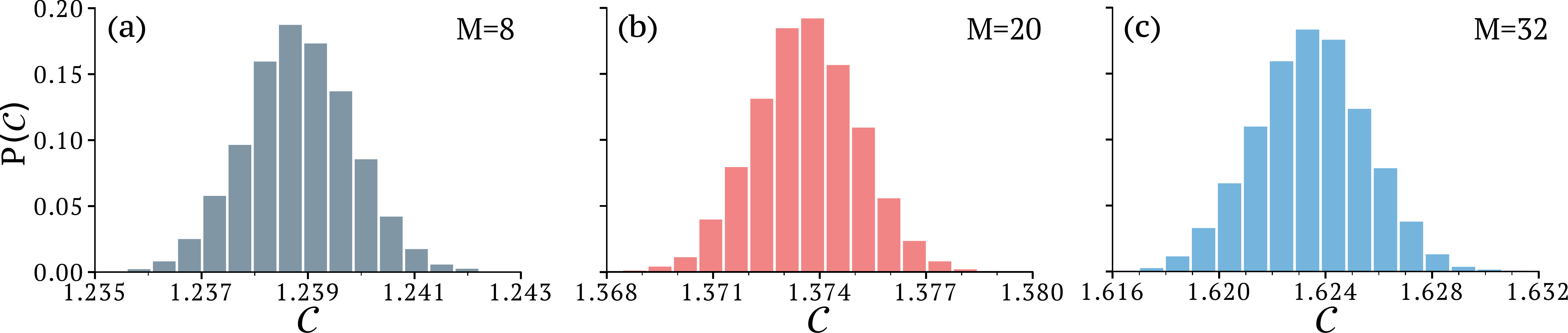}
	\caption{(Colour online) Distribution of complexity 
		$\mathcal{C}$ 
		over 10000 random
		independent node labellings for three synthetic 
		multiplex networks
		with $N=10000$, $\langle k \rangle = 6$, 
		$o\approx 0.6$, and
		different number of layers (\textbf{A}: $M=8$, 
		\textbf{B}: $M=20$,
		\textbf{C}: $M=32$).  The values of $\complM$ 
		are normally
		distributed, as confirmed by the 
		Anderson-Darling and Cramer Von
		Mises tests (resp. $p \approx 0.79$ and 
		$p\approx 0.71$ for $M=8$,
		$p \approx 0.97$ and $p\approx 0.93$ for 
		$M=20$, $p \approx 0.75$
		and $p\approx 0.76$ for $M=32$). In addition, 
		it is worth mentioning
		that the coefficient of variation for all these 
		distributions is
		approximately $0.2\%$, indicating a very low 
		dispersion around the
		mean. }
	\label{figure:histograms_node_relabelling}
\end{figure}

\subsection{Prime association}

In all the results reported in the main text, the complexity $\complM$
was obtained using the ``canonical prime association'', where prime
numbers are associated to layers in increasing order of their total
number of edges. In practice, we assign the prime 2 to the layer with
the smallest total number of edges $K\lay{\alpha}=\frac{1}{2}\sum_{ij}
a\lay{\alpha}$, the prime $3$ to the layer with the second-smallest
total number of edges, and so on.  In general, a multiplex with $M$
layers admits $M!$ different primes association, one for each of the
possible permutations of the primes associated to the M layers.

To support the choice of this particular prime association, we report
in Table~\ref{table:multiplex_prime_reshuffling} some statistics of
the distribution of complexity $\complM$ over 10000 prime associations
for all the real-world multiplex used in the main text. As expected,
the prime-weight matrix encoding depends on the chosen assignment of
primes to layers. Nevertheless, it is worth noting that the
distribution of complexity for all the multiplex considered is quite
peaked around the mean (very small standard deviation), and has a
coefficient of variation $cv=\frac{\sigma}{\mu}$ which is smaller than
$1 \%$ for almost all the distributions. In particular, 
the value of
complexity obtained using the prime canonical association
$\mathcal{C}_{CA}$ is normally located in the right-most tail of the
corresponding distribution, as confirmed by the value of the
associated z-score $Z$. Consequently, the value of complexity computed
using the canonical prime association is normally an upper-bound for
the actual value of complexity (i.e., the value corresponding to the
actual Kolmogorov Complexity of the multiplex and of the corresponding
aggregated graph).
%	\begin{equation}
%	\centering
%	W_{ij} =
%	\begin{cases}
%	\displaystyle max (\Omega_{ij}) \quad 
%	\quad\;\;\,\textrm{if 
%	}\exists\,\alpha \,:\, a_{ij}\lay{\alpha}=1
%	\qquad\\ \quad\;0 \qquad
%	\quad \qquad \,\textrm{if } a_{ij}\lay{\alpha}=0\quad 
%	\forall
%	\alpha=1,\ldots, M
%	\end{cases}
%	\vspace*{0.2 cm}
%	\end{equation}

\begin{table}[!ht]
	\footnotesize
	\begin{tabular}{l|cccccccc	}
		\hline
		\textbf{Multiplex} &  $\mathcal{C}_{CA}$ &    $\mu$ &   $\sigma$ &         $ste(\mu)$ &         $cv$ &  $\frac{\sigma}{\mu \cdot \sqrt(N)}$ &   $Z$ \\
		\hline
		London Tube &            1.128292 &  1.115452 &  0.007446 &  0.000074 &  0.006675 &                             0.000067 &  1.724483 \\
		Barcelona Tube &            1.126812 &  1.119959 &  0.007222 &  0.000072 &  0.006448 &                             0.000064 &  0.948881 \\
		Beijing Tube &            1.120737 &  1.122686 &  0.010934 &  0.000109 &  0.009739 &                             0.000097 & -0.178297 \\
		Berlin Tube &            1.107717 &  1.098907 &  0.009024 &  0.000090 &  0.008212 &                             0.000082 &  0.976258 \\
		Airports North America &            1.263749 &  1.250040 &  0.014373 &  0.000144 &  0.011498 &                             0.000115 &  0.953842 \\
		Airports Europe &            1.416298 &  1.352806 &  0.020549 &  0.000205 &  0.015190 &                             0.000152 &  3.089739 \\
		Airports Asia &            1.634686 &  1.591643 &  0.030954 &  0.000310 &  0.019448 &                             0.000194 &  1.390522 \\
		Airports South America &            1.306878 &  1.266330 &  0.019676 &  0.000197 &  0.015537 &                             0.000155 &  2.060841 \\
		Airports Oceania &            1.191390 &  1.139328 &  0.015222 &  0.000152 &  0.013361 &                             0.000134 &  3.420186 \\
		Airports Africa &            1.264403 &  1.230097 &  0.016762 &  0.000168 &  0.013626 &                             0.000136 &  2.046697 \\
		EU airlines &            1.218266 &  1.200618 &  0.016453 &  0.000165 &  0.013704 &                             0.000137 &  1.072666 \\
		Train UK &            1.019208 &  1.029434 &  0.017487 &  0.000175 &  0.016987 &                             0.000170 & -0.584796 \\
		APS countries &            1.625382 &  1.590660 &  0.015224 &  0.000152 &  0.009571 &                             0.000096 &  2.280844 \\
		Aarhus network &            1.294118 &  1.284601 &  0.008598 &  0.000086 &  0.006693 &                             0.000067 &  1.106912 \\
		Terrorist network &            1.156780 &  1.145212 &  0.006786 &  0.000068 &  0.005925 &                             0.000059 &  1.704832 \\
		Pierre Auger collab. &            0.989371 &  0.997252 &  0.008041 &  0.000080 &  0.008063 &                             0.000081 & -0.980117 \\
		Arabidopsis &            1.015857 &  1.017097 &  0.005778 &  0.000058 &  0.005681 &                             0.000057 & -0.214578 \\
		Candida &            1.029647 &  1.033560 &  0.008517 &  0.000085 &  0.008241 &                             0.000082 & -0.459470 \\
		Celegans &            1.020054 &  0.999806 &  0.010128 &  0.000101 &  0.010130 &                             0.000101 &  1.999236 \\
		Drosophila &            1.008709 &  1.021645 &  0.011899 &  0.000119 &  0.011647 &                             0.000116 & -1.087131 \\
		Gallus &            1.049566 &  1.054286 &  0.012510 &  0.000125 &  0.011866 &                             0.000119 & -0.377249 \\
		Human Herpes-4 &            1.066253 &  1.065862 &  0.004286 &  0.000043 &  0.004021 &                             0.000040 &  0.091231 \\
		Human HIV-1 &            1.033686 &  1.029318 &  0.016341 &  0.000163 &  0.015876 &                             0.000159 &  0.267315 \\
		Mus &            1.039883 &  1.039828 &  0.006973 &  0.000070 &  0.006706 &                             0.000067 &  0.007934 \\
		Oryctolagus &            1.009524 &  1.014351 &  0.007962 &  0.000080 &  0.007850 &                             0.000078 & -0.606175 \\
		Plasmodium &            0.988591 &  0.991987 &  0.006902 &  0.000069 &  0.006957 &                             0.000070 & -0.492102 \\
		Rattus &            1.048545 &  1.033855 &  0.011312 &  0.000113 &  0.010942 &                             0.000109 &  1.298544 \\
		S. Pombe &            1.061032 &  1.063236 &  0.007830 &  0.000078 &  0.007364 &                             0.000074 & -0.281524 \\
		Xenopus &            1.053121 &  1.060337 &  0.006209 &  0.000062 &  0.005856 &                             0.000059 & -1.162198 \\
		\hline
	\end{tabular}
	\caption{Statistics of multiplex complexity $\complM$ for different
		real-world multiplex over 10000 prime-layer associations. From
		left to right, we report the complexity obtained using the
		canonical prime association $\mathcal{C}_{CA}$, the mean of the
		distribution when re-shuffling the prime association $\mu$, the
		standard deviation $\sigma$, the standard error of the mean ${\rm
			ste}(\mu)$, the coefficient of variation
		$cv=\frac{\sigma}{\mu}$, and the coefficient of variation over the
		sample $\frac{\sigma}{\mu\cdot\sqrt(N)}$.  The last column reports
		the z-score of the prime canonical association
		$\mathcal{C}_{CA}$.}
	\label{table:multiplex_prime_reshuffling}
\end{table}

\subsection{Node relabelling}
Here we study the effect of node relabelling on the value of
complexity $\complM$, by computing the distribution of complexity over
10000 independent random node labellings for both synthetic and
real-world multiplex networks. In
Figure~\ref{figure:histograms_node_relabelling} we report the
distribution of complexity for three synthetic multiplex networks
having the same number of nodes and different number of layers. In all
the three cases, we find that the distribution of complexity has a
coefficient of variation ($cv$) less than $0.2\%$, indicating that the
distribution is indeed quite peaked around its mean.

The complexity distribution for real-world multiplex networks behaves
in a similar way. In Table~\ref{table:multiplex_node_reshuffling} we
report some statistics for the distribution of complexity under node
relabelling for the multiplex networks studied in the main text.  Also
in this case, the distribution of complexity is peaked around the mean
with a small variance for almost all the cases.  Hence, to account for
the intrinsic stochasticity due to node relabelling of the gzip
algorithm, in all our simulations of the main article we always refer
to the mean Complexity, which is obtained when averaging the value of
$\complM$ obtained over $10^3$ independent node 
labellings.

\begin{table}[!th]
	\small
	\centering
	\begin{tabular}{|l|ccccc|}
		\hline
		Multiplex &     $\mu$ &   $\sigma$ &       $ste(\mu)$ &        $cv$ & $ \frac{\sigma}{ \mu \cdot \sqrt(N)}$ \\
		\hline
		Synthetic 8 layers &  1.238824 &  0.001007 &  0.000010 &  0.000813 &                             0.000008 \\
		Synthetic 20 layers &  1.373632 &  0.001465 &  0.000015 &  0.001067 &                             0.000011 \\
		Synthetic 32 layers &  1.623441 &  0.002087 &  0.000021 &  0.001285 &                             0.000013 \\
		London Tube &  1.125283 &  0.009110 &  0.000091 &  0.008096 &                             0.000081 \\
		Barcelona Tube &  1.152427 &  0.013088 &  0.000131 &  0.011357 &                             0.000114 \\
		Beijing Tube &  1.140638 &  0.010211 &  0.000102 &  0.008952 &                             0.000090 \\
		Berlin Tube &  1.111365 &  0.011885 &  0.000119 &  0.010694 &                             0.000107 \\
		Airports North America &  1.270578 &  0.004499 &  0.000045 &  0.003541 &                             0.000035 \\
		Airports Europe &  1.412997 &  0.005673 &  0.000057 &  0.004015 &                             0.000040\\
		Airports Asia &  1.638689 &  0.005555 &  0.000056 &  0.003390 &                             0.000034 \\
		Airports South America &  1.324929 &  0.009317 &  0.000093 &  0.007032 &                             0.000070 \\
		Airports Oceania &  1.191850 &  0.010755 &  0.000108 &  0.009024 &                             0.000090 \\
		Airports Africa &  1.273712 &  0.010209 &  0.000102 &  0.008015 &                             0.000080 \\
		EU airlines &  1.232830 &  0.006486 &  0.000065 &  0.005261 &                             0.000053 \\
		Train UK &  1.018875 &  0.003015 &  0.000030 &  0.002959 &                             0.000030 \\
		APS countries &  1.617794 &  0.012755 &  0.000128 &  0.007884 &                             0.000079 \\
		Aarhus network &  1.291076 &  0.018428 &  0.000184 &  0.014274 &                             0.000143 \\
		Terrorist network &  1.165705 &  0.013456 &  0.000135 &  0.011543 &                             0.000115 \\
		Pierre Auger collab. &  1.018169 &  0.005196 &  0.000052 &  0.005103 &                             0.000051 \\
		Arabidopsis &  1.023403 &  0.001587 &  0.000016 &  0.001551 &                             0.000016 \\
		Candida &  1.030212 &  0.010781 &  0.000108 &  0.010465 &                             0.000105 \\
		Celegans &  1.022841 &  0.002089 &  0.000021 &  0.002043 &                             0.000020 \\
		Drosophila &  1.010760 &  0.000851 &  0.000009 &  0.000842 &                             0.000008 \\
		Gallus &  1.042861 &  0.008910 &  0.000089 &  0.008544 &                             0.000085 \\
		Human Herpes-4 &  1.063347 &  0.014228 &  0.000142 &  0.013381 &                             0.000134 \\
		Human HIV-1 &  1.072567 &  0.007249 &  0.000072 &  0.006759 &                             0.000068 \\
		Mus &  1.041120 &  0.001317 &  0.000013 &  0.001265 &                             0.000013 \\
		Oryctolagus &  1.025862 &  0.011836 &  0.000118 &  0.011538 &                             0.000115 \\
		Plasmodium &  0.986594 &  0.002003 &  0.000020 &  0.002030 &                             0.000020 \\
		Rattus &  1.040003 &  0.002909 &  0.000029 &  0.002797 &                             0.000028 \\
		S. Cerevisiae &  1.122325 &  0.000883 &  0.000009 &  0.000786 &                             0.000008 \\
		S. Pombe &  1.067023 &  0.001699 &  0.000017 &  0.001592 &                             0.000016 \\
		Xenopus &  1.071017 &  0.007500 &  0.000075 &  0.007002 &                             0.000070 \\
		\hline
	\end{tabular}
	\caption{Statistics of multiplex complexity $\complM$ for several
		synthetic and real-world multiplex over 10000 random independent
		node labellings. From left to right, we report the mean of the
		distribution $\mu$, the standard deviation $\sigma$, the standard
		error for the mean ${\rm ste}(\mu)$, the coefficient of variation
		$cv$, and the coefficient of variation over the sample
		$\frac{\sigma}{\mu \cdot \sqrt(N)}$.}
	\label{table:multiplex_node_reshuffling}
\end{table}
\newpage

\subsection{Average degree and topology}
\begin{figure}[!ht]
	\centering
	\includegraphics[width=0.9\textwidth]{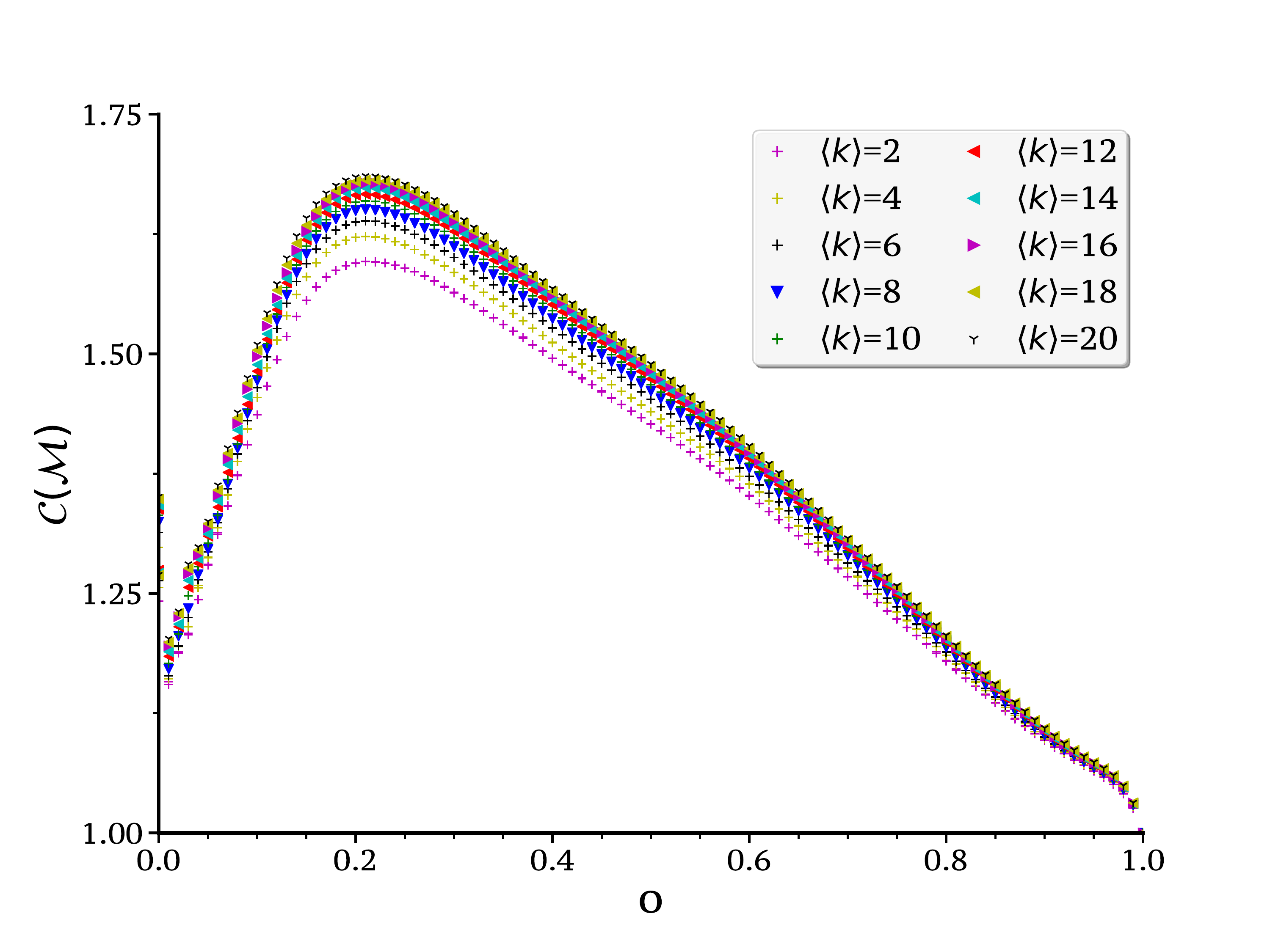}
	\caption{ (Colour online) Complexity $\complM$ in 
		ensembles of synthetic multiplex networks with $M=20$ layers,  variable average degree $\langle k \rangle$ and
		tunable structural overlap $o$. Each layer is an Erd\"{o}s-Renyi
		graph with $N = 10000$ nodes. Interestingly,
		$\complM$ proportionally increases as a function of $\langle k
		\rangle$ while maintaining the same optimal value of complexity
		around $o \approx 0.21$.}
	\label{figure:complexity_vs_overlap_avgdegree}
\end{figure}

In Figure~\ref{figure:complexity_vs_overlap_avgdegree} we report the
value of complexity in the ensembles of synthetic multiplex networks
with different average degree on each layer, where the total number of
nodes and the number of layers are kept fixed ($N = 10000$, $M = 20$),
while the the structural edge overlap $o$ is tunable. As expected, the
complexity increases slightly as a function of the average degree
$\langle k \rangle$, while the shape of the curve and the position of
the maximum remain unchanged.

\begin{figure}[!ht]
	\centering
	\includegraphics[width=0.9\textwidth]{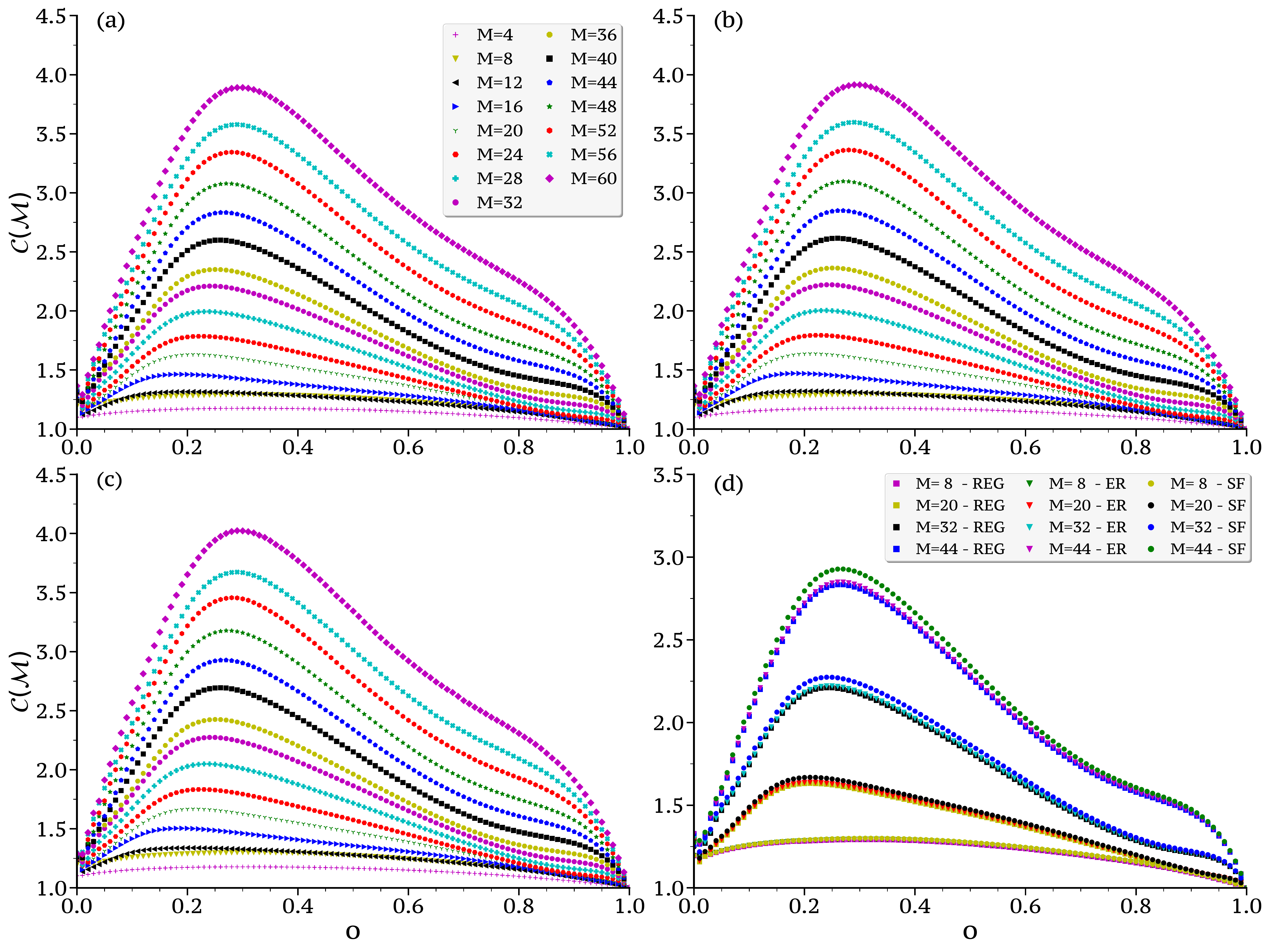}
	\caption{(Colour online) Complexity 
	  $\complM$ in 
		ensembles of synthetic multiplex
		networks with $\langle k \rangle =6$,
		tunable structural overlap $o$,  and different 
		topologies, namely, regular networks (a), ER 
		random graph (b), and scale free networks with 
		$\gamma = 2.7$ (c). We report a comparison  of 
		the complexity $\complM$ for the three 
		different topologies considered (d).	
		As one would expect, it appears 
		that scale-free multiplexes have a higher 
		complexity compared to multiplex having ER or 
		regular topologies.} 
	\label{figure:comparison_topologies}
\end{figure}

In addition, we  also
report in Figure \ref{figure:comparison_topologies} the 
comparison of complexity in the ensembles of synthetic 
multiplex networks with different topologies, namely, 
regular, Erd\"{o}s-R\'enyi,  and scale-free networks 
($\gamma = 2.7$) with the same average degree. In this 
case, we observe a slight difference in the values of 
complexity depending on the topology.  Interestingly,
scale-free multiplexes have a higher complexity 
compared to multiplex having ER or regular topologies.


\subsection{Role of the compression protocol}
\begin{figure}[hbt!]
	\includegraphics[width=1\textwidth]{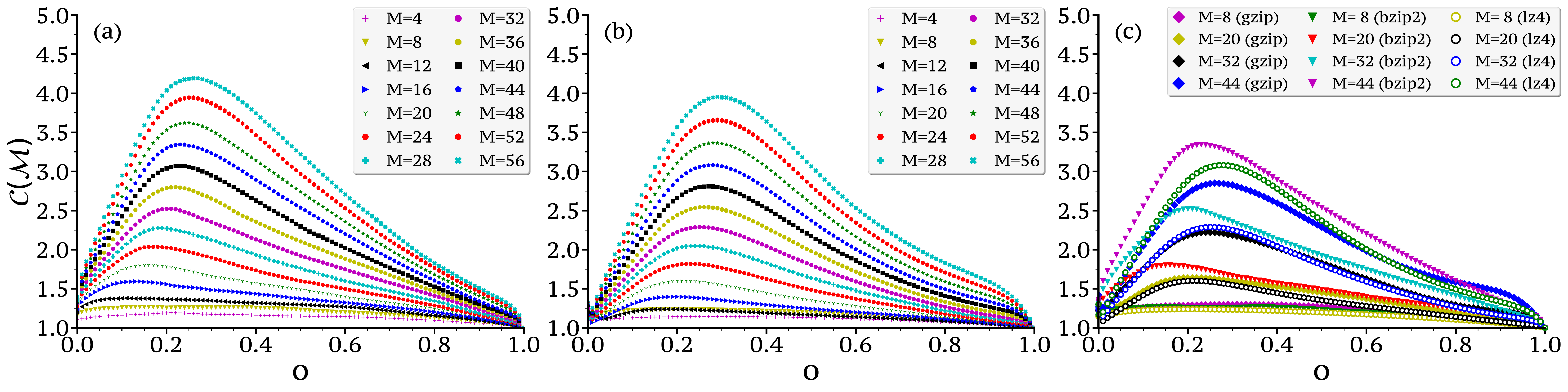}
	\caption{(Colour online) Impact of compression 
	  protocols in approximating the complexity $\complM$ 
	  of 
	  multiplex networks for an ensembles of 
	  synthetic multiplex networks with a variable average 
	  degree $\langle k \rangle$ and tunable structural 
	  overlap $o$. The bzip2 (a) and lz4 (b) compression 
	  protocols show a similar behaviour  when computing the 
	  Kolmogorov Complexity of binary strings, indeed also in 
	  this case the complexity 
	  $\complM$ has a non-monotonic behaviour as a function 
	  of the 
	  structural 
	  edge overlap, for any value of $M$. As expected, the 
	  values of complexity depend on the kind of 
	  compression protocols used. In (c) we report the 
	  comparison of all the three compression protocols used 
	  in this work, namely, gzip, bzip2 and lz4.  }
	\label{figure:compression_protocol}
\end{figure}

As discussed in the main text, the Kolmogorov 
Complexity of a binary string is in 
general incomputable. For this reason, to approximate 
the KC it is usually necessary to compute an upper 
bound of 
the KC  by means of a compression protocol. In our 
work, we 
used the gzip algorithm (https://www.gzip.org), even if 
this is not the only viable choice and 
other compression algorithms could have been used.
Here we considered the effect of two 
alternative compression protocols, namely bzip2 
(https://www.bzip.org) and 
lz4 (https://www.lz4.org), when computing the 
complexity function 
$\mathcal{C}$.\\
In Figure \ref{figure:compression_protocol} we 
report the results of our analysis when computing the  
Complexity $\mathcal{C}$ as a function of the 
structural edge 
overlap  and different number of layers for multiplex 
networks with $N=10000$, $<k>=6$ on each layers. 
Overall, for both the compression 
protocols, we still observe the same 
qualitative 
behaviour of the Complexity presented in the 
main paper using gzip, i.e. $\complM$ is a 
non-monotonic 
function of the structural edge overlap.
Yet, the compression protocols approximate in a 
different way the KC of binary 
strings, so that there exists a quantitative difference 
between the numerical values of the 
Complexity (see  Fig. 
\ref{figure:compression_protocol}(c) for a comparison 
between compression protocols).

\subsection{Role of the aggregate matrix W}
\begin{figure}[hbt!]
	\includegraphics[width=1\textwidth]{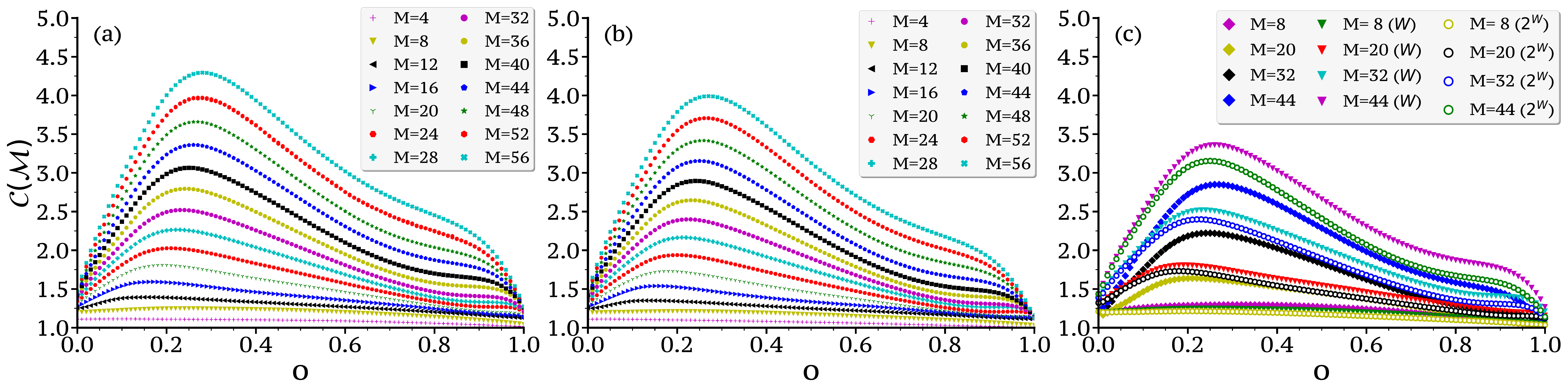}
	\caption{(Colour online) Complexity $\complM$ 
	  of multiplex networks for an ensembles of synthetic 
	  multiplex networks when considering different aggregate 
	  representations, with a variable average degree 
	  $\langle k \rangle$ and tunable structural overlap $o$. 
	  We respectively report in (a) the complexity values using
	  the definition of equation \ref{eq:weighted_aggregate_overlap},
	  while in (b) the results using  Eq. \ref{eq:weighted_aggregate_2*overlap}.
	  Also in this case the complexity $\complM$ has a 
	  non-monotonic behaviour as a function of the structural 
	  edge overlap, which is independent of the aggregate 
	  representation used when computing the complexity. As 
		expected, the numerical values of $\mathcal{C}$ 
		depend on 
		the definition of the aggregate. In (c) 
		we report 
		the comparison of all the three 
		aggregate representations.}
	\label{figure:aggregate_representation}
\end{figure}

Here we study the effect ot the aggregate matrix $W$ 
when computing the complexity $\mathcal{C}$. Indeed, as 
mentioned when introducing Eq. \ref{eq:complexity}, the 
value of the complexity  generally depends on the 
definition of $W$. Thus, we identified two alternative 
representations that could have been used in place of 
the one presented in the main paper (i.e. Eq. \ref{eq:weighted_aggregate}).
In formula:
\begin{equation}
	\centering
	W_{ij} = o_{ij} = \sum_{\alpha=1}^{M}a\lay{\alpha}_{ij} 
	\vspace*{0.2 cm}
	\label{eq:weighted_aggregate_overlap}
\end{equation}
or:
\begin{equation}
	\centering
	W_{ij} =
	\begin{cases}
	  \quad 2^{o_{ij}} \quad\quad \textrm{if } o_{ij} > 
	  0 \\ \quad 0 \quad
	  \qquad\textrm{otherwise}
	\end{cases}
	\vspace*{0.2 cm}
	\label{eq:weighted_aggregate_2*overlap}
\end{equation}
For the alternative definitions of $W$, we report in 
Figure 
\ref{figure:aggregate_representation} the complexity 
$\mathcal{C}$  as a  function of the structural edge  
overlap and different number of layers for multiplex 
networks with $N=10000$, $<k>=6$ on each layers.  
Notice that with  both the representations, the 
values of complexity are in general smaller then the 
ones presented in the main paper. This 
is a due to a difference in size (in terms of bit) of 
the 
symbols used in the aggregate. \\
However, for both the definitions, we still 
observe a similar qualitative behaviour of the one 
presented in the main paper, so that $\complM$ is a 
non-monotonic function of the structural edge overlap. 
Yet,  it appears that the maximum value of complexity, 
as well as the numerical values,  depend on 
the aggregate  representation used when 
computing  $\complM$. We report in  Figure
\ref{figure:aggregate_representation}(c) a comparison 
between the different representations of $W$.

\section{Reducibility on synthetic multiplex 
	networks}
\begin{figure}[t!]
	\centering
	\includegraphics[width=0.95\textwidth]{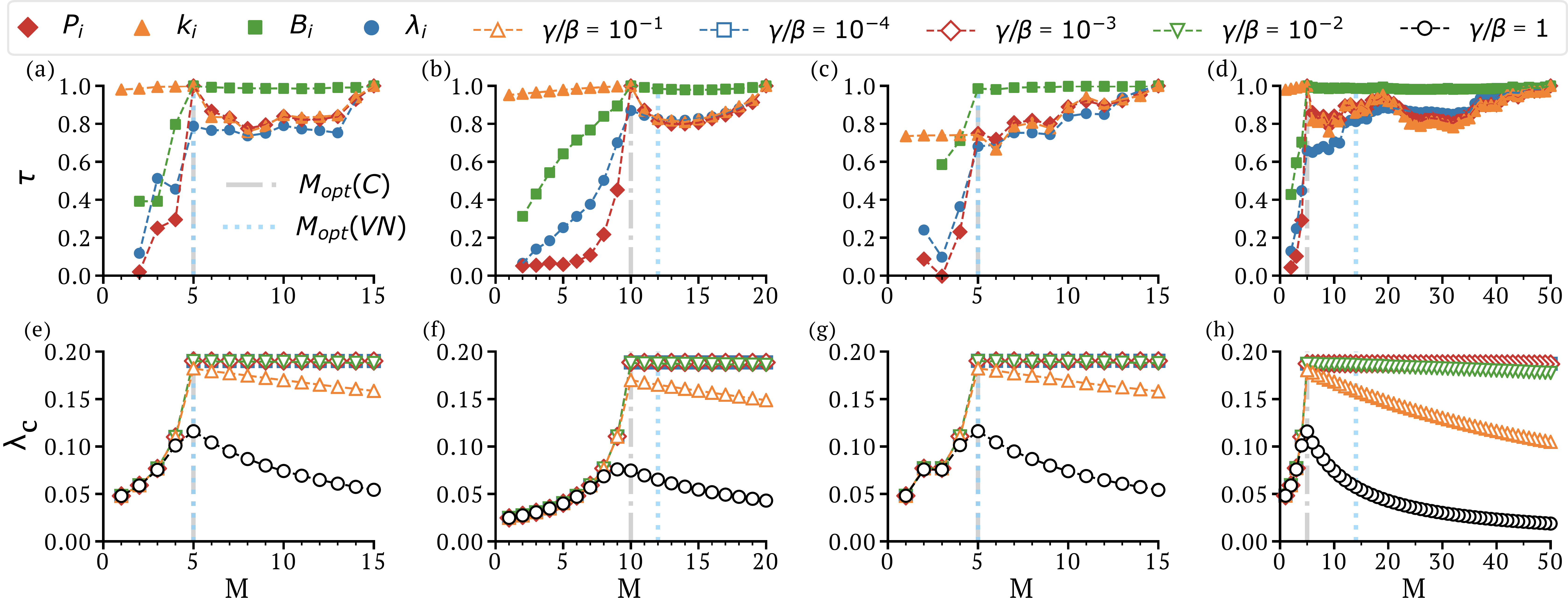}
	\caption{(Colour online) Impact of 
		reducibility in 
		altering structural and dynamical 
		properties of 
		four synthetic multiplex networks. For each 
		of the 
		four synthetic multiplex presented in Table 
		\ref{table:synthetic_benchmark_sametopology},
		we 
		report the Kendall's $\tau$ 
		correlation of the four structural 
		descriptors (top 
		row) and the critical threshold of the SIS 
		dynamic 
		(bottom row) as a function of the greedy 
		aggregation steps of the reducibility 
		procedure. 
		For each network we also indicate the 
		number of layers 
		in the optimal reduced system identified by 
		multiplex 
		complexity ($M_{opt}(C)$, grey	
		dash-and-dot line) and 
		by Von 	Neumann entropy ($M_{opt}(VN)$, 
		blue dotted 
		line)~\cite{DeDomenico_Nicosia_2015}. 
		Interestingly, the reducibility based on 
		Complexity 
		seems to correctly identify the optimal 
		aggregation 
		made of only distinct layers, therefore 
		outperforming the method based on the Von 
		Neumann entropy.}
	\label{figure:synthetic_benchmark_sametopology}	
\end{figure}

\begin{table}[b!]
	\centering
	\begin{tabular}{|c|ccccc|}
		\hline
		Benchmark ID & M  & $M_{distinct}$ & $pattern$ 
		& $M_{opt}(C)$ & $M_{opt}(VN)$ \\
		\hline
		1 & 15 & 5 & \{3,3,3,3,3\} & 5 & 5 \\
		2 & 20 & 10 & \{2,2,2,2,2,2,2,2,2,2\} & 10 & 12 
		\\
		3 & 15 & 5 & \{3,2,1,4,5\} & 5 & 5 \\
		4 & 50 & 5 & \{10,10,10,10,10\} & 5 & 14 \\
		\hline
	\end{tabular}
	\caption{Reducibility of four synthetic 
	  multiplex 
		benchmarks. From
		left to right, we report the total number of 
		layers $M$, the total
		number of distinct layers $M_{distinct}$, the 
		pattern of identical
		layers in the multiplex, and the number of 
		optimal layers
		$M_{opt}(C)$ obtained when maximising the 
		quality function
		$q(\bullet)$. The last column reports the 
		optimal number of layers
		$M_{opt}(VN)$ when using the multiplex 
		structural reducibility
		procedure described in 
		Ref.~\cite{DeDomenico_Nicosia_2015}. Notice
		that in all the cases $\complM$ correctly 
		identifies both the
		correct sequence of aggregation steps and the 
		best partition,
		outperforming the procedure based on the Von 
		Neumann entropy.}
	\label{table:synthetic_benchmark_sametopology}
\end{table}
To test the performance of our reducibility 
procedure, we created new ad-hoc synthetic multiplex 
networks by tuning the total number of layers and by 
considering different groups of identical layers. 
Indeed, as presented in the main paper, we constructed 
several benchmarks where pairs or group of
layers are identical to each other (i.e., whose layer 
adjacency matrices are identical). In this way, the 
number of truly distinct layers is by construction 
smaller than or equal to the total number of
layers $M$. In all the benchmarks layers are 
Erd\"os-Renyi random graphs with $N=1000$ and $\langle 
k \rangle = 4$.\\
We report in 
Table~\ref{table:synthetic_benchmark_sametopology}
four different synthetic benchmarks, along with the 
pattern of
identical layers in each multiplex. For instance, the 
pattern
$\{3,3,3,3,3\}$ in the first benchmark corresponds to a 
multiplex with
$M=15$ layers where every subsequent triplets of layers 
are 
identical,
i.e., \textit{layer1} is identical to \textit{layer2} 
and \textit{layer3}, 
\textit{layer4}
is identical to \textit{layer5} and \textit{layer6}, 
and so on.  We report 
in the third
column of the table the number $M_{distinct}$ of layers 
that are truly
distinct, in the sense mentioned above.\\
\indent
Notice that in all the benchmark considered, the 
reducibility measure
based on the multiplex complexity $\complM$ identifies 
both the
correct sequence of aggregation steps and the optimal
partition. Conversely, the reducibility procedure based 
on the Von	Neumann entropy introduced in 	
Ref.~\cite{DeDomenico_Nicosia_2015}	fails in some 
particular instances.
For the sake of clarity, we also report in
Figure~\ref{figure:synthetic_benchmark_sametopology} 
the Kendall's 
$\tau$ correlation coefficient of 
the rankings induced by total node degree, node 
activity, participation coefficient, and node
interdependence (Fig. 
\ref{figure:synthetic_benchmark_sametopology} - top 
row) and the critical threshold $\lambda_c$
of the SIS dynamic (Fig. 
\ref{figure:synthetic_benchmark_sametopology}
- bottom row) as a function of the greedy 
aggregation steps for the four synthetic 
benchmarks presented in Table 
\ref{table:synthetic_benchmark_sametopology}.

\subsection{Multiplex with different 
	topologies}
\begin{figure}[t!]
	\centering
	\includegraphics[width=0.95\textwidth]{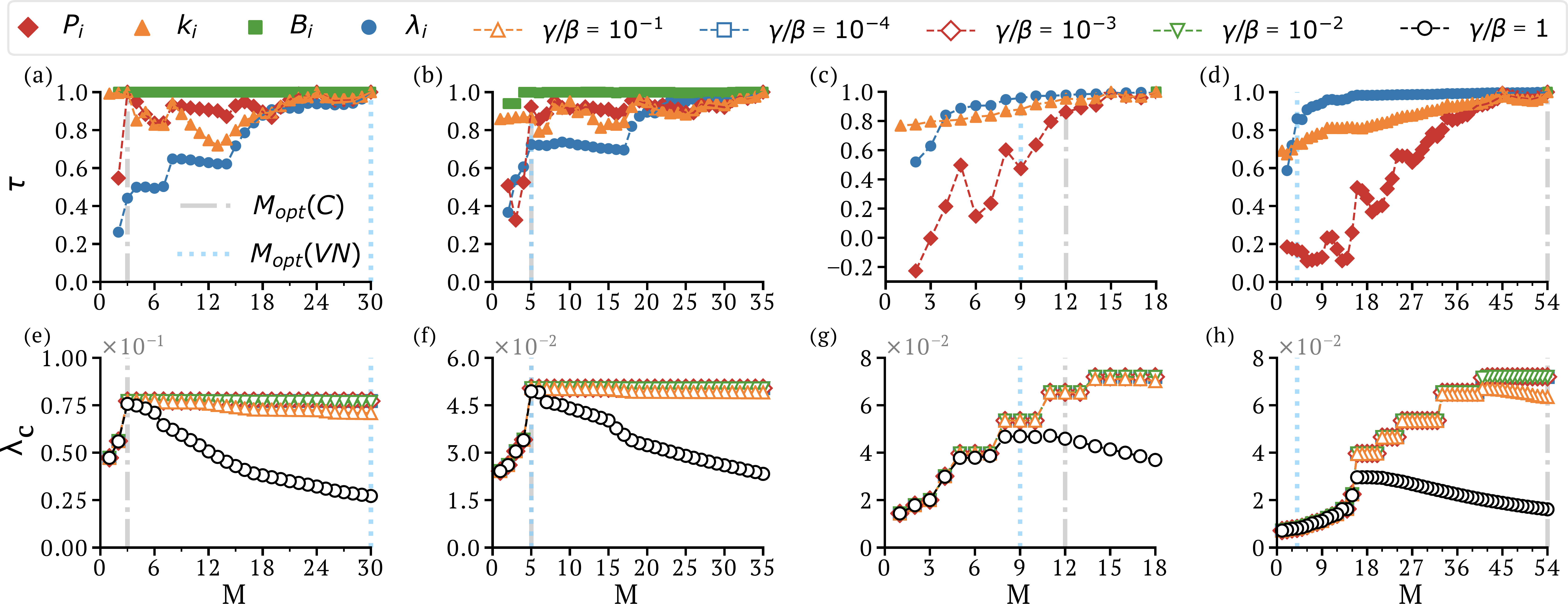}
	\caption{(Colour online) Impact of 
		reducibility in altering structural and 
		dynamical properties of four synthetic 
		multiplex networks made of layers with 
		different topologies and average degrees. 
		For each of the four synthetic multiplex, 
		namely made of 30, 35, 18 and 54 layers,
		we report the Kendall's $\tau$ correlation 
		of the four structural descriptors (top 
		row) and the critical threshold of the SIS  
		dynamic bottom row) as a function of the 
		greedy aggregation steps of the 
		reducibility procedure. For each network we 
		also indicate the number of layers in the 
		optimal reduced system identified by 
		multiplex complexity ($M_{opt}(C)$, grey	
		dash-and-dot line) and 
		by Von 	Neumann entropy ($M_{opt}(VN)$, 
		blue dotted 
		line)~\cite{DeDomenico_Nicosia_2015}. 
		Interestingly, also in this case the 
		quality function based on Complexity 
		shows a more conservative behaviour 
		compared to the one based on the Von 
		Neumann Entropy. In addition, notice 
		that the 
		greedy aggregation steps for both the 
		methods are almost identical, so that 
		couple of layers having the same topology 
		will be aggregated first in the procedure.}
	\label{figure:synthetic_benchmark_differenttopology}
\end{figure}

We additionally test the reducibility procedure on 
four other synthetic benchmarks, where the average 
degree and the topology of each layer are not kept 
fixed. In particular, we firstly consider two 
benchmarks with $N =1000$ nodes, constructed through 
two 
different models, namely, Barabasi-Albert linear 
preferential attachment graphs (BA), and ER graphs.  
We consider a multiplex consisting of M= 
30 layers, where every group composed of 10 layers is 
respectively identical to a BA with $m=3$, an ER with 
$\langle k \rangle =4$, and an ER with $\langle k 
\rangle=8$. Within this setting,  the number of 
distinct layers is by construction equal to 3. In the 
second 
benchmark, the number of total layers is equal to 
$M=35$. In this case, we vary the average degree 
for both the ER and BA models. The first five layers 
correspond to a BA  with $m=3$, next five layers are 
identical to a single realisation of a BA model
with $m=5$. The remaining layers are grouped in 7,6, 
and 10 layers respectively which are sampled from an ER 
graphs with $\langle k \rangle=4,6,8$. Thus, the number 
of truly distinct layers are only 5.\\
The last two benchmarks are similar to the one 
introduced in \cite{Nicosia_Latora_2015} (Supplementary
Note 1), with layers drawn from BA with $m=4$, ER 
with 
$p=0.05$ and Watts-Strogatz small-world models (WS, $m 
= 5, p = 0.2$). More precisely, the two benchmarks have
$N=200$ nodes with respectively one (three) realisations
for each model, named ``master layers'' and respectively
called BA 
1 (BA 2, BA 3), ER 1 (ER 2, ER 3), and  WS 1 (WS 2, WS 
3). For each realisation, we constructed 5 more 
layers, each characterised by an increasing
amount of edge intersection with the corresponding 
master layer, i.e 10\%, 25\%, 50\%, 75\%, 100\%. Thus, 
we obtained two synthetic benchmarks respectively made 
of 3 (9) groups of 6 layers each, for a total
of $M =18, 54$ layers.
We report in 
Figure~\ref{figure:synthetic_benchmark_differenttopology}
the Kendall's $\tau$ correlation coefficient of the 
rankings induced by total node degree, node activity, 
participation coefficient, and node interdependence 
(Fig. 
\ref{figure:synthetic_benchmark_differenttopology} - 
top row) and the critical threshold 
of the SIS dynamic (Fig. 
\ref{figure:synthetic_benchmark_differenttopology} - 
bottom row) as a function of the greedy 
aggregation steps for these four synthetic 
benchmarks.\\
For the first two synthetic benchmarks (Fig. 
\ref{figure:synthetic_benchmark_differenttopology}(a,b-e,f))
the reducibility measure based 
on Complexity seems to outperform the one based Von 
Neumann Entropy. Indeed, the optimal partition made of 
only distinct layers is only identified by the 
complexity reducibility. However, when considering the 
last two benchmarks, we again observe that the quality 
function $q(\bullet)$ based on Complexity seems to be 
very conservative (in the some cases, extremely 
conservative) compared to the procedure based on the 
Von Neumann entropy. Remarkably, in both cases, the 
aggregation steps are almost the same, so that layers 
having the same topology will be aggregated  first in 
the procedure.
\newpage
\section{Time-varying multiplex networks}
\begin{figure}[bt!]
	\centering
	\includegraphics[width=0.95 \textwidth]{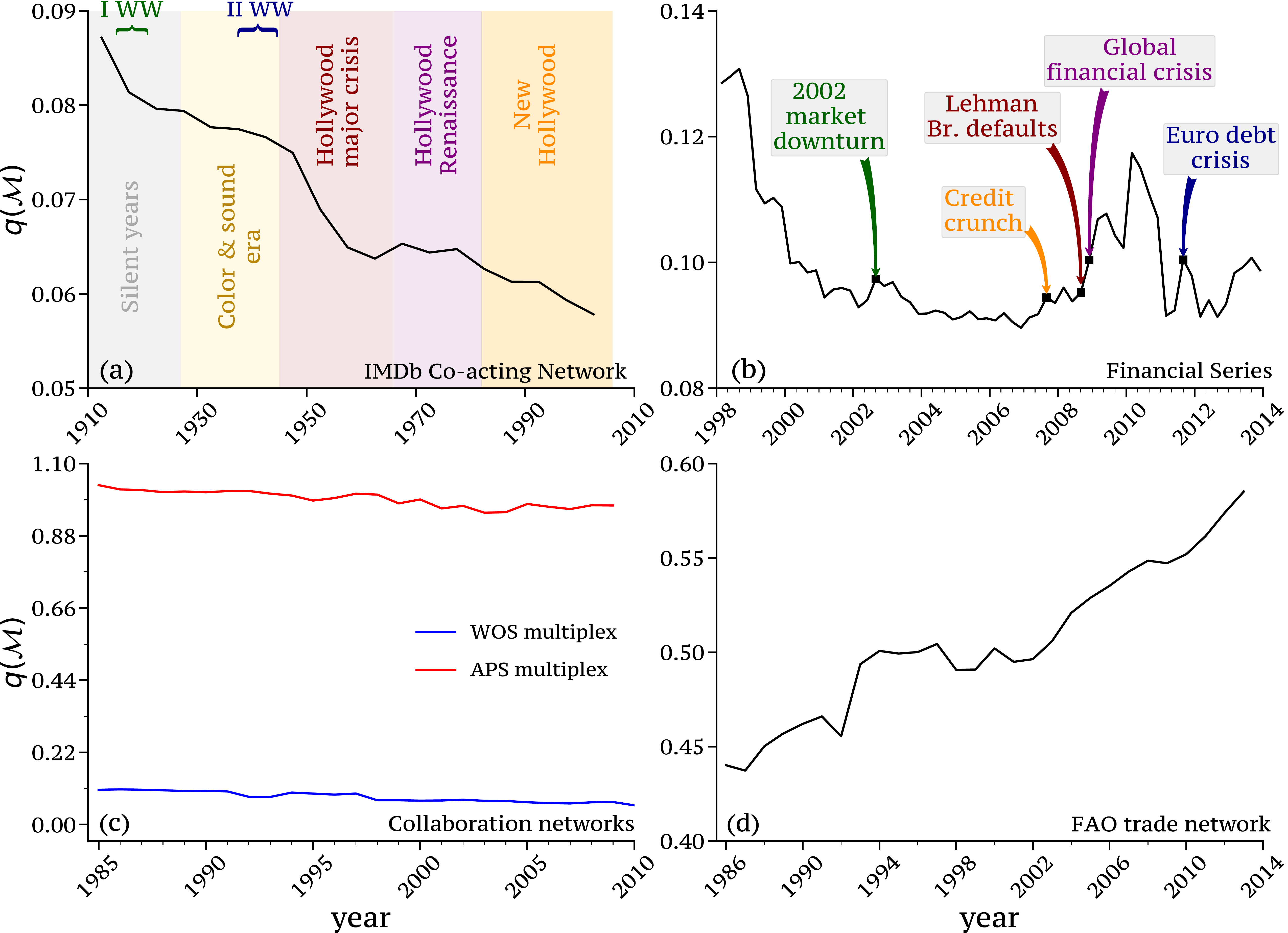}
	\caption{(Colour online) $q$ 
		as a function of time for four different 
		time-varying multiplex networks, namely, 
		(a) the IMDb co-starring network, 
		(b) the financial multiplex constructed 
		from price time series of 35 major assets in NYSE 
		and NASDAQ, (c) the physics collaboration 
		multiplex network of the American Physical Society 
		(APS) and Web of Science (WOS),and (d)
		the FAO food import/export multiplex network. 
		Unfortunately, the quality function $q$ does not 
		account for the variation of the number of edges in 
		the aggregate, so that the results obtained do not 
		show any significant patterns over time.}
	\label{figure:time_series_qe}
\end{figure}

Here, we analyse the impact of different 
functions in tracking the evolution of time-varying 
real multiplex network. We report in Figure 
\ref{figure:time_series_qe} the 
results obtained for the same four data sets presented 
in the main text using the quality function 
$q(\mathcal{M}) = \complM/\log{K_{\mathcal{M}}}$, where 
$K_{\mathcal{M}}$  represents the number of links in 
the multiplex. Notice that in this case the function 
$q$  does not perform well when describing  changes 
with time. This issue intrinsically lies in the 
definition of $q$, which does not consider the 
variation of the number of edges in the 
aggregate over time.

\begin{figure}[hbt!]
	\centering
	\includegraphics[width=0.95\textwidth]{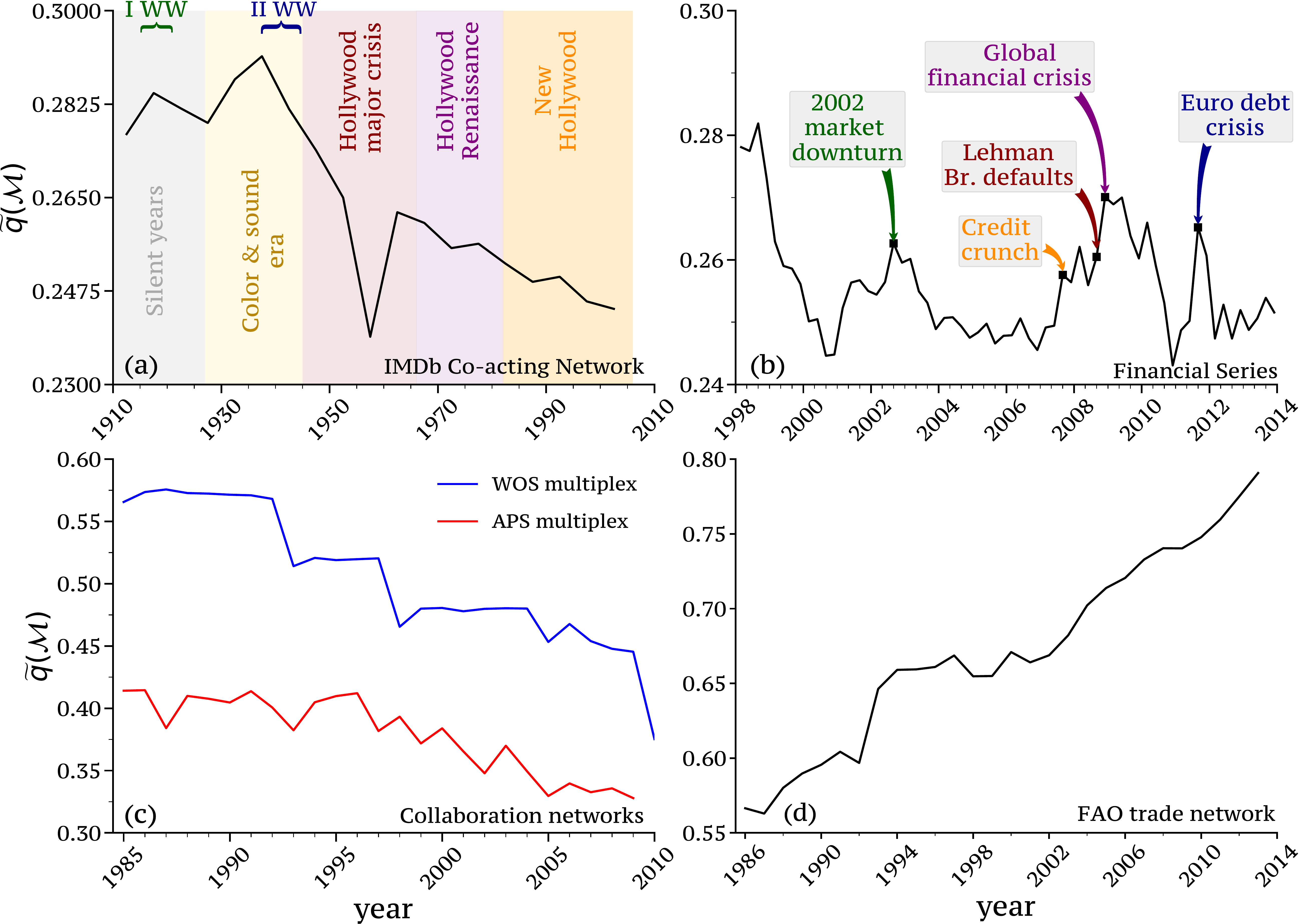}
	\caption{(Colour online) $\widetilde{q}$ as a 
	  function of time for four different time-varying 
	  multiplex networks, namely, 
		(a) the IMDb
		co-starring network, (b) the 
		financial 
		multiplex
		constructed from price time series of 35 major 
		assets in NYSE and
		NASDAQ, (c) the physics collaboration 
		multiplex network
		of the American Physical Society (APS) and Web 
		of Science (WOS),
		and (d) the FAO food import/export 
		multiplex
		network. Also in this case, we observe that the 
		most pronounced peaks of 
		the function $\widetilde{q}$ in (a) 
		and (b) 
		correspond to periods of
		instability and crises in the corresponding 
		systems. Conversely,
		the values of complexity in the physics 
		collaboration multiplex,
		for both the APS and WOS data sets 
		(c), 
		have remained
		pretty stable over time, and reveal that those 
		systems indeed
		benefit only marginally from a multi-layer
		representation. Finally, in the FAO food 
		import/export multiplex
		network the complexity has kept increasing 
		considerably over time
		(d), reflecting the relevant role 
		played by globalisation
		in the last twenty years in re-shaping the 
		international food
		market.}
	\label{figure:time_series_qe_withaggregate}
\end{figure}

To test our hypothesis, we therefore considered a 
modified version of the quality function $q$, defined 
as:
\begin{equation}
	\widetilde{q}(\mathcal{X}) = 
	\frac{\mathcal{C}(\mathcal{X})}{log (M * 
	  K_{\mathcal{X}}/K_\mathcal{A})}
\end{equation}
where $M$ represents the number of layers of the 
multiplex $\mathcal{X}$, while $K_{\mathcal{X}}$ and 
$K_\mathcal{A}$ respectively represent the number of 
link in the multiplex and in the aggregate.\\
In Figure \ref{figure:time_series_qe_withaggregate}, we 
report the values of the quality function 
$\widetilde{q}$ over time, for the same four 
time-varying multiplex networks. Interestingly, we 
observe a similar picture of the one presented in the 
main article, so that local maxima of complexity are 
consistent with the most notable periods of crisis in 
each data set, while local minima of complexity seem to 
be precursors of renaissance in IMDB or stability in 
the financial market.

\section{Carthography of real systems}
\begin{figure}[b!]
	\centering
	\includegraphics[width=0.95\textwidth]{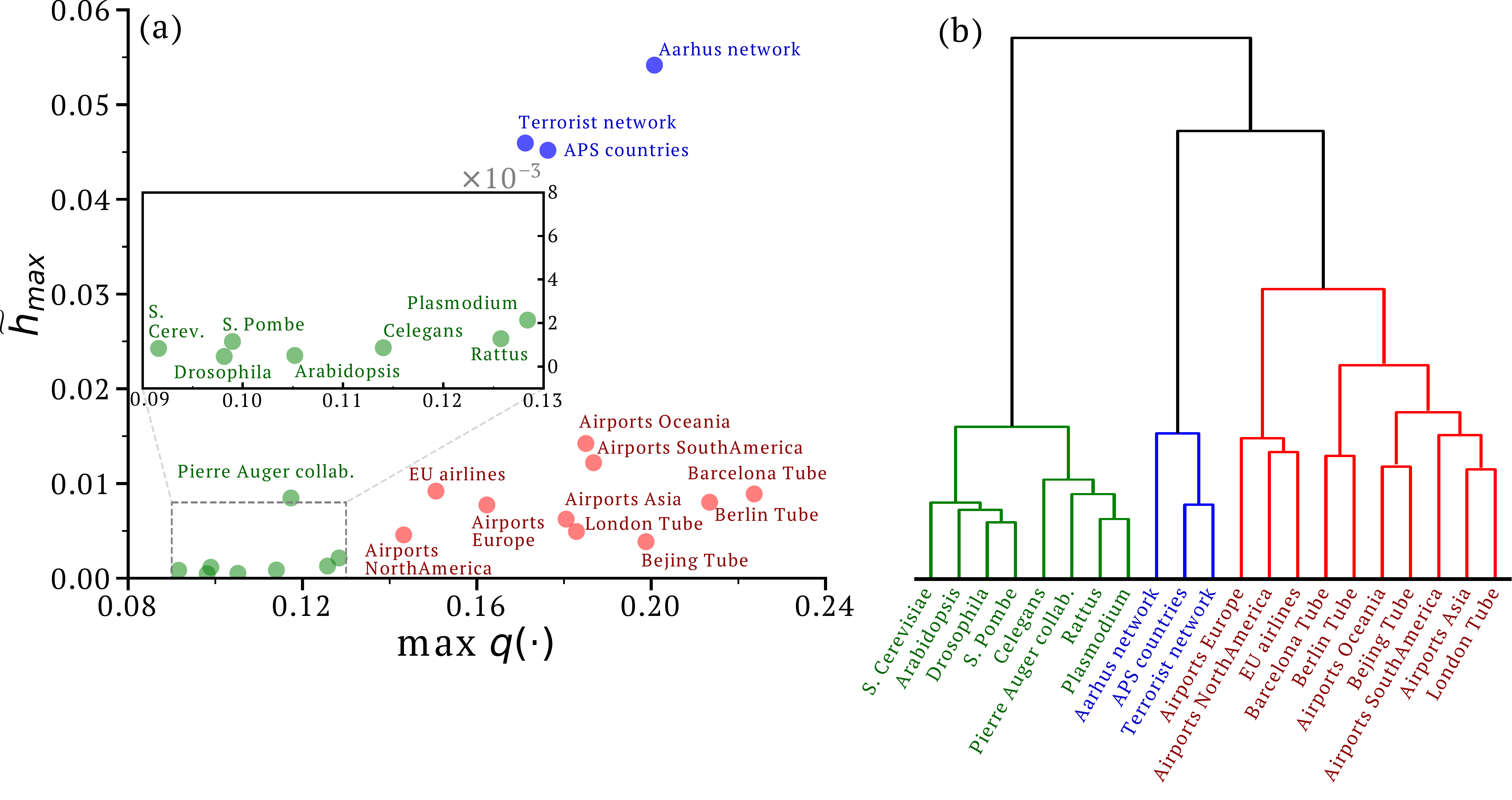}
	\caption{(Colour online) Multiplex cartography 
	  of 
		real-world systems in
		the plane $max \; q(\cdot) - \tilde{h}_{max}$ 
		(a) and the
		corresponding dendrogram obtained through 
		Ward's method hierarchical agglomerative 
		clustering. It 
		looks like these two structural descriptors 
		alone are 
		able to 
		identify three 
		classes of real-world multiplex networks, 
		namely, biological networks (green cluster), 
		social 
		systems (blue cluster), technological systems 
		(red 
		cluster).}
	\label{figure:multiplex_cartography}
\end{figure}
In this section we further show how multiplex complexity 
can be used to obtain a planar embedding of multiplex 
networks, in a similar fashion as presented in the main 
text. In this case, for each multiplex network in the data 
set, we use the maximum value of the quality function 
$\max q(\bullet)$ 
as one of the
coordinates, while we employ the maximal entropy rate per 
node 
$\tilde{h}_{\rm
	max}$ as the other one. 
Note that
$\tilde{h}_{\rm max}$ is linked to the dispersiveness 
of random walks
on a graph, and it thus carries information about the 
the large-scale
dynamical properties of a 
multiplex~\cite{Battiston_exploration_2016}.  Notice that 
for a multiplex $\mathcal{M}$, the maximal entropy 
rate is defined as:
\begin{equation}
  h_{max}= \log \lambda_{max}	
\end{equation}
where $\lambda_{max}$ is the maximum eigenvalue of the 
overlapping
matrix $O$ associated to
$\mathcal{M}$~\cite{Battiston_exploration_2016}. To 
account for the dependence of $\lambda_{max}$ on the total 
number of  nodes in the graph, we use the normalised 
maximal entropy rate:
\begin{equation}
  \tilde{h}_{\rm max} = \frac{h_{\rm max}}{N}
\end{equation}
Thus, as mentioned also in the main paper, also this 
measure provide a perspective on a system  that is 
orthogonal to that captured by multiplex complexity, which 
is instead a purely structural quantity. In  
Fig.~\ref{figure:multiplex_cartography}
we indicated with different colours the three largest 
group obtained through Ward's method~\cite{Ward1963hierarchical},
while in Fig.~\ref{figure:multiplex_cartography}(b) 
we show the corresponding dendrogram, highlighting 
all the aggregation steps. 
Interestingly, solely with these two measures we are able 
to group together biological, social and technological 
network in three different clusters.

\FloatBarrier

\section{Numerical approaches for node interdependence}
The encoding of a multiplex network through the prime-weight matrix
$\Omega$ allows to define an efficient algorithm for the computation
of node interdependence. The node interdependence is defined as
follows:
\begin{equation*}
	\lambda_i = \frac{1}{N-1}\sum_{\substack{j \in N \\ j \neq i}} \frac{\psi_{ij}}{\sigma_{ij}}
	\label{eq:node_interdependence}
\end{equation*}
where $\psi_{ij}$ represents the number of shortest paths from $i$ to
$j$ that use edges lying in at least two layers, while $\sigma_{ij}$
represents the total number of shortest paths between node $i$ and $j$
in the multiplex.  With a naive approach, the computation of such
quantity requires computing the number of shortest paths in the
multiplex, which usually scales exponentially with the number of
layers. If we consider the prime-weight matrix $\Omega$ associated to
a multiplex $\mathcal{M}$, we can compute the node interdependence of
each node using elementary properties of composite numbers. The
procedure is detailed below:

\begin{center}
	\label{alg:Interdependence}
	\fbox{
		\parbox{14cm}{
			\centering  
			%\begin{tcolorbox}[width=14 cm,left=0pt, right=0pt,top=4pt,colback=ocre!80,arc=0pt,outer arc=0pt,colframe=black,boxrule=0.0pt]
			\flushleft{\textbf{Algorithm: node interdependence} $\lambda_i$ }
			\normalsize
			%\vspace*{0.2 cm}
			\begin{enumerate}
			\item[\textbf{1.}] Calculate the unweighted aggregate ${W}$
				and prime-weight matrix ${\Omega} = \{ \Omega_{ij} \}$
				associated to the multiplex $\mathcal{M}$.\\ Enumerate all
				the shortest paths from node $i$ using the Breadth-First
				Search (BFS) algorithm \cite{newman2018networks} applied to
				the graph associated to W. Set $\lambda_i \leftarrow 0$.
			\item[\textbf{2.}]  Set $\sigma_{ij}\leftarrow 0$ and $\psi_{ij} \leftarrow 0$.
			\item[\textbf{3.}] We indicate the generic shortest path from $i$ to $j$ as the sequence $T=\{i,n_1,n_2, \ldots, n_k, j\}$, where $j \neq i$. We also indicate as $f(\Omega_{ij})$ the number of prime factors of $\Omega_{ij}$ .
			\item[\textbf{4.}] Compute $P= \displaystyle\prod_{s =i, s \in
				T}^{n_k} f(\Omega_{s,\,s+1}) $ and set $ \sigma_{ij}
				\leftarrow \sigma_{ij} + P$.
			\item[\textbf{5.}] Set $G \leftarrow GCD(\Omega_{i,\,n_1},
				\Omega_{n_1,\,n_2}, \ldots, \Omega_{n_k, \, j})$. If $G$ is
				not equal to 1, then $\psi_{ij} \leftarrow \psi_{ij} +
				f(G)$.  Repeat from step 3. for all the shortest paths from
				$i$ to $j$.
			\item[\textbf{6.}]  Set $\lambda_i \leftarrow \lambda_i +
				\frac{\sigma_{ij}- \psi_{ij}}{\sigma_{ij}}$. Repeat from
				step $2.$ for all the nodes $j$ different from $i$.
			\end{enumerate}
	}}
\end{center}

In the algorithm, $GCD(a, b)$ is the greatest common divisor of $a $
and $b$ (i.e., the largest positive integer that divides both $a$ and
$b$ without a remainder). The algorithm combines two major ingredients
from number theory, namely the unique factorisation theorem
(i.e. using the prime-weight matrix) and the properties of GCD. On the one
hand, with the prime-weight matrix it is possible to count all the
available shortest paths between two pairs of nodes in the multiplex
networks. On the other hand, the GCD properties lead to a fast
distinction between paths, selecting paths lying in just one layer to
the others.


\begin{thebibliography}{53}%
	\makeatletter
	\providecommand \@ifxundefined [1]{%
		\@ifx{#1\undefined}
	}%
	\providecommand \@ifnum [1]{%
		\ifnum #1\expandafter \@firstoftwo
		\else \expandafter \@secondoftwo
		\fi
	}%
	\providecommand \@ifx [1]{%
		\ifx #1\expandafter \@firstoftwo
		\else \expandafter \@secondoftwo
		\fi
	}%
	\providecommand \natexlab [1]{#1}%
	\providecommand \enquote  [1]{``#1''}%
	\providecommand \bibnamefont  [1]{#1}%
	\providecommand \bibfnamefont [1]{#1}%
	\providecommand \citenamefont [1]{#1}%
	\providecommand \href@noop [0]{\@secondoftwo}%
	\providecommand \href [0]{\begingroup \@sanitize@url \@href}%
	\providecommand \@href[1]{\@@startlink{#1}\@@href}%
	\providecommand \@@href[1]{\endgroup#1\@@endlink}%
	\providecommand \@sanitize@url [0]{\catcode `\\12\catcode `\$12\catcode
		`\&12\catcode `\#12\catcode `\^12\catcode `\_12\catcode `\%12\relax}%
	\providecommand \@@startlink[1]{}%
	\providecommand \@@endlink[0]{}%
	\providecommand \url  [0]{\begingroup\@sanitize@url \@url }%
	\providecommand \@url [1]{\endgroup\@href {#1}{\urlprefix }}%
	\providecommand \urlprefix  [0]{URL }%
	\providecommand \Eprint [0]{\href }%
	\providecommand \doibase [0]{https://doi.org/}%
	\providecommand \selectlanguage [0]{\@gobble}%
	\providecommand \bibinfo  [0]{\@secondoftwo}%
	\providecommand \bibfield  [0]{\@secondoftwo}%
	\providecommand \translation [1]{[#1]}%
	\providecommand \BibitemOpen [0]{}%
	\providecommand \bibitemStop [0]{}%
	\providecommand \bibitemNoStop [0]{.\EOS\space}%
	\providecommand \EOS [0]{\spacefactor3000\relax}%
	\providecommand \BibitemShut  [1]{\csname bibitem#1\endcsname}%
	\let\auto@bib@innerbib\@empty
	%</preamble>
\bibitem [{\citenamefont {Newman}(2010)}]{Newman_book2010}%
	\BibitemOpen
	\bibfield  {author} {\bibinfo {author} {\bibfnamefont {M.}~\bibnamefont
			{Newman}},\ }\href@noop {} {\emph {\bibinfo {title} {Networks: An
				Introduction}}}\ (\bibinfo  {publisher} {Oxford University Press},\ \bibinfo
	           {address} {New York},\ \bibinfo {year} {2010})\BibitemShut {NoStop}%
	         \bibitem [{\citenamefont {Latora}\ \emph {et~al.}(2017)\citenamefont {Latora},
		           \citenamefont {Nicosia},\ and\ \citenamefont
		                         {Russo}}]{Latora_Nicosia_Russo_book2017}%
	           \BibitemOpen
	           \bibfield  {author} {\bibinfo {author} {\bibfnamefont {V.}~\bibnamefont
			           {Latora}}, \bibinfo {author} {\bibfnamefont {V.}~\bibnamefont {Nicosia}},\
		           and\ \bibinfo {author} {\bibfnamefont {G.}~\bibnamefont {Russo}},\
	           }\href@noop {} {\emph {\bibinfo {title} {Complex Networks: Principles,
				           Methods and Applications}}}\ (\bibinfo  {publisher} {Cambridge University
		           Press},\ \bibinfo {year} {2017})\BibitemShut {NoStop}%
	         \bibitem [{\citenamefont {Pastor-Satorras}\ \emph {et~al.}(2015)\citenamefont
		           {Pastor-Satorras}, \citenamefont {Castellano}, \citenamefont {Van~Mieghem},\
		           and\ \citenamefont {Vespignani}}]{Pastor-Satorras2015}%
	           \BibitemOpen
	           \bibfield  {author} {\bibinfo {author} {\bibfnamefont {R.}~\bibnamefont
			           {Pastor-Satorras}}, \bibinfo {author} {\bibfnamefont {C.}~\bibnamefont
			           {Castellano}}, \bibinfo {author} {\bibfnamefont {P.}~\bibnamefont
			           {Van~Mieghem}},\ and\ \bibinfo {author} {\bibfnamefont {A.}~\bibnamefont
			           {Vespignani}},\ }\bibfield  {title} {\emph {\bibinfo {title} {Epidemic
				           processes in complex networks}},\ }\href
	                      {https://doi.org/10.1103/RevModPhys.87.925} {\bibfield  {journal} {\bibinfo
			                      {journal} {Rev. Mod. Phys.}\ }\textbf {\bibinfo {volume} {87}},\ \bibinfo
		                      {pages} {925} (\bibinfo {year} {2015})}\BibitemShut {NoStop}%
	                    \bibitem [{\citenamefont {Arenas}\ \emph {et~al.}(2008)\citenamefont {Arenas},
		                      \citenamefont {D{\'\i}az-Guilera}, \citenamefont {Kurths}, \citenamefont
		                                    {Moreno},\ and\ \citenamefont {Zhou}}]{Arenas2008}%
	                      \BibitemOpen
	                      \bibfield  {author} {\bibinfo {author} {\bibfnamefont {A.}~\bibnamefont
			                      {Arenas}}, \bibinfo {author} {\bibfnamefont {A.}~\bibnamefont
			                      {D{\'\i}az-Guilera}}, \bibinfo {author} {\bibfnamefont {J.}~\bibnamefont
			                      {Kurths}}, \bibinfo {author} {\bibfnamefont {Y.}~\bibnamefont {Moreno}},\
		                      and\ \bibinfo {author} {\bibfnamefont {C.}~\bibnamefont {Zhou}},\ }\bibfield
	                                 {title} {\emph {\bibinfo {title} {Synchronization in complex networks}},\
	                                 }\href {https://doi.org/10.1016/j.physrep.2008.09.002} {\bibfield  {journal}
		                                 {\bibinfo  {journal} {Phys. Rep}\ }\textbf {\bibinfo {volume} {469}},\
		                                 \bibinfo {pages} {93} (\bibinfo {year} {2008})}\BibitemShut {NoStop}%
	                               \bibitem [{\citenamefont {Jaynes}(1957)}]{Jaynes1957information}%
	                                 \BibitemOpen
	                                 \bibfield  {author} {\bibinfo {author} {\bibfnamefont {E.~T.}\ \bibnamefont
			                                 {Jaynes}},\ }\bibfield  {title} {\emph {\bibinfo {title} {Information theory
				                                 and statistical mechanics}},\ }\href
	                                            {https://doi.org/10.1103/PhysRev.106.620} {\bibfield  {journal} {\bibinfo
			                                            {journal} {Phys. Rev.}\ }\textbf {\bibinfo {volume} {106}},\ \bibinfo {pages}
		                                            {620} (\bibinfo {year} {1957})}\BibitemShut {NoStop}%
	                                          \bibitem [{\citenamefont {Bianconi}(2008)}]{Bianconi_2008}%
	                                            \BibitemOpen
	                                            \bibfield  {author} {\bibinfo {author} {\bibfnamefont {G.}~\bibnamefont
			                                            {Bianconi}},\ }\bibfield  {title} {\emph {\bibinfo {title} {The entropy of
				                                            randomized network ensembles}},\ }\href
	                                                       {https://doi.org/10.1209/0295-5075/81/28005} {\bibfield  {journal} {\bibinfo
			                                                       {journal} {Europhys. Lett.}\ }\textbf {\bibinfo {volume} {81}},\ \bibinfo
		                                                       {pages} {28005} (\bibinfo {year} {2008})}\BibitemShut {NoStop}%
	                                                     \bibitem [{\citenamefont {Anand}\ and\ \citenamefont
		                                                       {Bianconi}(2009)}]{Anand2009entropy}%
	                                                       \BibitemOpen
	                                                       \bibfield  {author} {\bibinfo {author} {\bibfnamefont {K.}~\bibnamefont
			                                                       {Anand}}\ and\ \bibinfo {author} {\bibfnamefont {G.}~\bibnamefont
			                                                       {Bianconi}},\ }\bibfield  {title} {\emph {\bibinfo {title} {Entropy measures
				                                                       for networks: Toward an information theory of complex topologies}},\ }\href
	                                                                  {https://doi.org/10.1103/PhysRevE.80.045102} {\bibfield  {journal} {\bibinfo
			                                                                  {journal} {Phys. Rev. E}\ }\textbf {\bibinfo {volume} {80}},\ \bibinfo
		                                                                  {pages} {045102} (\bibinfo {year} {2009})}\BibitemShut {NoStop}%
	                                                                \bibitem [{\citenamefont {Dehmer}(2008)}]{Dehmer2008information}%
	                                                                  \BibitemOpen
	                                                                  \bibfield  {author} {\bibinfo {author} {\bibfnamefont {M.}~\bibnamefont
			                                                                  {Dehmer}},\ }\bibfield  {title} {\emph {\bibinfo {title} {Information
				                                                                  processing in complex networks: Graph entropy and information functionals}},\
	                                                                  }\href {https://doi.org/https://doi.org/10.1016/j.amc.2007.12.010} {\bibfield
		                                                                  {journal} {\bibinfo  {journal} {Appl. Math. Comput.}\ }\textbf {\bibinfo
			                                                                  {volume} {201}},\ \bibinfo {pages} {82} (\bibinfo {year} {2008})}\BibitemShut
	                                                                             {NoStop}%
	                                                                           \bibitem [{\citenamefont {Passerini}\ and\ \citenamefont
		                                                                             {Severini}(2009)}]{Passerini2009}%
	                                                                             \BibitemOpen
	                                                                             \bibfield  {author} {\bibinfo {author} {\bibfnamefont {F.}~\bibnamefont
			                                                                             {Passerini}}\ and\ \bibinfo {author} {\bibfnamefont {S.}~\bibnamefont
			                                                                             {Severini}},\ }\bibfield  {title} {\emph {\bibinfo {title} {Quantifying
				                                                                             complexity in networks: the von Neumann entropy}},\ }\href
	                                                                                        {https://doi.org/10.4018/jats.2009071005} {\bibfield  {journal} {\bibinfo
			                                                                                        {journal} {Int. J. Agents Technol. Syst.}\ }\textbf {\bibinfo {volume} {1}},\
		                                                                                        \bibinfo {pages} {58} (\bibinfo {year} {2009})}\BibitemShut {NoStop}%
	                                                                                      \bibitem [{\citenamefont {Mowshowitz}\ and\ \citenamefont
		                                                                                        {Dehmer}(2012)}]{Mowshowitz2012entropy}%
	                                                                                        \BibitemOpen
	                                                                                        \bibfield  {author} {\bibinfo {author} {\bibfnamefont {A.}~\bibnamefont
			                                                                                        {Mowshowitz}}\ and\ \bibinfo {author} {\bibfnamefont {M.}~\bibnamefont
			                                                                                        {Dehmer}},\ }\bibfield  {title} {\emph {\bibinfo {title} {Entropy and the
				                                                                                        complexity of graphs revisited}},\ }\href {https://doi.org/10.3390/e14030559}
	                                                                                                   {\bibfield  {journal} {\bibinfo  {journal} {Entropy}\ }\textbf {\bibinfo
			                                                                                                   {volume} {14}},\ \bibinfo {pages} {559} (\bibinfo {year} {2012})}\BibitemShut
	                                                                                                   {NoStop}%
	                                                                                                 \bibitem [{\citenamefont {Dehmer}\ and\ \citenamefont
		                                                                                                   {Mowshowitz}(2011)}]{Dehmer_history_entropy_2011}%
	                                                                                                   \BibitemOpen
	                                                                                                   \bibfield  {author} {\bibinfo {author} {\bibfnamefont {M.}~\bibnamefont
			                                                                                                   {Dehmer}}\ and\ \bibinfo {author} {\bibfnamefont {A.}~\bibnamefont
			                                                                                                   {Mowshowitz}},\ }\bibfield  {title} {\emph {\bibinfo {title} {A history of
				                                                                                                   graph entropy measures}},\ }\href
	                                                                                                              {https://doi.org/https://doi.org/10.1016/j.ins.2010.08.041} {\bibfield
		                                                                                                              {journal} {\bibinfo  {journal} {Info. Sci.}\ }\textbf {\bibinfo {volume}
			                                                                                                              {181}},\ \bibinfo {pages} {57 } (\bibinfo {year} {2011})}\BibitemShut
	                                                                                                              {NoStop}%
	                                                                                                            \bibitem [{\citenamefont {Cimini}\ \emph {et~al.}(2019)\citenamefont {Cimini},
		                                                                                                              \citenamefont {Squartini}, \citenamefont {Saracco}, \citenamefont
		                                                                                                                            {Garlaschelli}, \citenamefont {Gabrielli},\ and\ \citenamefont
		                                                                                                                            {Caldarelli}}]{Cimini2019statistical}%
	                                                                                                              \BibitemOpen
	                                                                                                              \bibfield  {author} {\bibinfo {author} {\bibfnamefont {G.}~\bibnamefont
			                                                                                                              {Cimini}}, \bibinfo {author} {\bibfnamefont {T.}~\bibnamefont {Squartini}},
		                                                                                                              \bibinfo {author} {\bibfnamefont {F.}~\bibnamefont {Saracco}}, \bibinfo
		                                                                                                                       {author} {\bibfnamefont {D.}~\bibnamefont {Garlaschelli}}, \bibinfo {author}
		                                                                                                                       {\bibfnamefont {A.}~\bibnamefont {Gabrielli}},\ and\ \bibinfo {author}
		                                                                                                                       {\bibfnamefont {G.}~\bibnamefont {Caldarelli}},\ }\bibfield  {title} {\emph
		                                                                                                              {\bibinfo {title} {The statistical physics of real-world networks}},\ }\href
	                                                                                                                         {https://doi.org/https://doi.org/10.1038/s42254-018-0002-6} {\bibfield
		                                                                                                                         {journal} {\bibinfo  {journal} {Nat. Rev. Phys.}\ }\textbf {\bibinfo {volume}
			                                                                                                                         {1}},\ \bibinfo {pages} {58} (\bibinfo {year} {2019})}\BibitemShut {NoStop}%
	                                                                                                                       \bibitem [{\citenamefont {Morzy}\ \emph {et~al.}(2017)\citenamefont {Morzy},
		                                                                                                                         \citenamefont {Kajdanowicz},\ and\ \citenamefont {Kazienko}}]{Morzy2017}%
	                                                                                                                         \BibitemOpen
	                                                                                                                         \bibfield  {author} {\bibinfo {author} {\bibfnamefont {M.}~\bibnamefont
			                                                                                                                         {Morzy}}, \bibinfo {author} {\bibfnamefont {T.}~\bibnamefont {Kajdanowicz}},\
		                                                                                                                         and\ \bibinfo {author} {\bibfnamefont {P.}~\bibnamefont {Kazienko}},\
	                                                                                                                         }\bibfield  {title} {\emph {\bibinfo {title} {On measuring the complexity of
				                                                                                                                         networks: Kolmogorov complexity versus entropy}},\ }\href
	                                                                                                                                    {https://doi.org/https://doi.org/10.1155/2017/3250301} {\bibfield  {journal}
		                                                                                                                                    {\bibinfo  {journal} {Complexity}\ }\textbf {\bibinfo {volume} {2017}},\
		                                                                                                                                    \bibinfo {pages} {3250301} (\bibinfo {year} {2017})}\BibitemShut {NoStop}%
	                                                                                                                                  \bibitem [{\citenamefont {Zenil}\ \emph {et~al.}(2018)\citenamefont {Zenil},
		                                                                                                                                    \citenamefont {Kiani},\ and\ \citenamefont {Tegn{\'e}r}}]{Zenil2018review}%
	                                                                                                                                    \BibitemOpen
	                                                                                                                                    \bibfield  {author} {\bibinfo {author} {\bibfnamefont {H.}~\bibnamefont
			                                                                                                                                    {Zenil}}, \bibinfo {author} {\bibfnamefont {N.}~\bibnamefont {Kiani}},\ and\
		                                                                                                                                    \bibinfo {author} {\bibfnamefont {J.}~\bibnamefont {Tegn{\'e}r}},\ }\bibfield
	                                                                                                                                               {title} {\emph {\bibinfo {title} {A review of graph and network complexity
				                                                                                                                                               from an algorithmic information perspective}},\ }\href
	                                                                                                                                               {https://doi.org/https://doi.org/10.3390/e20080551} {\bibfield  {journal}
		                                                                                                                                               {\bibinfo  {journal} {Entropy}\ }\textbf {\bibinfo {volume} {20}},\ \bibinfo
		                                                                                                                                               {pages} {551} (\bibinfo {year} {2018})}\BibitemShut {NoStop}%
	                                                                                                                                             \bibitem [{\citenamefont {De~Domenico}\ \emph {et~al.}(2013)\citenamefont
		                                                                                                                                               {De~Domenico}, \citenamefont {Sol\'{e}-Ribalta}, \citenamefont {Cozzo},
		                                                                                                                                               \citenamefont {Kivel\"{a}}, \citenamefont {Moreno}, \citenamefont {Porter},
		                                                                                                                                               \citenamefont {G\'{o}mez},\ and\ \citenamefont {Arenas}}]{Arenas_2013}%
	                                                                                                                                               \BibitemOpen
	                                                                                                                                               \bibfield  {author} {\bibinfo {author} {\bibfnamefont {M.}~\bibnamefont
			                                                                                                                                               {De~Domenico}}, \bibinfo {author} {\bibfnamefont {A.~S.}\ \bibnamefont
			                                                                                                                                               {Sol\'{e}-Ribalta}}, \bibinfo {author} {\bibfnamefont {E.}~\bibnamefont
			                                                                                                                                               {Cozzo}}, \bibinfo {author} {\bibfnamefont {M.}~\bibnamefont {Kivel\"{a}}},
		                                                                                                                                               \bibinfo {author} {\bibfnamefont {Y.}~\bibnamefont {Moreno}}, \bibinfo
		                                                                                                                                                        {author} {\bibfnamefont {M.~A.}\ \bibnamefont {Porter}}, \bibinfo {author}
		                                                                                                                                                        {\bibfnamefont {S.}~\bibnamefont {G\'{o}mez}},\ and\ \bibinfo {author}
		                                                                                                                                                        {\bibfnamefont {A.}~\bibnamefont {Arenas}},\ }\bibfield  {title} {\emph
		                                                                                                                                               {\bibinfo {title} {Mathematical formulation of multilayer networks}},\ }\href
	                                                                                                                                                          {https://doi.org/10.1103/PhysRevX.3.041022} {\bibfield  {journal} {\bibinfo
			                                                                                                                                                          {journal} {Phys. Rev. X}\ }\textbf {\bibinfo {volume} {3}},\ \bibinfo {pages}
		                                                                                                                                                          {041022} (\bibinfo {year} {2013})}\BibitemShut {NoStop}%
	                                                                                                                                                        \bibitem [{\citenamefont {Boccaletti}\ \emph {et~al.}(2014)\citenamefont
		                                                                                                                                                          {Boccaletti}, \citenamefont {Bianconi}, \citenamefont {Criado}, \citenamefont
		                                                                                                                                                          {del Genio}, \citenamefont {G\'{o}mez-Garde\~{n}es}, \citenamefont {Romance},
		                                                                                                                                                          \citenamefont {Sendi\~{n}a Nadal}, \citenamefont {Wang},\ and\ \citenamefont
		                                                                                                                                                                        {Zanin}}]{Boccaletti_2014}%
	                                                                                                                                                          \BibitemOpen
	                                                                                                                                                          \bibfield  {author} {\bibinfo {author} {\bibfnamefont {S.}~\bibnamefont
			                                                                                                                                                          {Boccaletti}}, \bibinfo {author} {\bibfnamefont {G.}~\bibnamefont
			                                                                                                                                                          {Bianconi}}, \bibinfo {author} {\bibfnamefont {R.}~\bibnamefont {Criado}},
		                                                                                                                                                          \bibinfo {author} {\bibfnamefont {C.~I.}\ \bibnamefont {del Genio}}, \bibinfo
		                                                                                                                                                                   {author} {\bibfnamefont {J.}~\bibnamefont {G\'{o}mez-Garde\~{n}es}}, \bibinfo
		                                                                                                                                                                   {author} {\bibfnamefont {M.}~\bibnamefont {Romance}}, \bibinfo {author}
		                                                                                                                                                                   {\bibfnamefont {I.}~\bibnamefont {Sendi\~{n}a Nadal}}, \bibinfo {author}
		                                                                                                                                                                   {\bibfnamefont {Z.}~\bibnamefont {Wang}},\ and\ \bibinfo {author}
		                                                                                                                                                                   {\bibfnamefont {M.}~\bibnamefont {Zanin}},\ }\bibfield  {title} {\emph
		                                                                                                                                                          {\bibinfo {title} {The structure and dynamics of multilayer networks}},\
	                                                                                                                                                          }\href {https://doi.org/10.1016/j.physrep.2014.07.001} {\bibfield  {journal}
		                                                                                                                                                          {\bibinfo  {journal} {Phys. Rep.}\ }\textbf {\bibinfo {volume} {544}},\
		                                                                                                                                                          \bibinfo {pages} {1} (\bibinfo {year} {2014})}\BibitemShut {NoStop}%
	                                                                                                                                                        \bibitem [{\citenamefont {Bianconi}(2018)}]{Bianconi2018multilayer}%
	                                                                                                                                                          \BibitemOpen
	                                                                                                                                                          \bibfield  {author} {\bibinfo {author} {\bibfnamefont {G.}~\bibnamefont
			                                                                                                                                                          {Bianconi}},\ }\href@noop {} {\emph {\bibinfo {title} {Multilayer Networks:
				                                                                                                                                                          Structure and Function}}}\ (\bibinfo  {publisher} {Oxford University Press},\
	                                                                                                                                                          \bibinfo {address} {Oxford},\ \bibinfo {year} {2018})\BibitemShut {NoStop}%
	                                                                                                                                                        \bibitem [{\citenamefont {Cardillo}\ \emph {et~al.}(2013)\citenamefont
		                                                                                                                                                          {Cardillo}, \citenamefont {G\'{o}mez-Garde\~{n}es}, \citenamefont {Zanin},
		                                                                                                                                                          \citenamefont {Romance}, \citenamefont {Papo}, \citenamefont {Pozo},\ and\
		                                                                                                                                                          \citenamefont {Boccaletti}}]{Cardillo_2013}%
	                                                                                                                                                          \BibitemOpen
	                                                                                                                                                          \bibfield  {author} {\bibinfo {author} {\bibfnamefont {A.}~\bibnamefont
			                                                                                                                                                          {Cardillo}}, \bibinfo {author} {\bibfnamefont {J.}~\bibnamefont
			                                                                                                                                                          {G\'{o}mez-Garde\~{n}es}}, \bibinfo {author} {\bibfnamefont {M.}~\bibnamefont
			                                                                                                                                                          {Zanin}}, \bibinfo {author} {\bibfnamefont {M.}~\bibnamefont {Romance}},
		                                                                                                                                                          \bibinfo {author} {\bibfnamefont {D.}~\bibnamefont {Papo}}, \bibinfo {author}
		                                                                                                                                                                   {\bibfnamefont {F.}~\bibnamefont {Pozo}},\ and\ \bibinfo {author}
		                                                                                                                                                                   {\bibfnamefont {S.}~\bibnamefont {Boccaletti}},\ }\bibfield  {title} {\emph
		                                                                                                                                                          {\bibinfo {title} {Emergence of network features from multiplexity}},\ }\href
	                                                                                                                                                                     {https://doi.org/https://doi.org/10.1038/srep01344} {\bibfield  {journal}
		                                                                                                                                                                     {\bibinfo  {journal} {Sci. Rep.}\ }\textbf {\bibinfo {volume} {3}},\ \bibinfo
		                                                                                                                                                                     {pages} {1344} (\bibinfo {year} {2013})}\BibitemShut {NoStop}%
	                                                                                                                                                                   \bibitem [{\citenamefont {Gallotti}\ \emph {et~al.}(2016)\citenamefont
		                                                                                                                                                                     {Gallotti}, \citenamefont {Porter},\ and\ \citenamefont
		                                                                                                                                                                     {Barthelemy}}]{Gallotti2016}%
	                                                                                                                                                                     \BibitemOpen
	                                                                                                                                                                     \bibfield  {author} {\bibinfo {author} {\bibfnamefont {R.}~\bibnamefont
			                                                                                                                                                                     {Gallotti}}, \bibinfo {author} {\bibfnamefont {M.~A.}\ \bibnamefont
			                                                                                                                                                                     {Porter}},\ and\ \bibinfo {author} {\bibfnamefont {M.}~\bibnamefont
			                                                                                                                                                                     {Barthelemy}},\ }\bibfield  {title} {\emph {\bibinfo {title} {Lost in
				                                                                                                                                                                     transportation: Information measures and cognitive limits in multilayer
				                                                                                                                                                                     navigation}},\ }\href
	                                                                                                                                                                                {https://doi.org/https://doi.org/10.1126/sciadv.1500445} {\bibfield
		                                                                                                                                                                                {journal} {\bibinfo  {journal} {Sci. Adv.}\ }\textbf {\bibinfo {volume}
			                                                                                                                                                                                {2}},\ \bibinfo {pages} {e1500445} (\bibinfo {year} {2016})}\BibitemShut
	                                                                                                                                                                                {NoStop}%
	                                                                                                                                                                              \bibitem [{\citenamefont {De~Domenico}\ \emph
		                                                                                                                                                                                {et~al.}(2016{\natexlab{a}})\citenamefont {De~Domenico}, \citenamefont
		                                                                                                                                                                                {Sasai},\ and\ \citenamefont {Arenas}}]{DeDomenico2016mapping}%
	                                                                                                                                                                                \BibitemOpen
	                                                                                                                                                                                \bibfield  {author} {\bibinfo {author} {\bibfnamefont {M.}~\bibnamefont
			                                                                                                                                                                                {De~Domenico}}, \bibinfo {author} {\bibfnamefont {S.}~\bibnamefont {Sasai}},\
		                                                                                                                                                                                and\ \bibinfo {author} {\bibfnamefont {A.}~\bibnamefont {Arenas}},\
	                                                                                                                                                                                }\bibfield  {title} {\emph {\bibinfo {title} {Mapping multiplex hubs in human
				                                                                                                                                                                                functional brain networks}},\ }\href
	                                                                                                                                                                                           {https://doi.org/https://doi.org/10.3389/fnins.2016.00326} {\bibfield
		                                                                                                                                                                                           {journal} {\bibinfo  {journal} {Front. Neurosci.}\ }\textbf {\bibinfo
			                                                                                                                                                                                           {volume} {10}},\ \bibinfo {pages} {326} (\bibinfo {year}
		                                                                                                                                                                                           {2016}{\natexlab{a}})}\BibitemShut {NoStop}%
	                                                                                                                                                                                         \bibitem [{\citenamefont {Battiston}\ \emph
		                                                                                                                                                                                           {et~al.}(2017{\natexlab{a}})\citenamefont {Battiston}, \citenamefont
		                                                                                                                                                                                           {Nicosia}, \citenamefont {Chavez},\ and\ \citenamefont
		                                                                                                                                                                                           {Latora}}]{Battiston2017multilayer}%
	                                                                                                                                                                                           \BibitemOpen
	                                                                                                                                                                                           \bibfield  {author} {\bibinfo {author} {\bibfnamefont {F.}~\bibnamefont
			                                                                                                                                                                                           {Battiston}}, \bibinfo {author} {\bibfnamefont {V.}~\bibnamefont {Nicosia}},
		                                                                                                                                                                                           \bibinfo {author} {\bibfnamefont {M.}~\bibnamefont {Chavez}},\ and\ \bibinfo
		                                                                                                                                                                                                    {author} {\bibfnamefont {V.}~\bibnamefont {Latora}},\ }\bibfield  {title}
	                                                                                                                                                                                                      {\emph {\bibinfo {title} {Multilayer motif analysis of brain networks}},\
	                                                                                                                                                                                                      }\href {https://doi.org/https://doi.org/10.1063/1.4979282} {\bibfield
		                                                                                                                                                                                                      {journal} {\bibinfo  {journal} {Chaos}\ }\textbf {\bibinfo {volume} {27}},\
		                                                                                                                                                                                                      \bibinfo {pages} {047404} (\bibinfo {year} {2017}{\natexlab{a}})}\BibitemShut
	                                                                                                                                                                                                      {NoStop}%
	                                                                                                                                                                                                    \bibitem [{\citenamefont {Buldyrev}\ \emph {et~al.}(2010)\citenamefont
		                                                                                                                                                                                                      {Buldyrev}, \citenamefont {Parshani}, \citenamefont {Paul}, \citenamefont
		                                                                                                                                                                                                      {Stanley},\ and\ \citenamefont {Havlin}}]{Havlin_2010}%
	                                                                                                                                                                                                      \BibitemOpen
	                                                                                                                                                                                                      \bibfield  {author} {\bibinfo {author} {\bibfnamefont {S.~V.}\ \bibnamefont
			                                                                                                                                                                                                      {Buldyrev}}, \bibinfo {author} {\bibfnamefont {R.}~\bibnamefont {Parshani}},
		                                                                                                                                                                                                      \bibinfo {author} {\bibfnamefont {G.}~\bibnamefont {Paul}}, \bibinfo {author}
		                                                                                                                                                                                                               {\bibfnamefont {H.~E.}\ \bibnamefont {Stanley}},\ and\ \bibinfo {author}
		                                                                                                                                                                                                               {\bibfnamefont {S.}~\bibnamefont {Havlin}},\ }\bibfield  {title} {\emph
		                                                                                                                                                                                                      {\bibinfo {title} {Catastrophic cascade of failures in interdependent
				                                                                                                                                                                                                      networks}},\ }\href {https://doi.org/https://doi.org/10.1038/nature08932}
	                                                                                                                                                                                                                 {\bibfield  {journal} {\bibinfo  {journal} {Nature}\ }\textbf {\bibinfo
			                                                                                                                                                                                                                 {volume} {464}},\ \bibinfo {pages} {1025} (\bibinfo {year}
		                                                                                                                                                                                                                 {2010})}\BibitemShut {NoStop}%
	                                                                                                                                                                                                               \bibitem [{\citenamefont {G\'{o}mez}\ \emph {et~al.}(2013)\citenamefont
		                                                                                                                                                                                                                 {G\'{o}mez}, \citenamefont {D\'{\i}az-Guilera}, \citenamefont
		                                                                                                                                                                                                                 {G\'{o}mez-Garde\~{n}es}, \citenamefont {P\'{e}rez-Vicente}, \citenamefont
		                                                                                                                                                                                                                 {Moreno},\ and\ \citenamefont {Arenas}}]{Arenas_Diffusion_2013}%
	                                                                                                                                                                                                                 \BibitemOpen
	                                                                                                                                                                                                                 \bibfield  {author} {\bibinfo {author} {\bibfnamefont {S.}~\bibnamefont
			                                                                                                                                                                                                                 {G\'{o}mez}}, \bibinfo {author} {\bibfnamefont {A.}~\bibnamefont
			                                                                                                                                                                                                                 {D\'{\i}az-Guilera}}, \bibinfo {author} {\bibfnamefont {J.}~\bibnamefont
			                                                                                                                                                                                                                 {G\'{o}mez-Garde\~{n}es}}, \bibinfo {author} {\bibfnamefont {C.~J.}\
			                                                                                                                                                                                                                 \bibnamefont {P\'{e}rez-Vicente}}, \bibinfo {author} {\bibfnamefont
			                                                                                                                                                                                                                 {Y.}~\bibnamefont {Moreno}},\ and\ \bibinfo {author} {\bibfnamefont
			                                                                                                                                                                                                                 {A.}~\bibnamefont {Arenas}},\ }\bibfield  {title} {\emph {\bibinfo {title}
			                                                                                                                                                                                                                 {Diffusion dynamics on multiplex networks}},\ }\href
	                                                                                                                                                                                                                            {https://doi.org/10.1103/PhysRevLett.110.028701} {\bibfield  {journal}
		                                                                                                                                                                                                                            {\bibinfo  {journal} {Phys. Rev. Lett.}\ }\textbf {\bibinfo {volume} {110}},\
		                                                                                                                                                                                                                            \bibinfo {pages} {028701} (\bibinfo {year} {2013})}\BibitemShut {NoStop}%
	                                                                                                                                                                                                                          \bibitem [{\citenamefont {Nicosia}\ \emph {et~al.}(2017)\citenamefont
		                                                                                                                                                                                                                            {Nicosia}, \citenamefont {Skardal}, \citenamefont {Arenas},\ and\
		                                                                                                                                                                                                                            \citenamefont {Latora}}]{Nicosia_Skardal_2017}%
	                                                                                                                                                                                                                            \BibitemOpen
	                                                                                                                                                                                                                            \bibfield  {author} {\bibinfo {author} {\bibfnamefont {V.}~\bibnamefont
			                                                                                                                                                                                                                            {Nicosia}}, \bibinfo {author} {\bibfnamefont {P.~S.}\ \bibnamefont
			                                                                                                                                                                                                                            {Skardal}}, \bibinfo {author} {\bibfnamefont {A.}~\bibnamefont {Arenas}},\
		                                                                                                                                                                                                                            and\ \bibinfo {author} {\bibfnamefont {V.}~\bibnamefont {Latora}},\
	                                                                                                                                                                                                                            }\bibfield  {title} {\emph {\bibinfo {title} {Collective phenomena emerging
				                                                                                                                                                                                                                            from the interactions between dynamical processes in multiplex networks}},\
	                                                                                                                                                                                                                            }\href {https://doi.org/https://doi.org/10.1103/PhysRevLett.118.138302}
	                                                                                                                                                                                                                                       {\bibfield  {journal} {\bibinfo  {journal} {Phys. Rev. Lett.}\ }\textbf
		                                                                                                                                                                                                                                       {\bibinfo {volume} {118}},\ \bibinfo {pages} {138302} (\bibinfo {year}
		                                                                                                                                                                                                                                       {2017})}\BibitemShut {NoStop}%
	                                                                                                                                                                                                                                     \bibitem [{\citenamefont {Soriano-Pa\~nos}\ \emph {et~al.}(2018)\citenamefont
		                                                                                                                                                                                                                                       {Soriano-Pa\~nos}, \citenamefont {Lotero}, \citenamefont {Arenas},\ and\
		                                                                                                                                                                                                                                       \citenamefont {G\'omez-Garde\~nes}}]{Soriano_Gardenes_2018}%
	                                                                                                                                                                                                                                       \BibitemOpen
	                                                                                                                                                                                                                                       \bibfield  {author} {\bibinfo {author} {\bibfnamefont {D.}~\bibnamefont
			                                                                                                                                                                                                                                       {Soriano-Pa\~nos}}, \bibinfo {author} {\bibfnamefont {L.}~\bibnamefont
			                                                                                                                                                                                                                                       {Lotero}}, \bibinfo {author} {\bibfnamefont {A.}~\bibnamefont {Arenas}},\
		                                                                                                                                                                                                                                       and\ \bibinfo {author} {\bibfnamefont {J.}~\bibnamefont
			                                                                                                                                                                                                                                       {G\'omez-Garde\~nes}},\ }\bibfield  {title} {\emph {\bibinfo {title}
			                                                                                                                                                                                                                                       {Spreading processes in multiplex metapopulations containing different
				                                                                                                                                                                                                                                       mobility networks}},\ }\href {https://doi.org/10.1103/PhysRevX.8.031039}
	                                                                                                                                                                                                                                                  {\bibfield  {journal} {\bibinfo  {journal} {Phys. Rev. X}\ }\textbf {\bibinfo
			                                                                                                                                                                                                                                                  {volume} {8}},\ \bibinfo {pages} {031039} (\bibinfo {year}
		                                                                                                                                                                                                                                                  {2018})}\BibitemShut {NoStop}%
	                                                                                                                                                                                                                                                \bibitem [{\citenamefont {Granell}\ \emph {et~al.}(2013)\citenamefont
		                                                                                                                                                                                                                                                  {Granell}, \citenamefont {G{\'o}mez},\ and\ \citenamefont
		                                                                                                                                                                                                                                                  {Arenas}}]{Granell2013dynamical}%
	                                                                                                                                                                                                                                                  \BibitemOpen
	                                                                                                                                                                                                                                                  \bibfield  {author} {\bibinfo {author} {\bibfnamefont {C.}~\bibnamefont
			                                                                                                                                                                                                                                                  {Granell}}, \bibinfo {author} {\bibfnamefont {S.}~\bibnamefont {G{\'o}mez}},\
		                                                                                                                                                                                                                                                  and\ \bibinfo {author} {\bibfnamefont {A.}~\bibnamefont {Arenas}},\
	                                                                                                                                                                                                                                                  }\bibfield  {title} {\emph {\bibinfo {title} {Dynamical interplay between
				                                                                                                                                                                                                                                                  awareness and epidemic spreading in multiplex networks}},\ }\href
	                                                                                                                                                                                                                                                             {https://doi.org/https://doi.org/10.1103/PhysRevLett.111.128701} {\bibfield
		                                                                                                                                                                                                                                                             {journal} {\bibinfo  {journal} {Phys. Rev. Lett.}\ }\textbf {\bibinfo
			                                                                                                                                                                                                                                                             {volume} {111}},\ \bibinfo {pages} {128701} (\bibinfo {year}
		                                                                                                                                                                                                                                                             {2013})}\BibitemShut {NoStop}%
	                                                                                                                                                                                                                                                           \bibitem [{\citenamefont {Gleeson}\ \emph {et~al.}(2016)\citenamefont
		                                                                                                                                                                                                                                                             {Gleeson}, \citenamefont {O'Sullivan}, \citenamefont {Ba\~{n}os},\ and\
		                                                                                                                                                                                                                                                             \citenamefont {Moreno}}]{Gleeson_Moreno_2016}%
	                                                                                                                                                                                                                                                             \BibitemOpen
	                                                                                                                                                                                                                                                             \bibfield  {author} {\bibinfo {author} {\bibfnamefont {J.~P.}\ \bibnamefont
			                                                                                                                                                                                                                                                             {Gleeson}}, \bibinfo {author} {\bibfnamefont {K.~P.}\ \bibnamefont
			                                                                                                                                                                                                                                                             {O'Sullivan}}, \bibinfo {author} {\bibfnamefont {R.~A.}\ \bibnamefont
			                                                                                                                                                                                                                                                             {Ba\~{n}os}},\ and\ \bibinfo {author} {\bibfnamefont {Y.}~\bibnamefont
			                                                                                                                                                                                                                                                             {Moreno}},\ }\bibfield  {title} {\emph {\bibinfo {title} {Effects of network
				                                                                                                                                                                                                                                                             structure, competition and memory time on social spreading phenomena}},\
	                                                                                                                                                                                                                                                             }\href {https://doi.org/10.1103/PhysRevX.6.021019} {\bibfield  {journal}
		                                                                                                                                                                                                                                                             {\bibinfo  {journal} {Phys. Rev. X}\ }\textbf {\bibinfo {volume} {6}},\
		                                                                                                                                                                                                                                                             \bibinfo {pages} {021019} (\bibinfo {year} {2016})}\BibitemShut {NoStop}%
	                                                                                                                                                                                                                                                           \bibitem [{\citenamefont {De~Domenico}\ \emph
		                                                                                                                                                                                                                                                             {et~al.}(2016{\natexlab{b}})\citenamefont {De~Domenico}, \citenamefont
		                                                                                                                                                                                                                                                             {Granell}, \citenamefont {Porter},\ and\ \citenamefont
		                                                                                                                                                                                                                                                             {Arenas}}]{DeDomenico2016physics}%
	                                                                                                                                                                                                                                                             \BibitemOpen
	                                                                                                                                                                                                                                                             \bibfield  {author} {\bibinfo {author} {\bibfnamefont {M.}~\bibnamefont
			                                                                                                                                                                                                                                                             {De~Domenico}}, \bibinfo {author} {\bibfnamefont {C.}~\bibnamefont
			                                                                                                                                                                                                                                                             {Granell}}, \bibinfo {author} {\bibfnamefont {M.~A.}\ \bibnamefont
			                                                                                                                                                                                                                                                             {Porter}},\ and\ \bibinfo {author} {\bibfnamefont {A.}~\bibnamefont
			                                                                                                                                                                                                                                                             {Arenas}},\ }\bibfield  {title} {\emph {\bibinfo {title} {The physics of
				                                                                                                                                                                                                                                                             spreading processes in multilayer networks}},\ }\href
	                                                                                                                                                                                                                                                                        {https://doi.org/https://doi.org/10.1038/nphys3865} {\bibfield  {journal}
		                                                                                                                                                                                                                                                                        {\bibinfo  {journal} {Nat. Phys.}\ }\textbf {\bibinfo {volume} {12}},\
		                                                                                                                                                                                                                                                                        \bibinfo {pages} {901} (\bibinfo {year} {2016}{\natexlab{b}})}\BibitemShut
	                                                                                                                                                                                                                                                                        {NoStop}%
	                                                                                                                                                                                                                                                                      \bibitem [{\citenamefont {Lacasa}\ \emph {et~al.}(2018)\citenamefont {Lacasa},
		                                                                                                                                                                                                                                                                        \citenamefont {Mari{\~n}o}, \citenamefont {Miguez}, \citenamefont {Nicosia},
		                                                                                                                                                                                                                                                                        \citenamefont {Rold{\'a}n}, \citenamefont {Lisica}, \citenamefont {Grill},\
		                                                                                                                                                                                                                                                                        and\ \citenamefont {G{\'o}mez-Garde{\~n}es}}]{Lacasa_multiplex_2018}%
	                                                                                                                                                                                                                                                                        \BibitemOpen
	                                                                                                                                                                                                                                                                        \bibfield  {author} {\bibinfo {author} {\bibfnamefont {L.}~\bibnamefont
			                                                                                                                                                                                                                                                                        {Lacasa}}, \bibinfo {author} {\bibfnamefont {I.~P.}\ \bibnamefont
			                                                                                                                                                                                                                                                                        {Mari{\~n}o}}, \bibinfo {author} {\bibfnamefont {J.}~\bibnamefont {Miguez}},
		                                                                                                                                                                                                                                                                        \bibinfo {author} {\bibfnamefont {V.}~\bibnamefont {Nicosia}}, \bibinfo
		                                                                                                                                                                                                                                                                                 {author} {\bibfnamefont {{\'E}.}~\bibnamefont {Rold{\'a}n}}, \bibinfo
		                                                                                                                                                                                                                                                                                 {author} {\bibfnamefont {A.}~\bibnamefont {Lisica}}, \bibinfo {author}
		                                                                                                                                                                                                                                                                                 {\bibfnamefont {S.~W.}\ \bibnamefont {Grill}},\ and\ \bibinfo {author}
		                                                                                                                                                                                                                                                                                 {\bibfnamefont {J.}~\bibnamefont {G{\'o}mez-Garde{\~n}es}},\ }\bibfield
	                                                                                                                                                                                                                                                                                   {title} {\emph {\bibinfo {title} {Multiplex decomposition of non-markovian
				                                                                                                                                                                                                                                                                                   dynamics and the hidden layer reconstruction problem}},\ }\href
	                                                                                                                                                                                                                                                                                   {https://doi.org/https://doi.org/10.1103/PhysRevX.8.031038} {\bibfield
		                                                                                                                                                                                                                                                                                   {journal} {\bibinfo  {journal} {Phys. Rev. X}\ }\textbf {\bibinfo {volume}
			                                                                                                                                                                                                                                                                                   {8}},\ \bibinfo {pages} {031038} (\bibinfo {year} {2018})}\BibitemShut
	                                                                                                                                                                                                                                                                                   {NoStop}%
	                                                                                                                                                                                                                                                                                 \bibitem [{\citenamefont {Diakonova}\ \emph {et~al.}(2016)\citenamefont
		                                                                                                                                                                                                                                                                                   {Diakonova}, \citenamefont {Nicosia}, \citenamefont {Latora},\ and\
		                                                                                                                                                                                                                                                                                   \citenamefont {San~Miguel}}]{Diakonova_2016}%
	                                                                                                                                                                                                                                                                                   \BibitemOpen
	                                                                                                                                                                                                                                                                                   \bibfield  {author} {\bibinfo {author} {\bibfnamefont {M.}~\bibnamefont
			                                                                                                                                                                                                                                                                                   {Diakonova}}, \bibinfo {author} {\bibfnamefont {V.}~\bibnamefont {Nicosia}},
		                                                                                                                                                                                                                                                                                   \bibinfo {author} {\bibfnamefont {V.}~\bibnamefont {Latora}},\ and\ \bibinfo
		                                                                                                                                                                                                                                                                                            {author} {\bibfnamefont {M.}~\bibnamefont {San~Miguel}},\ }\bibfield  {title}
	                                                                                                                                                                                                                                                                                              {\emph {\bibinfo {title} {Irreducibility of multilayer network dynamics: the
				                                                                                                                                                                                                                                                                                              case of the voter model}},\ }\href
	                                                                                                                                                                                                                                                                                              {https://doi.org/10.1088/1367-2630/18/2/023010} {\bibfield  {journal}
		                                                                                                                                                                                                                                                                                              {\bibinfo  {journal} {New J. Phys.}\ }\textbf {\bibinfo {volume} {18}},\
		                                                                                                                                                                                                                                                                                              \bibinfo {pages} {023010} (\bibinfo {year} {2016})}\BibitemShut {NoStop}%
	                                                                                                                                                                                                                                                                                            \bibitem [{\citenamefont {{De Domenico}}\ \emph {et~al.}(2015)\citenamefont
		                                                                                                                                                                                                                                                                                              {{De Domenico}}, \citenamefont {Nicosia}, \citenamefont {Arenas},\ and\
		                                                                                                                                                                                                                                                                                              \citenamefont {Latora}}]{DeDomenico_Nicosia_2015}%
	                                                                                                                                                                                                                                                                                              \BibitemOpen
	                                                                                                                                                                                                                                                                                              \bibfield  {author} {\bibinfo {author} {\bibfnamefont {M.}~\bibnamefont {{De
					                                                                                                                                                                                                                                                                                              Domenico}}}, \bibinfo {author} {\bibfnamefont {V.}~\bibnamefont {Nicosia}},
		                                                                                                                                                                                                                                                                                              \bibinfo {author} {\bibfnamefont {A.}~\bibnamefont {Arenas}},\ and\ \bibinfo
		                                                                                                                                                                                                                                                                                                       {author} {\bibfnamefont {V.}~\bibnamefont {Latora}},\ }\bibfield  {title}
	                                                                                                                                                                                                                                                                                                         {\emph {\bibinfo {title} {Structural reducibility of multilayer networks}},\
	                                                                                                                                                                                                                                                                                                         }\href {https://doi.org/https://doi.org/10.1038/ncomms7864} {\bibfield
		                                                                                                                                                                                                                                                                                                         {journal} {\bibinfo  {journal} {Nat. Comm.}\ }\textbf {\bibinfo {volume}
			                                                                                                                                                                                                                                                                                                         {6}},\ \bibinfo {pages} {1} (\bibinfo {year} {2015})}\BibitemShut {NoStop}%
	                                                                                                                                                                                                                                                                                                       \bibitem [{\citenamefont {Iacovacci}\ \emph
		                                                                                                                                                                                                                                                                                                         {et~al.}(lack)\citenamefont {Iacovacci}, \citenamefont {Wu},\ and\
		                                                                                                                                                                                                                                                                                                         \citenamefont {Bianconi}}]{Iacovacci_Wu_2015}%
	                                                                                                                                                                                                                                                                                                         \BibitemOpen
	                                                                                                                                                                                                                                                                                                         \bibfield  {author} {\bibinfo {author} {\bibfnamefont {J.}~\bibnamefont
			                                                                                                                                                                                                                                                                                                         {Iacovacci}}, \bibinfo {author} {\bibfnamefont {Z.}~\bibnamefont
			                                                                                                                                                                                                                                                                                                         {Wu}},\ and\ \bibinfo {author} {\bibfnamefont {G.}~\bibnamefont {Bianconi}},\
	                                                                                                                                                                                                                                                                                                         }\bibfield  {title} {\emph {\bibinfo {title} {Mesoscopic structures reveal
				                                                                                                                                                                                                                                                                                                         the network between the layers of multiplex datasets.}},\ }\href
	                                                                                                                                                                                                                                                                                                                    {https://doi.org/https://doi.org/10.1103/PhysRevE.92.042806} {\bibfield
		                                                                                                                                                                                                                                                                                                                    {journal} {\bibinfo  {journal} {Phys. Rev. E}\ }\textbf {\bibinfo {volume}
			                                                                                                                                                                                                                                                                                                                    {92}},\ \bibinfo {pages} {042806} (\bibinfo {year}
		                                                                                                                                                                                                                                                                                                                    {2015})}\BibitemShut {NoStop}%
	                                                                                                                                                                                                                                                                                                                  \bibitem [{\citenamefont {Kao}\ and\ \citenamefont
		                                                                                                                                                                                                                                                                                                                    {Porter}(lack)}]{Kao2018layer}%
	                                                                                                                                                                                                                                                                                                                    \BibitemOpen
	                                                                                                                                                                                                                                                                                                                    \bibfield  {author} {\bibinfo {author} {\bibfnamefont {T.-C.}\ \bibnamefont
			                                                                                                                                                                                                                                                                                                                    {Kao}}\ and\ \bibinfo {author} {\bibfnamefont {M.~A.}\
			                                                                                                                                                                                                                                                                                                                    \bibnamefont {Porter}},\ }\bibfield  {title} {\emph {\bibinfo {title}
			                                                                                                                                                                                                                                                                                                                    {Layer communities in multiplex networks}},\ }\href
	                                                                                                                                                                                                                                                                                                                               {https://doi.org/https://doi.org/10.1007/s10955-017-1858-z} {\bibfield
		                                                                                                                                                                                                                                                                                                                               {journal} {\bibinfo  {journal} {J. Stat. Phys.}\ }\textbf {\bibinfo {volume}
			                                                                                                                                                                                                                                                                                                                               {173}},\ \bibinfo {pages} {1286} (\bibinfo {year}
		                                                                                                                                                                                                                                                                                                                               {2018})}\BibitemShut {NoStop}%
	                                                                                                                                                                                                                                                                                                                             \bibitem [{\citenamefont {Stanley}\ \emph
		                                                                                                                                                                                                                                                                                                                               {et~al.}(lack)\citenamefont {Stanley}, \citenamefont {Shai},
		                                                                                                                                                                                                                                                                                                                               \citenamefont {Taylor},\ and\ \citenamefont {Mucha}}]{Stanley2016clustering}%
	                                                                                                                                                                                                                                                                                                                               \BibitemOpen
	                                                                                                                                                                                                                                                                                                                               \bibfield  {author} {\bibinfo {author} {\bibfnamefont {N.}~\bibnamefont
			                                                                                                                                                                                                                                                                                                                               {Stanley}}, \bibinfo {author} {\bibfnamefont {S.}~\bibnamefont
			                                                                                                                                                                                                                                                                                                                               {Shai}}, \bibinfo {author} {\bibfnamefont {D.}~\bibnamefont {Taylor}},\ and\
		                                                                                                                                                                                                                                                                                                                               \bibinfo {author} {\bibfnamefont {P.~J.}\ \bibnamefont {Mucha}},\ }\bibfield
	                                                                                                                                                                                                                                                                                                                                          {title} {\emph {\bibinfo {title} {Clustering network layers with
				                                                                                                                                                                                                                                                                                                                                          the strata multilayer stochastic block model}},\ }\href
	                                                                                                                                                                                                                                                                                                                                          {https://doi.org/10.1109/TNSE.2016.2537545} {\bibfield  {journal} {\bibinfo
			                                                                                                                                                                                                                                                                                                                                          {journal} {IEEE Trans. Network Sci. Eng.}\ }\textbf {\bibinfo {volume} {3}},\
		                                                                                                                                                                                                                                                                                                                                          \bibinfo {pages} {95} (\bibinfo {year} {2016})}\BibitemShut
	                                                                                                                                                                                                                                                                                                                                          {NoStop}%
	                                                                                                                                                                                                                                                                                                                                        \bibitem [{\citenamefont {De~Bacco}\ \emph
		                                                                                                                                                                                                                                                                                                                                          {et~al.}(lack)\citenamefont {De~Bacco}, \citenamefont {Power},
		                                                                                                                                                                                                                                                                                                                                          \citenamefont {Larremore},\ and\ \citenamefont
		                                                                                                                                                                                                                                                                                                                                                        {Moore}}]{DeBacco2017community}%
	                                                                                                                                                                                                                                                                                                                                          \BibitemOpen
	                                                                                                                                                                                                                                                                                                                                          \bibfield  {author} {\bibinfo {author} {\bibfnamefont {C.}~\bibnamefont
			                                                                                                                                                                                                                                                                                                                                          {De~Bacco}}, \bibinfo {author} {\bibfnamefont {E.~A.}\
			                                                                                                                                                                                                                                                                                                                                          \bibnamefont {Power}}, \bibinfo {author} {\bibfnamefont {D.~B.}\ \bibnamefont
			                                                                                                                                                                                                                                                                                                                                          {Larremore}},\ and\ \bibinfo {author} {\bibfnamefont {C.}~\bibnamefont
			                                                                                                                                                                                                                                                                                                                                          {Moore}},\ }\bibfield  {title} {\emph {\bibinfo {title} {Community detection, link prediction, and layer interdependence in multilayer
				                                                                                                                                                                                                                                                                                                                                          networks}},\ }\href
	                                                                                                                                                                                                                                                                                                                                                     {https://doi.org/https://doi.org/10.1103/PhysRevE.95.042317} {\bibfield
		                                                                                                                                                                                                                                                                                                                                                     {journal} {\bibinfo  {journal} {Phys. Rev. E}\ }\textbf {\bibinfo {volume}
			                                                                                                                                                                                                                                                                                                                                                     {95}},\ \bibinfo {pages} {042317} (\bibinfo {year}
		                                                                                                                                                                                                                                                                                                                                                     {2017})}\BibitemShut {NoStop}%
	                                                                                                                                                                                                                                                                                                                                                   \bibitem [{\citenamefont {Kolmogorov}(1998)}]{Kolmogorov_1998}%
	                                                                                                                                                                                                                                                                                                                                                     \BibitemOpen
	                                                                                                                                                                                                                                                                                                                                                     \bibfield  {author} {\bibinfo {author} {\bibfnamefont {A.}~\bibnamefont
			                                                                                                                                                                                                                                                                                                                                                     {Kolmogorov}},\ }\bibfield  {title} {\emph {\bibinfo {title} {On tables of
				                                                                                                                                                                                                                                                                                                                                                     random numbers}},\ }\href
	                                                                                                                                                                                                                                                                                                                                                                {https://doi.org/https://doi.org/10.1016/S0304-3975(98)00075-9} {\bibfield
		                                                                                                                                                                                                                                                                                                                                                                {journal} {\bibinfo  {journal} {Theor. Comput. Sci.}\ }\textbf {\bibinfo
			                                                                                                                                                                                                                                                                                                                                                                {volume} {207}},\ \bibinfo {pages} {387 } (\bibinfo {year}
		                                                                                                                                                                                                                                                                                                                                                                {1998})}\BibitemShut {NoStop}%
	                                                                                                                                                                                                                                                                                                                                                              \bibitem [{Pap()}]{Paper_code}%
	                                                                                                                                                                                                                                                                                                                                                                \BibitemOpen
	                                                                                                                                                                                                                                                                                                                                                                \href@noop {} {}\bibinfo {howpublished} {\url{https://github.com/andresantoro/ALCOREM}}\BibitemShut {NoStop}%
	                                                                                                                                                                                                                                                                                                                                                              \bibitem [{\citenamefont {Delahaye}\ and\ \citenamefont
		                                                                                                                                                                                                                                                                                                                                                                {Zenil}(2012)}]{Delahaye2012numerical}%
	                                                                                                                                                                                                                                                                                                                                                                \BibitemOpen
	                                                                                                                                                                                                                                                                                                                                                                \bibfield  {author} {\bibinfo {author} {\bibfnamefont {J.-P.}\ \bibnamefont
			                                                                                                                                                                                                                                                                                                                                                                {Delahaye}}\ and\ \bibinfo {author} {\bibfnamefont {H.}~\bibnamefont
			                                                                                                                                                                                                                                                                                                                                                                {Zenil}},\ }\bibfield  {title} {\emph {\bibinfo {title} {Numerical evaluation
				                                                                                                                                                                                                                                                                                                                                                                of algorithmic complexity for short strings: A glance into the innermost
				                                                                                                                                                                                                                                                                                                                                                                structure of randomness}},\ }\href
	                                                                                                                                                                                                                                                                                                                                                                           {https://doi.org/https://doi.org/10.1016/j.amc.2011.10.006} {\bibfield
		                                                                                                                                                                                                                                                                                                                                                                           {journal} {\bibinfo  {journal} {Appl. Math Comput.}\ }\textbf {\bibinfo
			                                                                                                                                                                                                                                                                                                                                                                           {volume} {219}},\ \bibinfo {pages} {63} (\bibinfo {year} {2012})}\BibitemShut
	                                                                                                                                                                                                                                                                                                                                                                           {NoStop}%
	                                                                                                                                                                                                                                                                                                                                                                         \bibitem [{\citenamefont {Cozzo}\ \emph {et~al.}(lack)\citenamefont
		                                                                                                                                                                                                                                                                                                                                                                           {Cozzo}, \citenamefont {Banos}, \citenamefont {Meloni},\ and\
		                                                                                                                                                                                                                                                                                                                                                                           \citenamefont {Moreno}}]{Cozzo2013_SIS}%
	                                                                                                                                                                                                                                                                                                                                                                           \BibitemOpen
	                                                                                                                                                                                                                                                                                                                                                                           \bibfield  {author} {\bibinfo {author} {\bibfnamefont {E.}~\bibnamefont
			                                                                                                                                                                                                                                                                                                                                                                           {Cozzo}}, \bibinfo {author} {\bibfnamefont {R.~A.}\ \bibnamefont
			                                                                                                                                                                                                                                                                                                                                                                           {Banos}}, \bibinfo {author} {\bibfnamefont {S.}~\bibnamefont {Meloni}},\ and\
		                                                                                                                                                                                                                                                                                                                                                                           \bibinfo {author} {\bibfnamefont {Y.}~\bibnamefont {Moreno}},\ }\bibfield
	                                                                                                                                                                                                                                                                                                                                                                                      {title} {\emph {\bibinfo {title} {Contact-based social contagion
				                                                                                                                                                                                                                                                                                                                                                                                      in multiplex networks}},\ }\href
	                                                                                                                                                                                                                                                                                                                                                                                      {https://doi.org/https://doi.org/10.1103/PhysRevE.88.050801} {\bibfield
		                                                                                                                                                                                                                                                                                                                                                                                      {journal} {\bibinfo  {journal} {Phys. Rev. E}\ }\textbf {\bibinfo {volume}
			                                                                                                                                                                                                                                                                                                                                                                                      {88}},\ \bibinfo {pages} {050801} (\bibinfo {year} {2013
			                                                                                                                                                                                                                                                                                                                                                                                    })}\BibitemShut {NoStop}%
	                                                                                                                                                                                                                                                                                                                                                                                    \bibitem [{\citenamefont {Ward~Jr}(1963)}]{Ward1963hierarchical}%
	                                                                                                                                                                                                                                                                                                                                                                                      \BibitemOpen
	                                                                                                                                                                                                                                                                                                                                                                                      \bibfield  {author} {\bibinfo {author} {\bibfnamefont {J.~H.}\ \bibnamefont
			                                                                                                                                                                                                                                                                                                                                                                                      {Ward~Jr}},\ }\bibfield  {title} {\emph {\bibinfo {title} {Hierarchical
				                                                                                                                                                                                                                                                                                                                                                                                      grouping to optimize an objective function}},\ }\href
	                                                                                                                                                                                                                                                                                                                                                                                                 {https://doi.org/10.1080/01621459.1963.10500845} {\bibfield  {journal}
		                                                                                                                                                                                                                                                                                                                                                                                                 {\bibinfo  {journal} {J. Am. Stat. Assoc.}\ }\textbf {\bibinfo {volume}
			                                                                                                                                                                                                                                                                                                                                                                                                 {58}},\ \bibinfo {pages} {236} (\bibinfo {year} {1963})}\BibitemShut
	                                                                                                                                                                                                                                                                                                                                                                                                 {NoStop}%
	                                                                                                                                                                                                                                                                                                                                                                                               \bibitem [{\citenamefont {Battiston}\ \emph {et~al.}(2016)\citenamefont
		                                                                                                                                                                                                                                                                                                                                                                                                 {Battiston}, \citenamefont {Nicosia},\ and\ \citenamefont
		                                                                                                                                                                                                                                                                                                                                                                                                 {Latora}}]{Battiston_exploration_2016}%
	                                                                                                                                                                                                                                                                                                                                                                                                 \BibitemOpen
	                                                                                                                                                                                                                                                                                                                                                                                                 \bibfield  {author} {\bibinfo {author} {\bibfnamefont {F.}~\bibnamefont
			                                                                                                                                                                                                                                                                                                                                                                                                 {Battiston}}, \bibinfo {author} {\bibfnamefont {V.}~\bibnamefont {Nicosia}},\
		                                                                                                                                                                                                                                                                                                                                                                                                 and\ \bibinfo {author} {\bibfnamefont {V.}~\bibnamefont {Latora}},\
	                                                                                                                                                                                                                                                                                                                                                                                                 }\bibfield  {title} {\emph {\bibinfo {title} {Efficient exploration of
				                                                                                                                                                                                                                                                                                                                                                                                                 multiplex networks}},\ }\href {https://doi.org/10.1088/1367-2630/18/4/043035}
	                                                                                                                                                                                                                                                                                                                                                                                                            {\bibfield  {journal} {\bibinfo  {journal} {New J. Phys.}\ }\textbf {\bibinfo
			                                                                                                                                                                                                                                                                                                                                                                                                            {volume} {18}},\ \bibinfo {pages} {043035} (\bibinfo {year}
		                                                                                                                                                                                                                                                                                                                                                                                                            {2016})}\BibitemShut {NoStop}%
	                                                                                                                                                                                                                                                                                                                                                                                                          \bibitem [{\citenamefont {Rosvall}\ and\ \citenamefont
		                                                                                                                                                                                                                                                                                                                                                                                                            {Bergstrom}(2008)}]{Rosvall2008}%
	                                                                                                                                                                                                                                                                                                                                                                                                            \BibitemOpen
	                                                                                                                                                                                                                                                                                                                                                                                                            \bibfield  {author} {\bibinfo {author} {\bibfnamefont {M.}~\bibnamefont
			                                                                                                                                                                                                                                                                                                                                                                                                            {Rosvall}}\ and\ \bibinfo {author} {\bibfnamefont {C.~T.}\ \bibnamefont
			                                                                                                                                                                                                                                                                                                                                                                                                            {Bergstrom}},\ }\bibfield  {title} {\emph {\bibinfo {title} {Maps of random
				                                                                                                                                                                                                                                                                                                                                                                                                            walks on complex networks reveal community structure}},\ }\href
	                                                                                                                                                                                                                                                                                                                                                                                                                       {https://doi.org/10.1073/pnas.0706851105} {\bibfield  {journal} {\bibinfo
			                                                                                                                                                                                                                                                                                                                                                                                                                       {journal} {Proc. Natl. Acad. Sci. USA}\ }\textbf {\bibinfo {volume} {105}},\
		                                                                                                                                                                                                                                                                                                                                                                                                                       \bibinfo {pages} {1118} (\bibinfo {year} {2008})}\BibitemShut {NoStop}%
	                                                                                                                                                                                                                                                                                                                                                                                                                     \bibitem [{\citenamefont {Peixoto}(2015)}]{Peixoto2015}%
	                                                                                                                                                                                                                                                                                                                                                                                                                       \BibitemOpen
	                                                                                                                                                                                                                                                                                                                                                                                                                       \bibfield  {author} {\bibinfo {author} {\bibfnamefont {T.~P.}\ \bibnamefont
			                                                                                                                                                                                                                                                                                                                                                                                                                       {Peixoto}},\ }\bibfield  {title} {\emph {\bibinfo {title} {Model selection
				                                                                                                                                                                                                                                                                                                                                                                                                                       and hypothesis testing for large-scale network models with overlapping
				                                                                                                                                                                                                                                                                                                                                                                                                                       groups}},\ }\href {https://doi.org/10.1103/PhysRevX.5.011033} {\bibfield
		                                                                                                                                                                                                                                                                                                                                                                                                                       {journal} {\bibinfo  {journal} {Phys. Rev. X}\ }\textbf {\bibinfo {volume}
			                                                                                                                                                                                                                                                                                                                                                                                                                       {5}},\ \bibinfo {pages} {011033} (\bibinfo {year} {2015})}\BibitemShut
	                                                                                                                                                                                                                                                                                                                                                                                                                                  {NoStop}%
	                                                                                                                                                                                                                                                                                                                                                                                                                                \bibitem [{\citenamefont {Peixoto}(2018)}]{Peixoto2018}%
	                                                                                                                                                                                                                                                                                                                                                                                                                                  \BibitemOpen
	                                                                                                                                                                                                                                                                                                                                                                                                                                  \bibfield  {author} {\bibinfo {author} {\bibfnamefont {T.~P.}\ \bibnamefont
			                                                                                                                                                                                                                                                                                                                                                                                                                                  {Peixoto}},\ }\bibfield  {title} {\emph {\bibinfo {title} {Reconstructing
				                                                                                                                                                                                                                                                                                                                                                                                                                                  networks with unknown and heterogeneous errors}},\ }\href
	                                                                                                                                                                                                                                                                                                                                                                                                                                             {https://doi.org/10.1103/PhysRevX.8.041011} {\bibfield  {journal} {\bibinfo
			                                                                                                                                                                                                                                                                                                                                                                                                                                             {journal} {Phys. Rev. X}\ }\textbf {\bibinfo {volume} {8}},\ \bibinfo {pages}
		                                                                                                                                                                                                                                                                                                                                                                                                                                             {041011} (\bibinfo {year} {2018})}\BibitemShut {NoStop}%
	                                                                                                                                                                                                                                                                                                                                                                                                                                           \bibitem [{\citenamefont {Godoy-Lorite}\ \emph {et~al.}(2016)\citenamefont
		                                                                                                                                                                                                                                                                                                                                                                                                                                             {Godoy-Lorite}, \citenamefont {Guimer{\`a}}, \citenamefont {Moore},\ and\
		                                                                                                                                                                                                                                                                                                                                                                                                                                             \citenamefont {Sales-Pardo}}]{Godoy-Lorite2016}%
	                                                                                                                                                                                                                                                                                                                                                                                                                                             \BibitemOpen
	                                                                                                                                                                                                                                                                                                                                                                                                                                             \bibfield  {author} {\bibinfo {author} {\bibfnamefont {A.}~\bibnamefont
			                                                                                                                                                                                                                                                                                                                                                                                                                                             {Godoy-Lorite}}, \bibinfo {author} {\bibfnamefont {R.}~\bibnamefont
			                                                                                                                                                                                                                                                                                                                                                                                                                                             {Guimer{\`a}}}, \bibinfo {author} {\bibfnamefont {C.}~\bibnamefont {Moore}},\
		                                                                                                                                                                                                                                                                                                                                                                                                                                             and\ \bibinfo {author} {\bibfnamefont {M.}~\bibnamefont {Sales-Pardo}},\
	                                                                                                                                                                                                                                                                                                                                                                                                                                             }\bibfield  {title} {\emph {\bibinfo {title} {Accurate and scalable social
				                                                                                                                                                                                                                                                                                                                                                                                                                                             recommendation using mixed-membership stochastic block models}},\ }\href
	                                                                                                                                                                                                                                                                                                                                                                                                                                                        {https://doi.org/10.1073/pnas.1606316113} {\bibfield  {journal} {\bibinfo
			                                                                                                                                                                                                                                                                                                                                                                                                                                                        {journal} {Proc. Natl Acad. Sci. USA}\ }\textbf {\bibinfo {volume} {113}},\
		                                                                                                                                                                                                                                                                                                                                                                                                                                                        \bibinfo {pages} {14207} (\bibinfo {year} {2016})}\BibitemShut {NoStop}%
	                                                                                                                                                                                                                                                                                                                                                                                                                                                      \bibitem [{\citenamefont {Lambiotte}\ \emph {et~al.}(2019)\citenamefont
		                                                                                                                                                                                                                                                                                                                                                                                                                                                        {Lambiotte}, \citenamefont {Rosvall},\ and\ \citenamefont
		                                                                                                                                                                                                                                                                                                                                                                                                                                                        {Scholtes}}]{Lambiotte2019}%
	                                                                                                                                                                                                                                                                                                                                                                                                                                                        \BibitemOpen
	                                                                                                                                                                                                                                                                                                                                                                                                                                                        \bibfield  {author} {\bibinfo {author} {\bibfnamefont {R.}~\bibnamefont
			                                                                                                                                                                                                                                                                                                                                                                                                                                                        {Lambiotte}}, \bibinfo {author} {\bibfnamefont {M.}~\bibnamefont {Rosvall}},\
		                                                                                                                                                                                                                                                                                                                                                                                                                                                        and\ \bibinfo {author} {\bibfnamefont {I.}~\bibnamefont {Scholtes}},\
	                                                                                                                                                                                                                                                                                                                                                                                                                                                        }\bibfield  {title} {\emph {\bibinfo {title} {From networks to optimal
				                                                                                                                                                                                                                                                                                                                                                                                                                                                        higher-order models of complex systems}},\ }\href
	                                                                                                                                                                                                                                                                                                                                                                                                                                                                   {https://doi.org/https://doi.org/10.1038/s41567-019-0459-y} {\bibfield
		                                                                                                                                                                                                                                                                                                                                                                                                                                                                   {journal} {\bibinfo  {journal} {Nat. Phys.}\ ,\ \bibinfo {pages} {1}}
		                                                                                                                                                                                                                                                                                                                                                                                                                                                                   (\bibinfo {year} {2019})}\BibitemShut {NoStop}%
	                                                                                                                                                                                                                                                                                                                                                                                                                                                                 \bibitem [{\citenamefont {Battiston}\ \emph {et~al.}(2014)\citenamefont
		                                                                                                                                                                                                                                                                                                                                                                                                                                                                   {Battiston}, \citenamefont {Nicosia},\ and\ \citenamefont
		                                                                                                                                                                                                                                                                                                                                                                                                                                                                   {Latora}}]{Battiston_2014}%
	                                                                                                                                                                                                                                                                                                                                                                                                                                                                   \BibitemOpen
	                                                                                                                                                                                                                                                                                                                                                                                                                                                                   \bibfield  {author} {\bibinfo {author} {\bibfnamefont {F.}~\bibnamefont
			                                                                                                                                                                                                                                                                                                                                                                                                                                                                   {Battiston}}, \bibinfo {author} {\bibfnamefont {V.}~\bibnamefont {Nicosia}},\
		                                                                                                                                                                                                                                                                                                                                                                                                                                                                   and\ \bibinfo {author} {\bibfnamefont {V.}~\bibnamefont {Latora}},\
	                                                                                                                                                                                                                                                                                                                                                                                                                                                                   }\bibfield  {title} {\emph {\bibinfo {title} {Structural measures for
				                                                                                                                                                                                                                                                                                                                                                                                                                                                                   multiplex networks}},\ }\href
	                                                                                                                                                                                                                                                                                                                                                                                                                                                                              {https://doi.org/https://doi.org/10.1103/PhysRevE.89.032804} {\bibfield
		                                                                                                                                                                                                                                                                                                                                                                                                                                                                              {journal} {\bibinfo  {journal} {Phys. Rev. E}\ }\textbf {\bibinfo {volume}
			                                                                                                                                                                                                                                                                                                                                                                                                                                                                              {89}},\ \bibinfo {pages} {032804} (\bibinfo {year} {2014})}\BibitemShut
	                                                                                                                                                                                                                                                                                                                                                                                                                                                                              {NoStop}%
	                                                                                                                                                                                                                                                                                                                                                                                                                                                                            \bibitem [{\citenamefont {Chaitin}(1969)}]{Chaitin_1969}%
	                                                                                                                                                                                                                                                                                                                                                                                                                                                                              \BibitemOpen
	                                                                                                                                                                                                                                                                                                                                                                                                                                                                              \bibfield  {author} {\bibinfo {author} {\bibfnamefont {G.~J.}\ \bibnamefont
			                                                                                                                                                                                                                                                                                                                                                                                                                                                                              {Chaitin}},\ }\bibfield  {title} {\emph {\bibinfo {title} {On the simplicity
				                                                                                                                                                                                                                                                                                                                                                                                                                                                                              and speed of programs for computing infinite sets of natural numbers}},\
	                                                                                                                                                                                                                                                                                                                                                                                                                                                                              }\href {https://doi.org/10.1145/321526.321530} {\bibfield  {journal}
		                                                                                                                                                                                                                                                                                                                                                                                                                                                                              {\bibinfo  {journal} {J. ACM}\ }\textbf {\bibinfo {volume} {16}},\ \bibinfo
		                                                                                                                                                                                                                                                                                                                                                                                                                                                                              {pages} {407} (\bibinfo {year} {1969})}\BibitemShut {NoStop}%
	                                                                                                                                                                                                                                                                                                                                                                                                                                                                            \bibitem [{\citenamefont {Battiston}\ \emph
		                                                                                                                                                                                                                                                                                                                                                                                                                                                                              {et~al.}(2017{\natexlab{b}})\citenamefont {Battiston}, \citenamefont
		                                                                                                                                                                                                                                                                                                                                                                                                                                                                              {Nicosia}, \citenamefont {Latora},\ and\ \citenamefont
		                                                                                                                                                                                                                                                                                                                                                                                                                                                                              {San~Miguel}}]{Battiston_Axelrod_2017}%
	                                                                                                                                                                                                                                                                                                                                                                                                                                                                              \BibitemOpen
	                                                                                                                                                                                                                                                                                                                                                                                                                                                                              \bibfield  {author} {\bibinfo {author} {\bibfnamefont {F.}~\bibnamefont
			                                                                                                                                                                                                                                                                                                                                                                                                                                                                              {Battiston}}, \bibinfo {author} {\bibfnamefont {V.}~\bibnamefont {Nicosia}},
		                                                                                                                                                                                                                                                                                                                                                                                                                                                                              \bibinfo {author} {\bibfnamefont {V.}~\bibnamefont {Latora}},\ and\ \bibinfo
		                                                                                                                                                                                                                                                                                                                                                                                                                                                                                       {author} {\bibfnamefont {M.}~\bibnamefont {San~Miguel}},\ }\bibfield  {title}
	                                                                                                                                                                                                                                                                                                                                                                                                                                                                                         {\emph {\bibinfo {title} {Layered social influence promotes multiculturality
				                                                                                                                                                                                                                                                                                                                                                                                                                                                                                         in the axelrod model}},\ }\href
	                                                                                                                                                                                                                                                                                                                                                                                                                                                                                         {https://doi.org/https://doi.org/10.1038/s41598-017-02040-4} {\bibfield
		                                                                                                                                                                                                                                                                                                                                                                                                                                                                                         {journal} {\bibinfo  {journal} {Sci. Rep.}\ }\textbf {\bibinfo {volume}
			                                                                                                                                                                                                                                                                                                                                                                                                                                                                                         {7}},\ \bibinfo {pages} {1809} (\bibinfo {year}
		                                                                                                                                                                                                                                                                                                                                                                                                                                                                                         {2017}{\natexlab{b}})}\BibitemShut {NoStop}%
	                                                                                                                                                                                                                                                                                                                                                                                                                                                                                       \bibitem [{\citenamefont {Bell}(lack)}]{Bell1934}%
	                                                                                                                                                                                                                                                                                                                                                                                                                                                                                         \BibitemOpen
	                                                                                                                                                                                                                                                                                                                                                                                                                                                                                         \bibfield  {author} {\bibinfo {author} {\bibfnamefont {E.~T.}\ \bibnamefont
			                                                                                                                                                                                                                                                                                                                                                                                                                                                                                         {Bell}},\ }\bibfield  {title} {\emph {\bibinfo {title}
			                                                                                                                                                                                                                                                                                                                                                                                                                                                                                         {Exponential numbers}},\ }\href {https://doi.org/10.2307/2300300}
	                                                                                                                                                                                                                                                                                                                                                                                                                                                                                                    {\bibfield  {journal} {\bibinfo  {journal} {Am. Math. Monthly}\ }\textbf
		                                                                                                                                                                                                                                                                                                                                                                                                                                                                                                    {\bibinfo {volume} {41}},\ \bibinfo {pages} {411} (\bibinfo {year}
		                                                                                                                                                                                                                                                                                                                                                                                                                                                                                                    {1934})}\BibitemShut {NoStop}%
	                                                                                                                                                                                                                                                                                                                                                                                                                                                                                                  \bibitem [{\citenamefont {Nicosia}\ and\ \citenamefont
		                                                                                                                                                                                                                                                                                                                                                                                                                                                                                                    {Latora}(2015)}]{Nicosia_Latora_2015}%
	                                                                                                                                                                                                                                                                                                                                                                                                                                                                                                    \BibitemOpen
	                                                                                                                                                                                                                                                                                                                                                                                                                                                                                                    \bibfield  {author} {\bibinfo {author} {\bibfnamefont {V.}~\bibnamefont
			                                                                                                                                                                                                                                                                                                                                                                                                                                                                                                    {Nicosia}}\ and\ \bibinfo {author} {\bibfnamefont {V.}~\bibnamefont
			                                                                                                                                                                                                                                                                                                                                                                                                                                                                                                    {Latora}},\ }\bibfield  {title} {\emph {\bibinfo {title} {Measuring and
				                                                                                                                                                                                                                                                                                                                                                                                                                                                                                                    modeling correlations in multiplex networks}},\ }\href
	                                                                                                                                                                                                                                                                                                                                                                                                                                                                                                               {https://doi.org/https://doi.org/10.1103/PhysRevE.92.032805} {\bibfield
		                                                                                                                                                                                                                                                                                                                                                                                                                                                                                                               {journal} {\bibinfo  {journal} {Phys. Rev. E}\ }\textbf {\bibinfo {volume}
			                                                                                                                                                                                                                                                                                                                                                                                                                                                                                                               {92}},\ \bibinfo {pages} {032805} (\bibinfo {year} {2015})}\BibitemShut
	                                                                                                                                                                                                                                                                                                                                                                                                                                                                                                               {NoStop}%
	                                                                                                                                                                                                                                                                                                                                                                                                                                                                                                             \bibitem [{\citenamefont {Musmeci}\ \emph {et~al.}(2017)\citenamefont
		                                                                                                                                                                                                                                                                                                                                                                                                                                                                                                               {Musmeci}, \citenamefont {Nicosia}, \citenamefont {Aste}, \citenamefont
		                                                                                                                                                                                                                                                                                                                                                                                                                                                                                                               {Di~Matteo},\ and\ \citenamefont {Latora}}]{financial_series}%
	                                                                                                                                                                                                                                                                                                                                                                                                                                                                                                               \BibitemOpen
	                                                                                                                                                                                                                                                                                                                                                                                                                                                                                                               \bibfield  {author} {\bibinfo {author} {\bibfnamefont {N.}~\bibnamefont
			                                                                                                                                                                                                                                                                                                                                                                                                                                                                                                               {Musmeci}}, \bibinfo {author} {\bibfnamefont {V.}~\bibnamefont {Nicosia}},
		                                                                                                                                                                                                                                                                                                                                                                                                                                                                                                               \bibinfo {author} {\bibfnamefont {T.}~\bibnamefont {Aste}}, \bibinfo {author}
		                                                                                                                                                                                                                                                                                                                                                                                                                                                                                                                        {\bibfnamefont {T.}~\bibnamefont {Di~Matteo}},\ and\ \bibinfo {author}
		                                                                                                                                                                                                                                                                                                                                                                                                                                                                                                                        {\bibfnamefont {V.}~\bibnamefont {Latora}},\ }\bibfield  {title} {\emph
		                                                                                                                                                                                                                                                                                                                                                                                                                                                                                                               {\bibinfo {title} {The multiplex dependency structure of financial
				                                                                                                                                                                                                                                                                                                                                                                                                                                                                                                               markets}},\ }\href {https://doi.org/10.1155/2017/9586064} {\bibfield
		                                                                                                                                                                                                                                                                                                                                                                                                                                                                                                               {journal} {\bibinfo  {journal} {Complexity}\ }\textbf {\bibinfo {volume}
			                                                                                                                                                                                                                                                                                                                                                                                                                                                                                                               {2017}},\ \bibinfo {pages} {9586064} (\bibinfo {year} {2017})}\BibitemShut
	                                                                                                                                                                                                                                                                                                                                                                                                                                                                                                                          {NoStop}%
	                                                                                                                                                                                                                                                                                                                                                                                                                                                                                                                        \bibitem [{\citenamefont {Santoro}\ \emph {et~al.}(2018)\citenamefont
		                                                                                                                                                                                                                                                                                                                                                                                                                                                                                                                          {Santoro}, \citenamefont {Latora}, \citenamefont {Nicosia},\ and\
		                                                                                                                                                                                                                                                                                                                                                                                                                                                                                                                          \citenamefont {Nicosia}}]{Santoro_2018}%
                                                                                                                                                                                                                                                                                                                                                                                                                                                                                                                            \BibitemOpen
                                                                                                                                                                                                                                                                                                                                                                                                                                                                                                                            \bibfield  {author} {\bibinfo {author} {\bibfnamefont {A.}~\bibnamefont
		                                                                                                                                                                                                                                                                                                                                                                                                                                                                                                                            {Santoro}}, \bibinfo {author} {\bibfnamefont {V.}~\bibnamefont {Latora}},
	                                                                                                                                                                                                                                                                                                                                                                                                                                                                                                                            \bibinfo {author} {\bibfnamefont {G.}~\bibnamefont {Nicosia}},\ and\ \bibinfo
	                                                                                                                                                                                                                                                                                                                                                                                                                                                                                                                                     {author} {\bibfnamefont {V.}~\bibnamefont {Nicosia}},\ }\bibfield  {title}
                                                                                                                                                                                                                                                                                                                                                                                                                                                                                                                                       {\emph {\bibinfo {title} {Pareto optimality in multilayer network growth}},\
                                                                                                                                                                                                                                                                                                                                                                                                                                                                                                                                       }\href {https://doi.org/10.1103/PhysRevLett.121.128302} {\bibfield  {journal}
	                                                                                                                                                                                                                                                                                                                                                                                                                                                                                                                                       {\bibinfo  {journal} {Phys. Rev. Lett.}\ }\textbf {\bibinfo {volume} {121}},\
	                                                                                                                                                                                                                                                                                                                                                                                                                                                                                                                                       \bibinfo {pages} {128302} (\bibinfo {year} {2018})}\BibitemShut {NoStop}%
                                                                                                                                                                                                                                                                                                                                                                                                                                                                                                                                     \bibitem [{Ded()}]{Dedomenico_website}%
                                                                                                                                                                                                                                                                                                                                                                                                                                                                                                                                       \BibitemOpen
                                                                                                                                                                                                                                                                                                                                                                                                                                                                                                                                       \href@noop {} {}\bibinfo {howpublished}
                                                                                                                                                                                                                                                                                                                                                                                                                                                                                                                                                  {\url{https://comunelab.fbk.eu/data.php}}\BibitemShut {NoStop}%
                                                                                                                                                                                                                                                                                                                                                                                                                                                                                                                                                \bibitem [{\citenamefont {Domenico}\ \emph {et~al.}(2015)\citenamefont
	                                                                                                                                                                                                                                                                                                                                                                                                                                                                                                                                                  {Domenico}, \citenamefont {Lancichinetti}, \citenamefont {Arenas},\ and\
	                                                                                                                                                                                                                                                                                                                                                                                                                                                                                                                                                  \citenamefont {Rosvall}}]{DeDomenico_Arenas_Rosvall_2015}%
                                                                                                                                                                                                                                                                                                                                                                                                                                                                                                                                                  \BibitemOpen
                                                                                                                                                                                                                                                                                                                                                                                                                                                                                                                                                  \bibfield  {author} {\bibinfo {author} {\bibfnamefont {M.~D.}\ \bibnamefont
		                                                                                                                                                                                                                                                                                                                                                                                                                                                                                                                                                  {Domenico}}, \bibinfo {author} {\bibfnamefont {A.}~\bibnamefont
		                                                                                                                                                                                                                                                                                                                                                                                                                                                                                                                                                  {Lancichinetti}}, \bibinfo {author} {\bibfnamefont {A.}~\bibnamefont
		                                                                                                                                                                                                                                                                                                                                                                                                                                                                                                                                                  {Arenas}},\ and\ \bibinfo {author} {\bibfnamefont {M.}~\bibnamefont
		                                                                                                                                                                                                                                                                                                                                                                                                                                                                                                                                                  {Rosvall}},\ }\bibfield  {title} {\emph {\bibinfo {title} {Identifying
			                                                                                                                                                                                                                                                                                                                                                                                                                                                                                                                                                  modular flows on multilayer networks reveals highly overlapping organization
			                                                                                                                                                                                                                                                                                                                                                                                                                                                                                                                                                  in social systems.}},\ }\href
                                                                                                                                                                                                                                                                                                                                                                                                                                                                                                                                                             {https://doi.org/https://doi.org/10.1103/PhysRevX.5.011027} {\bibfield
	                                                                                                                                                                                                                                                                                                                                                                                                                                                                                                                                                             {journal} {\bibinfo  {journal} {Phys. Rev. X}\ }\textbf {\bibinfo {volume}
		                                                                                                                                                                                                                                                                                                                                                                                                                                                                                                                                                             {5}},\ \bibinfo {pages} {011027} (\bibinfo {year} {2015})}\BibitemShut
                                                                                                                                                                                                                                                                                                                                                                                                                                                                                                                                                             {NoStop}%
\end{thebibliography}

\newpage

\begin{thebibliography}{1}
	
\bibitem{DeDomenico_Nicosia_2015}
	M.~{De Domenico}, V.~Nicosia, A.~Arenas, V.~Latora, Structural reducibility of
	multilayer networks.
	\newblock {\it Nat. Comm.\/} {\bf 6}, 1--9 (2015).
	
\bibitem{Battiston_2014}
	F.~Battiston, V.~Nicosia, V.~Latora, Structural measures for multiplex
	networks.
	\newblock {\it Phys. Rev. E\/} {\bf 89}, 032804 (2014).
	
\bibitem{Nicosia_Latora_2015}
	V.~Nicosia, V.~Latora, Measuring and modeling correlations in multiplex
	networks.
	\newblock {\it Phys. Rev. E\/} {\bf 92}, 032805 (2015).
	
\bibitem{Battiston_exploration_2016}
	F.~Battiston, V.~Nicosia, V.~Latora, Efficient 
	exploration of multiplex networks
	\newblock {\it New J. Phys.\/} {\bf 18}, 043035 
	(2016).
	
\bibitem{Ward1963hierarchical}
	J.H.~Ward, Hierarchical grouping to optimize
	an objective function
	\newblock{\it J. Am. Stat. Assoc.} {\bf 58}, 
	236 (1963).
	
\bibitem{newman2018networks}
	M.~Newman, {\it Networks\/} (Oxford Univ. Press, 
	2018).
	
\end{thebibliography}
\end{document}